# OPERADS, HOMOTOPY ALGEBRA, AND ITERATED INTEGRALS FOR DOUBLE LOOP SPACES

E. GETZLER AND J.D.S. JONES

Chen's theory of iterated integrals provides a remarkable model for the differential forms on the based loop space $\Omega M$ of a differentiable manifold $M$ (Chen [10]; see also Hain-Tondeur [23] and Getzler-Jones-Petrack [21]). This article began as an attempt to find an analogous model for the complex of differentiable forms on the double loop space $\Omega^2 M$, motivated in part by the hope that this might provide an algebraic framework for understanding two-dimensional topological field theories.

Our approach is to use the formalism of operads. Operads can be defined in any symmetric monoidal category, although we will mainly be concerned with dg-operads (differential graded operads), that is, operads in the category of chain complexes with monoidal structure defined by the graded tensor product. An operad is a sequence of objects $\mathbf{a}(k)$, $k \geq 0$, carrying an action of the symmetric group $\mathbb{S}_k$, with products

$$\mathbf{a}(k) \otimes \mathbf{a}(j_1) \otimes \ldots \otimes \mathbf{a}(j_k) \to \mathbf{a}(j_1 + \cdots + j_k)$$

which are equivariant and associative — we give a precise definition in Section 1.2. An operad such that $\mathbf{a}(k) = 0$ for $k \neq 1$ is a monoid: in this sense, operads are a non-linear generalization of monoids.

If $V$ is a chain complex, we may define an operad with

$$\mathcal{E}_V(k) = \mathrm{Hom}(V^{(k)}, V),$$

where $V^{(k)}$ is the $k$-th tensor power of $V$. The symmetric group $\mathbb{S}_k$ acts on $\mathcal{E}_V(k)$ through its action on $V^{(k)}$, and the structure maps of $\mathcal{E}_V$ are the obvious ones. This operad plays the same role in the theory of operads that the algebra $\mathrm{End}(V)$ does in the theory of associative algebras.

An algebra over an operad $\mathbf{a}$ (or $\mathbf{a}$-algebra) is a chain complex $A$ together with a morphism of operads $\rho: \mathbf{a} \to \mathcal{E}_A$. In other words, $A$ is equipped with structure maps

$$\rho_k: \mathbf{a}(k) \otimes_{\mathbb{S}_k} A^{(k)} \to A$$

which are compatible in a natural sense. The space $\mathbf{a}(k)$ should be thought of as parametrizing all of the ways of combining $k$ elements of an $\mathbf{a}$-algebra $A$ to obtain an element of $A$, while the structure maps of $\mathbf{a}$ express the way in which these operations compose. If the operad $\mathbf{a}$ is such that $\mathbf{a}(k) = 0$ for $k \neq 1$, so that $\mathbf{a}(1)$ is an associative algebra, then an $\mathbf{a}$-algebra is the same thing as an $\mathbf{a}(1)$-module.

Many familiar algebraic structures, such as commutative, associative, Lie and Poisson algebras, can be described by dg-operads. (Restricted Lie algebras and fields are examples of algebraic structures which cannot.) In a sense, discussing algebraic structures without mentioning the associated operads leaves out half the story, since one cannot then formulate theorems involving morphisms of operads.

Many of our results are expressed in the language of Quillen's homotopical algebra [40]. This is a non-linear generalization of homological algebra, allowing the construction of derived functors in categories with some of the structure of homotopy theory. Such categories are called closed model categories, and in Chapter 4, we prove that the category of algebras over a dg-operad $\mathbf{a}$ is a closed model category, provided the $\mathbb{S}_k$-modules $\mathbf{a}(k)$ are flat for all $k > 1$ (of course, this is always true over a field of characteristic 0).

If $\Phi : \mathbf{a} \to \mathbf{b}$ is a map of operads, there is a functor $\Phi_*$ from $\mathbf{a}$-algebras to $\mathbf{b}$-algebras, analogous to base change in categories of modules (Section 1.5). In the special case that $\mathbf{a}$ is an augmented operad with augmentation $\varepsilon$, the functor $\varepsilon_*$ from $\mathbf{a}$-algebras to chain complexes is the functor of indecomposables $Q$.

The homology functor on $\mathbf{a}$-algebras is the left derived functor $\mathsf{L}\varepsilon_*$ of the indecomposables functor $\varepsilon_*$. An almost free $\mathbf{a}$-algebra $\tilde{A}$ is one which is free when we forget the differentials on $\tilde{A}$ and $\mathbf{a}$. The homology $\mathsf{L}\varepsilon_* A$ may be calculated by resolving $A$ by an almost free $\mathbf{a}$-algebra $\tilde{A}$: the homology of $A$ is the homology of the complex of generators $\varepsilon_* \tilde{A}$ of $\tilde{A}$. For associative, commutative and Lie algebras (the last two cases in characteristic zero), this homology theory may be identified, up to a suspension, with Hochschild, Harrison and Lie algebra homology respectively.

One of the main applications of topological operads is to the study of iterated loop spaces; for two excellent overviews of this subject, see Adams [2] and May [34]. Boardman and Vogt [9] introduced a sequence of operads, the little $n$-cubes operads $\mathcal{F}_n(k)$, which have the homotopy type of the configuration space $\mathbb{F}_n(k)$ of $k$ distinct points in $\mathbb{R}^n$. A space with an action of the operad $\mathcal{F}_n$ is called an $E_n$-space, and a connected $E_n$-space has the homotopy type of an $n$-fold loop space. Certain cases are of particular interest: $n = 1$ recovers the theory of $A_\infty$-spaces [49], $n = \infty$ leads to infinite loop spaces, and $n = 2$ is intimately related to the braid groups $\mathbb{B}_k$, since $\mathbb{F}_2(k)/\mathbb{S}_k \simeq K(\mathbb{B}_k, 1)$.

The homology $\mathbf{e}_n(k) = H_\bullet(\mathbb{F}_n(k), K)$ of the topological operad $\mathbb{F}_n$ over a field $K$ of characteristic zero is a dg-operad with trivial differential, which we will view as an algebraic model for the structure of $n$-fold loop spaces. Algebras over $\mathbf{e}_1$ are associative dg-algebras, while $\mathbf{e}_\infty$-algebras are commutative dg-algebras. We adopt the term $n$-algebra for an $\mathbf{e}_n$-algebra.

For $n > 1$, the structure of $n$-algebras was explicitly determined by F. Cohen in his thesis [11]. An $n$-algebra is a commutative dg-algebra $A_\bullet$ with a graded Lie bracket $[a, b]$ of degree $n - 1$ (that is, a graded Lie bracket on the $(n-1)$-fold suspension $\Sigma^{n-1} A$ of $A$) satisfying the Poisson relation

$$[a, bc] = (-1)^{(|a|-1)|b|} b[a, c] + [a, b]c.$$

If $A$ is an $n$-algebra, let $\mathsf{B}_n A$ be the chain complex

$$\mathsf{B}_n A = \bigoplus_{k=1}^{\infty} \Sigma^{n(k-1)} \mathbf{e}_n^*(k) \otimes_{\mathbb{S}_k} A^{(k)}.$$

In particular, $\mathsf{B}_1 A = \bigoplus_{k=1}^{\infty} \Sigma^{-1}(\Sigma A)^{(k)}$ is closely related to the usual bar construction for associative algebras

$$\mathsf{B} A = \bigoplus_{k=0}^{\infty} (\Sigma A)^{(k)},$$

while $\mathsf{B}_2 A$ may be expressed in terms of cohomology of braid groups:

$$\mathsf{B}_2 A = \bigoplus_{k=1}^{\infty} \Sigma^{-2} H^\bullet(\mathbb{B}_k, (\Sigma^2 A)^{(k)}).$$

The functor $\mathsf{B}_n$ takes values in the category of cofree $n$-coalgebras, that is, chain complexes with Lie cobracket of degree 1 and a cocommutative coproduct of degree $n$, satisfying the Poisson co-relation. (The $n$-fold suspension of the dual of an $n$-coalgebra is an $n$-algebra.) The following theorem is one of the main results of this paper.

**Theorem.** *Over a field of characteristic zero, the bar construction $\mathsf{B}_n A$ of an $n$-algebra $A$ is an $n$-coalgebra whose image in the homotopy category of $n$-coalgebras equals the left derived functor $\mathsf{L}\varepsilon_* A$.*

The functor $\mathsf{B}_n$ has an adjoint functor $\Omega_n$, and this pair of functors gives an equivalence of homotopy categories. For $n = 1$ and $n = \infty$, this result is familiar from the work of Quillen: there are homotopy equivalences between the categories of associative dg-algebras and connected coassociative dg-coalgebras, and between the categories of commutative dg-algebras and connected Lie dg-coalgebras [42].

Our strategy for proving the above theorem is to generalize work of Stasheff [49]. He introduced a sequence of convex $(k-2)$-dimensional polyhedra $K(k)$, the associahedra, such that the products $K(k) \times \mathbb{S}_k$ form a topological operad, and proved that a connected space $X$ has the homotopy type of a loop space if $X$ is an algebra over this operad; such a space is called an $A_\infty$-space.

We construct an operad $\mathsf{F}_n$ with the same homotopy type as $\mathcal{F}_n$, such that $\mathsf{F}_n(k)$ is a manifold with corners of dimension $n(k-1) - 1$, generalizing the sequence of polyhedra $K(k)$, which is the case $n = 1$. The construction of $\mathsf{F}_n(k)$ makes use of a compactification of the configuration space $\mathbb{F}_n(k)$, or rather, its quotient by translations and dilatations, due to Fulton and MacPherson [15]. The operad $\mathsf{F}_n$ is obtained by gluing together faces corresponding to components of the free operad generated by the top-dimensional strata of the spaces $\mathsf{F}_n$. (In a sense which we do not make precise in this article, $\mathsf{F}_n$ is a cofibrant operad.)

The theorem that the bar construction $\mathsf{B}_n$ realizes the derived functor $\mathsf{L}\varepsilon_*$ for $n$-algebras now follow from the study of the spectral sequences associated to the manifolds with corners $\mathsf{F}_n(k)$, and an application of Lefschetz duality. This method of proof was inspired by an article of Beilinson-Ginburg [8]; in effect, they considered the case $n = \infty$.

In the language of Ginzburg-Kapranov [22], we show that the operad $\mathbf{e}_n$ is Koszul. In our exposition, we make use of their bar construction for dg-operads, which we view as a functor from dg-operads to dg-cooperads. We also make use of the free operad triple, which we believe is new: this provides a clear explanation for the role of trees in the theory of operads, since the free operad may be expressed as a sum over trees.

Over the integers, the operad $\mathbf{e}_n$ is not well-behaved, and must be replaced by a dg-operad $\mathcal{E}_n$, which we study in Chapter 5: in the special case of $n = 1$, this is just the operad $A_\infty$. Let $S^\bullet(X)$ be the dg-algebra of singular cochains on a topological space $X$, with integral coefficients. By a theorem of Adams [1], if $X$ is simply connected there is a natural isomorphism in homology between the bar construction $\mathsf{B}S^\bullet(X)$ and $S^\bullet(\Omega X)$. Many authors have attempted to find an analogue of this theorem for iterated loop spaces; however, it is not obvious that the bar construction $\mathsf{B}$ can be iterated even twice, since $S^\bullet(X)$ is not a commutative dg-algebra, and thus it is not evident that $\mathsf{B}S^\bullet(X)$ has an associative product. For further iterations, the problem only becomes worse. Nevertheless, Baues [7] has constructed an associative product on $\mathsf{B}S^\bullet(X)$ using Steenrod's $\cup_1$ operation and certain multilinear analogues, which allow him to construct the structure of a dg-bialgebra on $\mathsf{B}S^\bullet(X)$. He shows that if $X$ is 2-connected, the bar complex of this algebra is a model for the cohomology of the double loop space $\Omega^2 X$.

In Chapter 5, we take a different approach to this problem. From a regular $\mathbb{S}_k$-equivariant cell decomposition of the Fulton-MacPherson space $\mathsf{F}_n(k)$ which we define using the lexicographical ordering of points in $\mathbb{R}^n$, we obtain an almost free operad $\mathcal{E}_n$ resolving $\mathbf{e}_n$. We prove that $S^\bullet(X)$ carries a natural $\mathcal{E}_\infty$-algebra structure; in a sequel to this paper, we will prove that if $X$ is $n$-connected, there is natural isomorphism in homology between $\mathsf{L}\varepsilon_* S^\bullet(X)$, where $S^\bullet(X)$ is considered as an $\mathcal{E}_n$-algebra, and $\Sigma^{-n} S^\bullet(\Omega^n X)$. For a different extension of Adam's theorem which uses simplicial methods and thus avoids the introduction of almost free operads, see Smirnov [46], [47].

In Chapter 6, we examine the same questions from the point of view of rational homotopy and iterated integrals. Let $\mathcal{A}^\bullet(M)$ be the dg-algebra of differential forms on a smooth manifold $M$, negatively graded ($A_{-i} = \mathcal{A}^i(M)$) in order that the differential lowers degree. Quillen's and Sullivan's theories of rational homotopy show that if $M$ is a simply connected manifold, the bar construction $\mathsf{B}_\infty \mathcal{A}^\bullet(M)$ provides a model for $\pi^\bullet(M, \mathbb{C}) = \mathrm{Hom}(\pi_\bullet(M), \mathbb{C})$, the dual of the homotopy of $M$ with complex coefficients. As we saw above, the bar construction $\mathsf{B}\mathcal{A}^\bullet(M) \cong \Sigma \mathsf{B}_1 \mathcal{A}^\bullet(M) \oplus \mathbb{C}$ provides a model for the cohomology of the loop space $\Omega M$. In fact, Chen's theory of iterated integrals gives a natural map from $\mathsf{B}\mathcal{A}^\bullet(M)$ to $\mathcal{A}^\bullet(\Omega M)$, inducing an isomorphism in homology if $M$ is simply connected.

Now, the algebra of differential forms $\mathcal{A}^\bullet(M)$ may be considered as a 2-algebra, by imposing on it a vanishing Lie bracket $[a,b]$. In Chapter 6, we construct an iterated integral map from $\Sigma^2 \mathsf{B}_2 \mathcal{A}^\bullet(M)$ to the complex of differential forms $\mathcal{A}^\bullet(\Omega^2 M)$ on the double loop space $\Omega^2 M$. We show that this map induces an isomorphism in homology if $M$ is 2-connected. In fact, it was our discovery of this iterated integral map which led us to the bar complex $\mathsf{B}_2$ on 2-algebras, and to the conjecture that $\mathbf{e}_2$ is Koszul.

The construction of the iterated integral map

$$\Sigma^2 \mathsf{B}_2 \mathcal{A}^\bullet(M) \to \mathcal{A}^\bullet(\Omega^2 M)$$

relies on certain currents on the spaces $\mathbb{C}^k$, which are the principal values of the differential forms

$$\omega_{ij} = \frac{1}{2\pi i} \frac{d(z_i - z_j)}{z_i - z_j},$$

with logarithmic singularities along the divisors $z_i = z_j$. These currents play the role which the fundamental class of the simplex $\Delta^k$ plays in Chen's theory of iterated integrals.

When $n > 2$, there is also an isomorphism in homology between $\Sigma^n \mathsf{B}_n \mathcal{A}^\bullet(M)$ and $\mathcal{A}^\bullet(\Omega^n M)$ for $n$-connected $M$. Although it would be desirable to realize this by means of an iterated integral map, this is a far more difficult problem than the case $n = 2$: as Kontsevich has observed, the configuration spaces $\mathbb{F}_n(k)$ are formal only if $n \in \{1, 2, \infty\}$.

Recently, it has emerged that the deformation theory of homotopy Batalin-Vilkovisky algebras (which are closely related to 2-algebras) is central to string field theory, or at least to its genus zero approximation (Zwiebach [51] and Getzler [19]). In particular, the notion of a homotopy Batalin-Vilkovisky algebra, analogous to the notion of homotopy 2-algebra of Section 4.4, is precisely the algebraic structure induced by genus zero correlation functions on the state space of a two-dimensional topological field theory. The Batalin-Vilkovisky operad $\mathbf{bv}$ is a Koszul operad (in a generalized sense, since $\mathbf{bv}(1)$ is an exterior algebra on a generator of degree 1) and the bar construction for homotopy Batalin-Vilkovisky algebras appears to play a basic role in string field theory.

**Acknowledgments.** We began the study of iterated integrals for double loop spaces at the Max Planck Institute in Bonn in May, 1989, and we are grateful to G. Segal for organizing this stimulating workshop.

It is a pleasure to thank D. Quillen for an inviting the first author to the Mathematical Institute at Oxford University in the winter and spring of 1992, and the SERC and the Sloan Foundation for their support of his visit.

The first author would also like to thank the Master and Fellows of Balliol College, and especially K. Hannabus, for the welcome which they extended him within their College during this period, and to the Coolidge Fund for its support of his residence in the College.

The exposition of this article was profoundly influenced by conversations with V. Ginzburg and M. Kapranov on their work on Koszul operads, and with M. Hopkins on his work on triples and the homotopy theory of algebras.

The first author is also partially suppported by the NSF.

## 1. Operads and Algebras

In this chapter, we give an exposition of the theory of operads in a symmetric monoidal category $\mathcal{C}$. Let $\mathbb{S}$ be the groupoid $\coprod_{k=0}^{\infty} \mathbb{S}_k$ obtained by taking the disjoint union of the symmetric groups $\mathbb{S}_k$. The category of $\mathbb{S}$-objects, or functors from $\mathbb{S}$ to $\mathcal{C}$, is a monoidal category with respect to a certain tensor product, and operads are monoids in this monoidal category. (This tensor product was studied by Joyal [26], who calls $\mathbb{S}$-objects species, and by Smirnov [46].)

Representations of operads are called algebras: for example, associative and commutative algebras are algebras over certain operads $\mathbf{e}_1$ and $\mathbf{e}_\infty$ in the category of vector spaces. We generalize these examples in Section 1.3, introducing the $n$-algebra operads $\mathbf{e}_n$, $1 < n < \infty$, in the category of graded vector spaces, which are related to the homology of configuration spaces of $\mathbb{R}^n$. One of the objectives of this paper is to study the homotopical algebra of the category of algebras over the operads $\mathbf{e}_n$.

In Section 1.4, we construct a triple on the category of $\mathbb{S}$-objects, involving a sum over trees, such that operads are algebras for this triple. This triple may be viewed as a non-linear analogue of the space of tensors on a vector space.

In Section 1.5, we prove that the category of operads has all limits and colimits: the interesting part of the proof is the explicit construction of coproducts, and more generally, pushouts.

In Section 1.6, we prove that the category of algebras over an operad has all limits and colimits: again, the interesting part of the proof is the explicit construction of pushouts. Given a morphism of operads $\Phi : \mathbf{a} \to \mathbf{b}$, we construct a left adjoint $\Phi_*$ of the natural functor $\Phi^*$ from $\mathbf{b}$-algebras to $\mathbf{a}$-algebras, called the direct image: this functor generalize the base change $M \mapsto B \otimes_A M$ along a morphism of algebras $f : A \to B$.

In Section 1.7, we describe the dual theory, of cooperads and their coalgebras. Just as the theory of coalgebras is not perfectly dual to that of algebras, so with cooperads: the difficulty is that in general, the functor of $\mathbb{S}_k$-covariants is left but not right exact. However, in the category of chain complexes over a field of characteristic zero this difficulty does not arise.

**1.1. Triples.** Triples arise in category theory as an abstraction of the notion of an algebraic structure on an underlying space. For example, there are triples associated to such notions as groups, associative algebras, Lie algebras, commutative algebras and modules over a ring.

If $\mathcal{C}$ is a category, we denote by $\mathcal{C}(A, B)$ the set of arrows between objects $A$ and $B$. Similarly, if $\mathcal{C}_1$ and $\mathcal{C}_2$ are categories, we denote by $\mathrm{Cat}(\mathcal{C}_1, \mathcal{C}_2)$ the category of functors from $\mathcal{C}_1$ to $\mathcal{C}_2$.

Let $\mathcal{C}$ be a category, and let $\mathrm{End}(\mathcal{C}) = \mathrm{Cat}(\mathcal{C}, \mathcal{C})$ be the category of endofunctors $R : \mathcal{C} \to \mathcal{C}$, with morphisms the natural transformations. The category $\mathrm{End}(\mathcal{C})$ is a strictly monoidal category, with tensor product given by composition. The unit of $\mathrm{End}(\mathcal{C})$ is the identity functor, which we will denote by $\mathbb{1}$.

Monoids in the monoidal category $\mathrm{End}(\mathcal{C})$ are called triples: let us make this definition more explicit.

**Definition 1.1.** A triple $(\mathsf{T}, \mu, \eta)$ on a category $\mathcal{C}$ is a functor $\mathsf{T} : \mathcal{C} \to \mathcal{C}$, together with natural transformations $\mu : \mathsf{TT} \to \mathsf{T}$ and $\eta : \mathbb{1} \to \mathsf{T}$, such that the following diagrams commute:

$$\begin{array}{ccc} \mathsf{TTT} & \xrightarrow{\mathsf{T}\mu} & \mathsf{TT} \\ \mu\mathsf{T} \downarrow & & \downarrow \mu \\ \mathsf{TT} & \xrightarrow{\mu} & \mathsf{T} \end{array} \qquad \begin{array}{ccccc} \mathsf{T} & \xrightarrow{\mathsf{T}\eta} & \mathsf{TT} & \xleftarrow{\eta\mathsf{T}} & \mathsf{T} \\ & \searrow & \downarrow \mu & \swarrow & \\ & & \mathsf{T} & & \end{array}$$

Triples arise when one has an adjoint pair of functors $U : \mathcal{C} \rightleftarrows \mathrm{Alg} : F$. (In this notation, the functor associated to the right pointing arrow is the left adjoint.) The composition $FU : \mathcal{C} \to \mathcal{C}$ is a triple, with product $\mu = F\varepsilon U : FUFU \to FU$, where $\varepsilon : UF \to \mathbb{1}$ is the counit of the adjunction, while the unit $\eta : \mathbb{1} \to FU$ of the triple is the unit of the adjunction.

In fact, any triple may be factored as $\mathsf{T} = FU$, where $U$ is left adjoint to $F$: this is proved by introducing the category of algebras over the triple, following Eilenberg-Moore (see MacLane [32], Section VI.2).

**Definition 1.2.** An algebra over a triple $(\mathsf{T}, \mu, \eta)$ is a pair $(X, \rho)$ where $X$ is an object of the category $\mathcal{C}$ and $\rho : \mathsf{T}X \to X$ is a morphism, such that the composition $X \xrightarrow{\eta X} \mathsf{T}X \xrightarrow{\rho} X$ is the identity, and the following diagram commutes:

$$\begin{array}{ccc} \mathsf{T}\mathsf{T}X & \xrightarrow{\mathsf{T}\rho} & \mathsf{T}X \\ \mu X \downarrow & & \downarrow \rho \\ \mathsf{T}X & \xrightarrow{\rho} & X \end{array}$$

A morphism $f : (X, \rho_X) \to (Y, \rho_Y)$ of $\mathsf{T}$-modules is a morphism $f : X \to Y$ in $\mathcal{C}$ such that the diagram

$$\begin{array}{ccc} \mathsf{T}X & \xrightarrow{\mathsf{T}f} & \mathsf{T}Y \\ \rho_X \downarrow & & \downarrow \rho_Y \\ X & \xrightarrow{f} & Y \end{array}$$

commutes. The category of $\mathsf{T}$-algebras is denoted $\mathcal{C}^\mathsf{T}$.

If $X$ is an object of $\mathcal{C}$, then $\mathsf{T}X$ is the underlying object of a $\mathsf{T}$-algebra, called the free $\mathsf{T}$-algebra: its structure map $\mu X : \mathsf{T}\mathsf{T}X \to \mathsf{T}X$ is induced by the product $\mu : \mathsf{T}\mathsf{T} \to \mathsf{T}$ of $\mathsf{T}$. We obtain a functor $F^\mathsf{T} : \mathcal{C} \to \mathcal{C}^\mathsf{T}$ by mapping an object $X$ of $\mathcal{C}$ to the free $\mathsf{T}$-algebra $(\mathsf{T}X, \mu X)$; this functor has a left adjoint, the functor $U^\mathsf{T} : \mathcal{C}^\mathsf{T} \to \mathcal{C}$ which sends $(X, \rho)$ to the object $X$, and $\mathsf{T}$ may be factored $\mathsf{T} = U^\mathsf{T} F^\mathsf{T}$.

**1.2. Operads.** From now on, we will restrict attention to a category $\mathcal{C}$ with the following properties:
(1) $\mathcal{C}$ is a symmetric monoidal category with tensor product functor $- \otimes -$ and unit $\mathbb{1}$;
(2) $\mathcal{C}$ has all small limits and colimits;
(3) for any object $X$, the functor $X \otimes -$ preserves colimits.

Denote by 0 the initial object of $\mathcal{C}$, which exists by the assumption that $\mathcal{C}$ has colimits. We will eventually specialize to two categories.
(1) Let $\mathcal{T}$ be the category of compactly generated topological spaces, with tensor product the product of spaces. An operad in $\mathcal{T}$ is called a topological operad (May [33]): this generalizes the notion of a topological monoid.
(2) Let $\mathcal{M}$ be the category of chain complexes $V_\bullet$ over a field $K$ with $V_i = 0$ for $i < 0$, and with tensor product the graded tensor product. An operad in $\mathcal{M}$ is called a differential graded operad (dg-operad): this generalizes the notion of a dg-algebra.

A symmetric monoidal category such as $\mathcal{T}$, in which the tensor product equals the product, is called cartesian.

Denote the $k$-fold tensor product of an object $V$ of $\mathcal{C}$ by $V^{(k)}$; it carries an action of the symmetric group $\mathbb{S}_k$.

Let $\mathbb{S}$ be the symmetric groupoid whose objects are all finite sets, including the empty set, and whose morphisms are bijections between sets. If $k$ is a natural number, we denote the object $\{1, \ldots, k\}$ of $\mathbb{S}$ by $k$: the set of all such objects and their automorphisms form a skeleton of $\mathbb{S}$, the disjoint union of the symmetric groups $\coprod_{k=0}^{\infty} \mathbb{S}_k$.

Let $\mathrm{Cat}(\mathbb{S}, \mathcal{C})$ be the category of functors from $\mathbb{S}$ to $\mathcal{C}$; an object of $\mathrm{Cat}(\mathbb{S}, \mathcal{C})$ is called an $\mathbb{S}$-object, or, in the differential graded case where $\mathcal{C}$ is the category of chain complexes $\mathcal{M}$, an $\mathbb{S}$-module. An $\mathbb{S}$-object **v** determines a sequence **v**$(k)$ of objects in $\mathcal{C}$ with action of $\mathbb{S}_k$, and conversely, from such a sequence of objects we may reconstruct an $\mathbb{S}$-object by setting $\mathbb{S}(S)$ to be the colimit

$$\mathbf{v}(S) = \left( \bigoplus_{f \in \mathbb{S}(k,S)} \mathbf{v}(k) \right)_{\mathbb{S}_k}.$$

To an $\mathbb{S}$-object **v**, we may associate an endofunctor $\mathsf{S}(\mathbf{v})$ on the category $\mathcal{C}$, called the Schur functor associated to **v**, by the formula

$$\mathsf{S}(\mathbf{v}, V) = \bigoplus_{k=0}^{\infty} \mathbf{v}(k) \otimes_{\mathbb{S}_k} V^{(k)};$$

it is here that we have used the hypothesis that $\mathcal{C}$ has small colimits. The functor $\mathsf{S}(\mathbf{v})$ is analytic, with the $\mathbb{S}$-object **v** as its Taylor coefficients.

If **a** and **b** are $\mathbb{S}$-objects, the composition $\mathsf{S}(\mathbf{a})\mathsf{S}(\mathbf{b})$ is again a Schur functor: Joyal [26] and Smirnov [46] give an explicit formula for its Taylor coefficients, which we now recall.

Denote by $\Pi(S,T)$ the set of maps from $S$ to $T$: we think of an element $\pi \in \Pi(S,T)$ as a partition of $S$ into a finite number of disjoint subsets $\{\pi^{-1}(i) \mid i \in T\}$, possibly empty, labelled by elements of $T$. Given an $\mathbb{S}$-object **b** and $\pi \in \Pi(S,T)$, we define

$$\mathbf{b}(\pi) = \bigotimes_{i \in T} \mathbf{b}(\pi^{-1}(i)).$$

Define the tensor product on the category of $\mathbb{S}$-objects by the formula

$$(\mathbf{a} \circ \mathbf{b})(S) = \bigoplus_{k=0}^{\infty} \mathbf{a}(k) \otimes_{\mathbb{S}_k} \left( \bigoplus_{\pi \in \Pi(S,k)} \mathbf{b}(\pi) \right).$$

This tensor product gives $\mathrm{Cat}(\mathbb{S}, \mathcal{C})$ the structure of a monoidal category, with unit $\mathbb{1}$: to see the associativity, we use the fact that the functor $V \otimes -$ preserves colimits. Of course, this monoidal structure is not symmetric.

The inclusion of categories $(\mathcal{C}, \otimes) \hookrightarrow (\mathrm{Cat}(\mathbb{S}, \mathcal{C}), \circ)$ defined by sending an object $V$ of $\mathcal{C}$ to the $\mathbb{S}$-object, also denoted by $V$, such that

$$V(S) = \begin{cases} V, & |S| = 1, \\ 0, & |S| \neq 1, \end{cases}$$

is a monoidal functor; that is, the functor $- \circ -$ on $\mathbb{S}$-objects is an extension of the usual tensor product $- \otimes -$ on $\mathcal{C}$.

**Proposition 1.3.** *The functor* $\mathbf{a} \mapsto \mathsf{S}(\mathbf{a}) : \mathrm{Cat}(\mathbb{S}, \mathcal{C}) \to \mathrm{End}(\mathcal{C})$ *is a monoidal functor:*

$$\mathsf{S}(\mathbf{a})\mathsf{S}(\mathbf{b}) \cong \mathsf{S}(\mathbf{a} \circ \mathbf{b}).$$

The following definition of an operad is due to Smirnov [46].

**Definition 1.4.** An operad is a monoid in the monoidal category $\mathrm{Cat}(\mathbb{S}, \mathcal{C})$, that is, an $\mathbb{S}$-object **a** together with morphisms $\mu : \mathbf{a} \circ \mathbf{a} \to \mathbf{a}$ and $\eta : \mathbb{1} \to \mathbf{a}$ satisfying the associativity and unit axioms. Denote the category of operads in $\mathcal{C}$ by $\mathrm{Op}(\mathcal{C})$.

Note that a morphism of $\mathbb{S}$-objects $\mathbf{a} \circ \mathbf{a} \to \mathbf{a}$ is determined by maps

$$\mathbf{a}(k) \otimes \mathbf{a}(j_1) \otimes \ldots \otimes \mathbf{a}(j_k) \to \mathbf{a}(j_1 + \cdots + j_k),$$

corresponding to the partition

$$\pi(j) = i \text{ if } j_1 + \cdots + j_{i-1} < i \leq j_1 + \cdots + j_i.$$

The equivariance axiom in the original definition of an operad (May [33]) amounts to the assumption that $\mathbf{a} \circ \mathbf{a} \to \mathbf{a}$ is a morphism of $\mathbb{S}$-objects.

If $\mathbf{a}$ is an operad, Proposition 1.3 shows that the associated Schur functor $\mathsf{S}(\mathbf{a})$ is the underlying functor of a triple, which we denote by $\mathsf{T}(\mathbf{a})$. If $\mathbf{a}$ is an operad, an algebra over $\mathsf{T}(\mathbf{a})$ is called an $\mathbf{a}$-algebra; the category of $\mathbf{a}$-algebras will be denoted $\mathcal{C}^{\mathbf{a}}$, and the space of morphisms between two $\mathbf{a}$-algebras $A$ and $B$ will be denoted $\mathrm{Hom}_{\mathbf{a}}(A, B)$. The $\mathbf{a}$-algebra structure is determined by a sequence of maps
$$\rho_k : \mathbf{a}(k) \otimes V^{(k)} \to V$$
invariant under the action of $\mathbb{S}_k$ on $\mathbf{a}(k) \otimes V^{(k)}$ and compatible with the products of $\mathbf{a}$; the space $\mathbf{a}(k)$ parametrizes the $k$-linear products which make $V$ into an $\mathbf{a}$-algebra. If $V$ is an object of $\mathcal{C}$, denote by $[V]$ the $\mathbb{S}$-object
$$[V](S) = \begin{cases} V, & |S| = 0, \\ 0, & |S| \geq 1. \end{cases}$$
Note that $\mathbf{a} \circ [V] \cong [\mathsf{S}(\mathbf{a}, V)]$; thus, if $\mathbf{a}$ is an operad, an $\mathbf{a}$-algebra $A$ is the same thing as a module $[A]$ for the monoid $\mathbf{a}$ in the category of $\mathbb{S}$-objects.

Let $\mathbb{1}$ be the unit for the tensor product $- \otimes -$ in the symmetric monoidal category $\mathcal{C}$, and also denote by $\mathbb{1}$ the corresponding $\mathbb{S}$-object. The $\mathbb{S}$-object $\mathbb{1}$ is an operad in an obvious way, since the functor $\mathsf{S}(\mathbb{1})$ is just the identity functor. An augmented operad is an operad together with a morphism of operads $\varepsilon : \mathbf{a} \to \mathbb{1}$ to the trivial operad.

An operad $\mathbf{a}$ such that $\mathbf{a}(S) = 0$ for $|S| \neq 1$ is the same thing as a monoid in $\mathcal{C}$, and an algebra over such an operad is the same thing as a representation of this monoid: the triple $\mathsf{T}(\mathbf{a})$ is the functor
$$\mathsf{T}(\mathbf{a}, X) = \mathbf{a}(1) \otimes X.$$
A monoid in the category of vector spaces is an associative algebra; thus, the theory of dg-operads is a generalization of the theory of associative algebras.

If $\mathbf{a}$ is an operad, we denote the free $\mathbf{a}$-algebra functor by $F^{\mathbf{a}} : \mathcal{C} \to \mathcal{C}^{\mathbf{a}}$, and its left adjoint, the forgetful functor, by $U^{\mathbf{a}} : \mathcal{C}^{\mathbf{a}} \to \mathcal{C}$.

**1.3. The little $n$-cubes operads.** In this section, we introduce a sequence of operads in the category $\mathcal{M}$ of chain complexes over a field $K$ of characteristic zero, which interpolate between the associative and commutative operads.

Let us state here our conventions on chain complexes. Unless otherwise stated, all chain complexes will be concentrated in positive degree. If $V$ is a chain complex, we denote by $\Sigma V$ its suspension, with
$$(\Sigma V)_i = V_{i-1},$$
and $\delta(\Sigma v) = (-1)^{|v|} \Sigma(\delta v)$. A map $V \to W$ of chain complexes inducing an isomorphism in homology is called a weak equivalence. If $V$ is a chain complex, its dual $V^*$ is the chain complex such that
$$(V^*)_i = (V_{-i})^*.$$
Clearly, if $V$ is concentrated in positive degree, the chain complex $V^*$ is concentrated in negative degree.

If $\mathbf{v}$ is an $\mathbb{S}$-module, denote by $\Lambda \mathbf{v}$ the $\mathbb{S}$-module
$$(\Lambda \mathbf{v})(k) = \Sigma^{1-k} \mathbf{v}(k) \otimes \mathrm{sgn}_k,$$
where $\mathrm{sgn}_k$ is the sign representation of $\mathbb{S}_k$; this definition is motivated by the natural isomorphism
$$\Sigma \mathsf{S}(\mathbf{a}, V) \cong \mathsf{S}(\Lambda \mathbf{a}, \Sigma V).$$

It follows that $(\Lambda \mathbf{v}) \circ (\Lambda \mathbf{w}) \cong \Lambda(\mathbf{v} \circ \mathbf{w})$. Thus, if $\mathbf{a}$ is an operad, then $\Lambda \mathbf{a}$ is again an operad, while if $\mathbf{z}$ is a cooperad, $\Lambda \mathbf{z}$ is again a cooperad. Furthermore, a $\Lambda \mathbf{a}$-algebra structure on a chain complex $A$ is the same thing as an $\mathbf{a}$-algebra structure on $\Sigma A$.

We will say that a map $f : V \to W$ of chain complexes is a weak equivalence if it induces an isomorphism in homology (this is sometimes called a quasi-isomorphism). Similarly, a map $\Phi : \mathbf{a} \to \mathbf{b}$ of $\mathbb{S}$-modules is a weak equivalence if $\Phi(S) : \mathbf{a}(S) \to \mathbf{b}(S)$ is a weak equivalence for all finite sets $S$. If $\Phi : \mathbf{a} \to \mathbf{b}$ is a map of operads, $\Phi$ is a weak equivalence if the underlying map of $\mathbb{S}$-modules is a weak equivalence.

Denote by $\mathbb{F}_X(S)$ the configuration space of embeddings of the finite set $S$ into the topological space $X$,
$$\mathbb{F}_X(S) = \{x \in X^S \mid x(s) \neq x(t) \text{ for } s \neq t\}.$$
As $S$ varies over all finite sets, we obtain an $\mathbb{S}$-space $\mathbb{F}_X$ on which $\mathbb{S}$ acts freely. Abbreviate $\mathbb{F}_{\mathbb{R}^n}$ by $\mathbb{F}_n$.

Denote by $\mathbf{e}_n^+$ the $\mathbb{S}$-module $S \mapsto H_\bullet(\mathbb{F}_n(S), K)$, and by $\mathbf{e}_n$ the $\mathbb{S}$-module
$$\mathbf{e}_n(S) = \begin{cases} \mathbf{e}_n^+(S), & |S| > 0, \\ 0, & |S| = 0. \end{cases}$$

The $\mathbb{S}$-modules $\mathbf{e}_n^+$ and $\mathbf{e}_n$ have natural structures of operads, as we will now show. Call an algebra for the operad $\mathbf{e}_n$ an $n$-algebra, and denote the associated triple by $\mathsf{T}_n$; call an algebra for the operad $\mathbf{e}_n^+$ unital $n$-algebra, and denote the associated triple by $\mathsf{T}_n^+$. It is easily seen that a 1-algebra is a differential graded algebra in the usual sense, while in characteristic zero, an $\infty$-algebra is a commutative dg-algebra.

In order to define the operad structure of $\mathbf{e}_n^+$, we will construct a topological operad $\mathcal{F}_n$ such that the space $\mathcal{F}_n(k)$ is homotopy equivalent to $\mathbb{F}_n(k)$: this operad is Boardman and Vogt's "little $n$-cubes operad" [9]. Let $I$ denote the open subset $(-1, 1)$ of $\mathbb{R}$. If $S$ is a finite set, let $\mathcal{F}_n(S)$ be the space of all maps
$$\coprod_{i \in S} d(i) : \coprod_{i \in S} I^n \to I^n$$
such that each map $d(i) : I^n \to I^n$ is a composition of a translation and a dilatation, and the maps $d(i)$ have disjoint images. The map from $\mathcal{F}_n(k)$ to the configuration space $\mathbb{F}_n(k)$ defined by sending a map $I^n \to I^n$ to the image of the point $(0, \ldots, 0) \in I^n$ is a homotopy equivalence, since it is a fibration with contractible fibres (May [33], page 34). Thus $H_\bullet(\mathcal{F}_n(k)) \cong \mathbf{e}_n^+(k)$.

The operad structure of $\mathcal{F}_n$ is given by maps
$$\mathcal{F}_n(T) \times \prod_{i \in T} \mathcal{F}_n(\pi^{-1}(i)) \to \mathcal{F}_n(S), \quad \pi \in \Pi(S, T),$$
defined for a set of disjoint cubes $c \in \mathcal{F}_n(T)$ labelled by the set $T$ and sets of disjoint cubes $c_i \in \mathcal{F}_n(\pi^{-1}(i))$ labelled by the set $\pi^{-1}(i)$ by attaching to the label $j \in \pi^{-1}(i) \subset S$ the cube obtained by composing the embeddings $d_i(j) : I^n \to I^n$ and $c(i) : I^n \to I^n$. Taking the homology of the little $n$-cubes operad $\mathcal{F}_n$, we see that $\mathbf{e}_n^+$ is an operad in the category of graded vector spaces.

Arnold [3] and F. Cohen [11] have calculated $H_\bullet(\mathbb{F}_n(k), \mathbb{Z})$ for $n = 2$, and $n \geq 2$, respectively: the answer is summarized in the following proposition, proved by induction on $k$.

**Proposition 1.5.** *Let $1 < n < \infty$. For $1 \leq i \neq j \leq k$, let $\omega_{ij} \in H^{n-1}(\mathbb{F}_n(k), \mathbb{Z})$ be the inverse image of the generator of $H^{n-1}(S^{n-1}, \mathbb{Z})$ under the map $\mathbb{F}_n(k) \to S^{n-1}$ defined by the formula*
$$(x_1, \ldots, x_k) \mapsto \frac{x_i - x_j}{|x_i - x_j|}.$$
*The homology group $H_\bullet(\mathbb{F}_n(k), \mathbb{Z})$ is torsion free, and its dual $H^\bullet(\mathbb{F}_n(k), \mathbb{Z})$ is the graded commutative ring with generators $\omega_{ij} \in H^{n-1}(\mathbb{F}_n(k), \mathbb{Z})$, and relations*

(1) $\omega_{ji} = (-1)^n \omega_{ij}$;
(2) $\omega_{ij}\omega_{jk} + \omega_{jk}\omega_{ki} + \omega_{ki}\omega_{ij} = 0$;
(3) $\omega_{ij}^2 = 0$, *for $n$ odd*.

*The symmetric group $\mathbb{S}_k$ acts on $H^\bullet(\mathbb{F}_n(k), \mathbb{Z})$ by $\pi \cdot \omega_{ij} = \omega_{\pi(i)\pi(j)}$.*

In the special case $n = 2$, the cohomology class $\omega_{ij}$ is represented in de Rham cohomology by the closed one-form
$$\omega_{ij} = \frac{1}{2\pi i} \frac{d(z_i - z_j)}{z_i - z_j},$$
where $z_i$ is the complex coordinate of the $i$-th point. The differential forms $\omega_{ij}$ satisfy the above relations without any correction terms involving exact differential forms — there is no choice of closed forms $\omega_{ij}$ for which this is true for $n > 2$ (Kontsevich, unpublished).

The Poincaré series of a dg-operad $\mathbf{a}$ is the power series in two variables
$$g_{\mathbf{a}}(x, t) = \sum_{k=0}^\infty \frac{x^k}{k!} \sum_{i=0}^\infty t^i \dim H_i(\mathbf{a}(k)).$$

For example, the Poincaré series of the operads $\mathbf{e}_1$ and $\mathbf{e}_\infty$ are given by
$$g_{\mathbf{e}_1}(x, t) = \frac{x}{1-x}, \qquad g_{\mathbf{e}_\infty}(x, t) = \exp(x) - 1.$$

Using Proposition 1.5, we see that the Poincaré series of the operad $\mathbf{e}_n$ equals
$$\begin{aligned} g_{\mathbf{e}_n}(x, t) &= \sum_{k=1}^\infty (1 + t^{n-1})(1 + 2t^{n-1})\ldots(1 + (k-1)t^{n-1}) \frac{x^k}{k!} \\ &= \sum_{k=1}^\infty (-t^{1-n})(-t^{1-n} - 1)(-t^{1-n} - 2)\ldots(-t^{1-n} - (k-1)) \frac{(-t^{n-1}x)^k}{k!} \\ &= \sum_{k=1}^\infty \binom{-t^{1-n}}{k}(-t^{n-1}x)^k = (1 - t^{n-1}x)^{-t^{1-n}} - 1. \end{aligned}$$

As a final example, let $\mathcal{L}$ be the Lie operad: a basis of $\mathcal{L}(k)$ is given by the words in the free Lie algebra generated by letters $a_1, \ldots, a_k$ involving each letter exactly once. The vector space $\mathcal{L}(k)$ has dimension $(k-1)!$, and thus the Poincaré series of the Lie operad equals
$$g_{\mathcal{L}}(x, t) = -\log(1-x).$$

Let us make explicit the structure of an $n$-algebra. A Lie bracket of degree $m$ on a chain complex $V$ is defined to be a Lie bracket on $\Sigma^m V$, that is, a bracket $[-,-] : V_i \otimes V_j \to V_{i+j+m}$ such that
$$\begin{aligned} [v, w] &= -(-1)^{(|v|+m)(|w|+m)}[w, v], \\ \delta[v, w] &= [\delta v, w] + (-1)^{|v|+m}[v, \delta w], \\ [u, [v, w]] &= [[u, v], w] + (-1)^{(|u|+m)(|v|+m)}[v, [u, w]]. \end{aligned}$$

A Poisson algebra of degree $m$ is a graded commutative algebra together with a graded Lie bracket $[u, v]$ of degree $m$ satisfying the Poisson relation
$$[u, vw] = [u, v]w + (-1)^{(|u|+m)|v|} v[u, w].$$

Let $\mathbf{p}_n$ be operad whose algebras are Poisson algebras of degree $n-1$: $\mathbf{p}_n(k)$ is the vector space spanned by the words in letters $a_1, \ldots, a_k$ constructed by means of the two products $ab$ and $[a, b]$, involving each letter exactly once, and such a word is assigned a degree equal to $n-1$ times the number of bracket operations in it. We call $\mathbf{p}_n$ the $n$-Poisson operad.

The following result is contained in the thesis of F. Cohen [11]; we will give a different proof in Chapter 3, generalizing work of Beilinson and Ginzburg [8].

**Theorem 1.6.** *If $n \geq 2$, the operads $\mathbf{e}_n$ and $\mathbf{p}_n$ are naturally isomorphic.*

The two products on an $n$-algebra correspond to the two homology classes in $H_\bullet(\mathbb{F}_n(2)) \cong H_\bullet(S^{n-1})$, since $\mathbb{F}_n(2) \simeq S^{n-1}$. The symmetry or antisymmetry of these two products is determined by the action of the involution in $\mathbb{S}_2$ on $H_\bullet(S^{n-1})$. The operad $\mathbf{e}_n$ is quadratic, in the sense of Ginzburg and Kapranov [22] (see Section 2.4): associativity, the Jacobi relation and the Poisson relation are quadratic in the two operations.

We can also show that the unital analogue $\mathbf{e}_n^+$ of the operad $\mathbf{e}_n$ is isomorphic to the unital version of the $n$-Poisson operad. If 1 is the distinguished element of $\mathbf{e}_n^+(0)$, we must show that $1a = a1 = a$ and that $\{1, a\} = \{a, 1\} = 0$. The first formula follows from the fact that the commutative product is represented by any point of $\mathcal{F}_n(2)$, thought of as a 0-chain, while the identity is represented by any point of $\mathcal{F}_n(1)$. The second formula is clear, since any element of degree $n - 1$ of $\mathbf{e}_n(1)$ must vanish.

Denote by $\mathsf{T}_\mathcal{L}$ the free Lie algebra functor $\mathsf{T}(\mathcal{L}, V)$.

**Corollary 1.7.** *There is an isomorphism of functors $\mathsf{T}_n(V) \cong \mathsf{T}_\infty(\Sigma^{1-n}\mathsf{T}_\mathcal{L}(\Sigma^{n-1}V))$.*

*Proof.* For $n = 1$, this is the Poincaré-Birkhoff-Witt theorem, while for $n > 1$, it is an immediate consequence of Theorem 1.6. □

An entirely different approach to Theorem 1.6 when $n = 2$ is contained in the work of Schechtman and Varchenko [45]: they prove the more general result, that if $S$ is a finite subset of the complex plane, then

$$\bigoplus_{k=0}^\infty H_\bullet(\mathbb{F}_{\mathbb{C}\setminus S}(k)) \otimes_{\mathbb{S}_k} V^{(k)} \cong \mathsf{T}_\infty^+(\Sigma^{-1}\mathsf{T}_\mathcal{L}(\Sigma V)) \otimes \mathsf{T}_1^+(\Sigma V)^{(S)}.$$

It also follows from Theorem 1.6 that for any $n > 1$, the operad $\mathcal{L}$ may be identified with the operad consisting of the components in zero degree of $\Lambda^{n-1}\mathbf{e}_n$: thus $\mathcal{L}(k)$ may be identified with the top-dimensional homology group $H_{(k-1)(n-1)}(\mathbb{F}_n(k))$, tensored with the $(n-1)$-th power of the sign representation $\mathrm{sgn}_k$.

We also see that the augmentation $A^+$ of an $n$-algebra is a unital $n$-algebra in a natural way: if 1 is the basis vector of the subspace $\mathbb{1} \subset A^+$, we define the product with 1 to be the identity, and the bracket with 1 to be zero.

The following lemma provides evidence for the validity of Theorem 1.6, and is used in our proof of this theorem in Chapter 3.

**Lemma 1.8.** *The Poincaré series of the operads $\mathbf{p}_n$ and $\mathbf{e}_n$ are equal.*

*Proof.* Using the Poisson relation, we see that, for $k > 0$, the words in $\mathbf{p}_n(k)$ may be written as a commutative product of the form $A_1 \ldots A_m$, where $\pi \in \Pi(k, m)$ and $A_i$ is a word in the letters $\{a_j \mid \pi(j) = i\}$ involving only the bracket. Taking into account the fact that such the monomial $A_i$ has degree $(n - 1)(|\pi^{-1}(i)| - 1)$, we see that the Poincaré series of $\mathbf{p}_n$ equals the exponential of $t^{1-n}g_\mathcal{L}(t^{n-1}x)$, where $g_\mathcal{L}(x) = -\log(1 - x)$ is the Poincaré series of the Lie operad $\mathcal{L}$. In other words,

$$g_{\mathbf{p}_n}(x, t) = e^{-t^{1-n}\log(1-t^{n-1}x)} - 1,$$

which does indeed equal the Poincaré series of $\mathbf{e}_n$. □

The usual Poisson operad is obtained by setting $n = 1$ in the definition of $\mathbf{p}_n$. Of course, the operads $\mathbf{e}_1$ and $\mathbf{p}_1$ are not isomorphic: associative and Poisson algebras are quite different. In Section 5.3, we will show that $\mathbf{e}_1$ is naturally a filtered operad, and that the associated graded operad is the Poisson operad.

The homology of an $n$-fold loop space $H_\bullet(\Omega^n X, K)$ over a field of characteristic zero is an $n$-algebra: the product is the Pontryagin product, and comes from the H-space structure of $\Omega^n X$, while the Lie bracket is the Browder operation. (For more details, see Cohen [11].) Furthermore, Cohen shows that $H_\bullet(\Omega^n\Sigma^n X, K)$ is the free unital $n$-algebra $\mathsf{T}_n^+(\tilde{H}_\bullet(X))$ generated by the reduced homology of $X$: this is familiar in the special cases $n=1$ and $n=\infty$.

The cases $n=1$, $2$ and $\infty$ are of particular interest. The configuration space $\mathbb{F}_1(k)$ is a disjoint union of contractible spaces; the components correspond to permutations $\sigma \in \mathbb{S}_k$ by the rule

$$\sigma \mapsto \{(x_{\sigma(1)}, \ldots, x_{\sigma(k)}) \in \mathbb{R}^k \mid x_{\sigma(1)} < \cdots < x_{\sigma(k)}\}.$$

It follows that the underlying $\mathbb{S}$-module of $\mathbf{e}_1^+$ is the free $\mathbb{S}$-module $\mathbf{e}_1^+(k) = K[\mathbb{S}_k]$. The associated triple is the space of tensors

$$\mathsf{T}_1^+(V) = \bigoplus_{k=0}^\infty V^{(k)}.$$

Algebras over the operad $\mathbf{e}_1^+$ are just unital associative algebras; the space $\mathsf{T}_1^+(V)$ is of course the vector space underlying the free unital associative algebra generated by $V$. Indeed, the distinct ways of multiplying elements $(x_1, \ldots, x_k)$ of a unital associative algebra are parametrized by elements of $\mathbb{S}_k$: to $\sigma \in \mathbb{S}_k$, we associate the multiple product $x_{\sigma(1)} \ldots x_{\sigma(k)}$. The natural transformation $\mu : \mathsf{T}_1^+ \mathsf{T}_1^+ \to \mathsf{T}_1^+$ is defined by erasing parentheses. Finally, the unit $\eta$ is given by the canonical inclusion of the vector space $V$ in $\mathsf{T}_1^+(V)$.

The configuration spaces of $\mathbb{R}^\infty$ are contractible. It follows that the underlying $\mathbb{S}$-module of $\mathbf{e}_\infty^+$ has $\mathbf{e}_\infty^+(k) = \mathbb{1}$, where $\mathbb{1}$ is the trivial representation of $\mathbb{S}_k$. The associated triple is the space of symmetric tensors on $V$

$$\mathsf{T}_\infty^+(V) = \bigoplus_{k=0}^\infty (V^{(k)})_{\mathbb{S}_k}.$$

Algebras over the operad $\mathbf{e}_\infty^+$ are just unital commutative algebras; the symmetric tensor space $\mathsf{T}_\infty^+(V)$ is of course the vector space underlying the free unital commutative algebra generated by $V$. The structure maps $\mu : \mathsf{T}_\infty^+ \mathsf{T}_\infty^+ \to \mathsf{T}_\infty^+$ and $\eta : \mathbb{1} \to \mathsf{T}_\infty^+$ are defined in much the same way as for the triple $\mathsf{T}_1^+$.

We now turn to the case $n=2$. The braid group on $k$ strands $\mathbb{B}_k$ is the fundamental group of the quotient space $\mathbb{F}_2(k)/\mathbb{S}_k$. The pure braid group $\mathbb{P}_k$ on $k$ strands is defined to be the fundamental group of $\mathbb{F}_2(k)$. Thus, there is an exact sequence of groups

$$1 \to \mathbb{P}_k \to \mathbb{B}_k \to \mathbb{S}_k \to 1.$$

By induction on $k$, we see that $\mathbb{F}_2(k)$ is an Eilenberg-Maclane space. It follows that, at least over a field of characteristic zero, the triple associated to $\mathbf{e}_2^+$ has the following expression in terms of the braid groups $\mathbb{B}_k$:

$$\mathsf{T}_2^+(V) = \bigoplus_{k=0}^\infty H_\bullet(\mathbb{B}_k, V^{(k)}).$$

Here, the braid group $\mathbb{B}_k$ acts on the tensor power $V^{(k)}$ through its quotient the symmetric group.

**1.4. Free operads.** We will now construct a triple $\mathbb{T}$ on the category of $\mathbb{S}$-objects such that an operad is an algebra over this triple. We use what Boardman and Vogt call "the language of trees."

We work with graphs which are not necessarily compact: an edge of a graph may be terminated by a vertex at only one end (or none). Such an edge is called external. If $\mathcal{S}$ is a graph, its set of vertices is denoted $\mathrm{v}(\mathcal{S})$.

An orientation of a graph is an orientation of each edge. If $s$ is a vertex of an oriented graph, we denote by $\mathrm{in}(s)$ and $\mathrm{out}(s)$ the sets of incoming and outgoing edges incident to $s$. We also denote by $\mathrm{in}(\mathcal{S})$ and $\mathrm{out}(\mathcal{S})$ the sets of incoming and outgoing external edges of $\mathcal{S}$. (If $\mathcal{S}$ is the unique graph with one external edge and no vertices, then $\mathrm{in}(\mathcal{S}) = \mathrm{out}(\mathcal{S})$; in all other cases, these two sets are disjoint.)

**Definition 1.9.** A tree $\mathcal{S}$ is an oriented, contractible graph, such that out($\mathcal{S}$) has one element (the root of the tree), and out($s$) has one element for each vertex $s \in$ v($\mathcal{S}$).

An ordered tree is a tree $\mathcal{S}$ together with a total order on its set v($\mathcal{S}$) of vertices.

A tree on a finite set $S$ is a tree $\mathcal{S}$ together with a bijection between in($\mathcal{S}$) and $S$.

The cardinality $|\operatorname{in}(s)|$ of the set of incoming edges incident to a vertex $s$ is called its valence. We have the formula
$$\sum_{s \in \mathcal{S}}(|\operatorname{in}(s)| - 1) = |\operatorname{in}(\mathcal{S})| - 1.$$
A parent of a vertex $t$ is a vertex $s$ whose outgoing edge $e$ terminates at $t$. A vertex with no parents is called maximal: this is equivalent to in($t$) $\subset$ in($\mathcal{S}$).

Denote the infinite set of isomorphism classes of trees on $S$ by $\mathcal{T}(S)$; it is a functor from $\mathbb{S}$ to the category of sets.

If $\mathcal{S}$ is a tree with no vertices of valence 0 or 1, we say that $\mathcal{S}$ is a nest. To a vertex of a nest, we may associate the set of elements of $\in (\mathcal{S})$ lying above it. The resulting collection n($\mathcal{S}$) of subsets of in($\mathcal{S}$) is a nest in the sense of Fulton and MacPherson [15]:

(1) if $S \in$ n($\mathcal{S}$), $|S| \geq 2$;
(2) for any two sets $S_1, S_2 \in$ n($\mathcal{S}$), either $S_1 \cap S_2 = \emptyset$ or $S_1 \subset S_2$ or $S_2 \subset S_1$;
(3) in($\mathcal{S}$) $\in$ n($\mathcal{S}$).

The set of nests $\mathcal{N}(S)$ on a finite set $S$ is finite.

We often employ a graphical notation to denote trees, with the incoming edges at the top and the root at the bottom: for example, here are two trees with $\in (\mathcal{S}) = 5$, the first of which is a nest and the second of which is not.

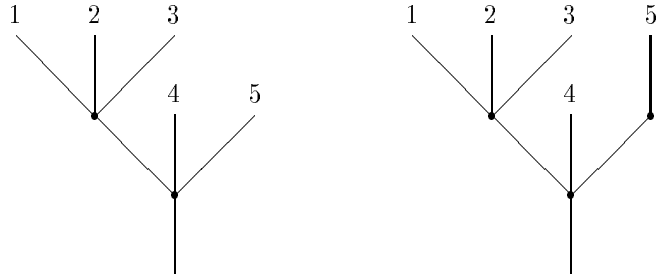

The first of these trees has one maximal vertex, while the second has two.

The number of trees with $k$ incoming external edges and a given underlying graph may be counted by dividing $k!$ by the number of automorphisms of the graph which fix the root. As an example, we may enumerate the 26 nests with four incoming external edges:

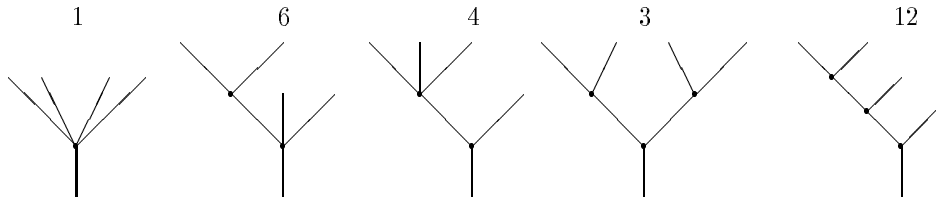

If $\pi \in \Pi(S, T)$ is a partition of the set $S$ into disjoint non-empty subsets labelled by $T$, and $\mathcal{S}_0 \in \mathcal{T}(T)$ and $\mathcal{S}_i \in \mathcal{T}(\pi^{-1}(i))$ for $i \in T$, define the tree $\mathcal{S}_0[\mathcal{S}_1, \ldots, \mathcal{S}_k] \in \mathcal{T}(S)$ by attaching the tree $\mathcal{S}_i$ to the external edges of $\mathcal{S}_0$ labelled by $i \in T$. This operation is called grafting. If the trees $\mathcal{S}_i$ are ordered, then the grafted tree $\mathcal{S}_0[\mathcal{S}_1, \ldots, \mathcal{S}_k]$ has a natural order, obtained by defining vertices of $\mathcal{S}_i$ to come before vertices of $\mathcal{S}_j$ if $i < j$.

If $\mathbf{v}$ is an $\mathbb{S}$-object in a symmetric monoidal category $\mathcal{C}$ and $\mathcal{S}$ is an ordered tree on $S$, let $\mathbf{v}(\mathcal{S}) = \bigotimes_{s \in V(\mathcal{S})} \mathbf{v}(\operatorname{in}(s))$; the ordering on the vertices of $\mathcal{S}$ is needed so that we know in what order to take the tensor products. If $\mathcal{S}_1$ and $\mathcal{S}_2$ are two ordered trees with the same underlying tree but different orderings of the vertices, then $\mathbf{v}(\mathcal{S}_1)$ and $\mathbf{v}(\mathcal{S}_2)$ are naturally isomorphic, since $\mathcal{C}$ is a symmetric monoidal category. If $\mathcal{S}$ is the tree with no vertices and one external edge, $\mathbf{v}(\mathcal{S}) \cong \mathbb{1}$.

Let $\mathbf{v}$ be an $\mathbb{S}$-object in a symmetric monoidal category $\mathcal{C}$. The $\mathbb{S}$-object of tensors $\mathbb{T}\mathbf{v}$ over $\mathbf{v}$ is defined to be the sum over trees

$$\mathbb{T}\mathbf{v}(S) = \bigoplus_{\mathcal{S} \in \mathcal{T}(S)} \mathbf{v}(\mathcal{S}),$$

where an ordering is chosen for each tree. There is a more invariant way of defining $\mathbb{T}\mathbf{v}(S)$: if $\mathcal{T}_{\mathrm{ord}}(S)$ is the category whose objects are ordered trees on $S$ and whose morphisms are pairs $(\mathcal{S}_1, \mathcal{S}_2)$ of ordered trees with the same underlying tree, then $\mathbb{T}\mathbf{v}(S)$ is the colimit of the functor from $\mathcal{T}_{\mathrm{ord}}(S)$ to $\mathcal{C}$ which sends an ordered tree to $\mathbf{v}(\mathcal{S})$.

Note that if $V$ is an object of $\mathcal{C}$, which we may consider to be an $\mathbb{S}$-object as in Section 1.2, then

$$\mathbb{T}V \cong \bigoplus_{k=0}^{\infty} V^{(k)}$$

is just the usual space of tensors on $V$. It is for this reason that we think of $\mathbb{T}\mathbf{v}$ as generalizing the notion of tensors to the category of $\mathbb{S}$-objects.

If $\mathbf{v}$ is an $\mathbb{S}$-object, the $\mathbb{S}$-object $\mathbb{T}\mathbf{v}$ may be given an operad structure in a natural way. The unit $\eta : \mathbb{1} \to \mathbb{T}\mathbf{v}$ is given by the injection of the summand $\mathbb{1}$ of $\mathbb{T}\mathbf{v}(1)$, while the product $\mu : (\mathbb{T}\mathbf{v}) \circ (\mathbb{T}\mathbf{v}) \to \mathbb{T}\mathbf{v}$ is defined by grafting of trees. The summands of $((\mathbb{T}\mathbf{v}) \circ (\mathbb{T}\mathbf{v}))(S)$ are labelled by a partition $\pi \in \Pi(S, k)$ of $S$, a tree $\mathcal{S}_0 \in \mathcal{T}(k)$ with $k$ leaves, and $k$ trees $\mathcal{S}_i \in \mathcal{T}(\pi_i)$, $1 \leq i \leq k$, labelled by the sets $\pi_i = \pi^{-1}(i) \subset S$. Such a summand is isomorphic to

$$\mathbf{v}(\mathcal{S}_0) \otimes \mathbf{v}(\mathcal{S}_1) \otimes \ldots \otimes \mathbf{v}(\mathcal{S}_k),$$

which in turn is naturally isomorphic to the summand $\mathbf{v}(\mathcal{S}_0[\mathcal{S}_1, \ldots, \mathcal{S}_k])$ of $\mathbb{T}\mathbf{v}$: the product $\mu : (\mathbb{T}\mathbf{v}) \circ (\mathbb{T}\mathbf{v}) \to \mathbb{T}\mathbf{v}$ is defined by means of this identification.

We will denote this operad by $\mathbb{T}\mathbf{v}$, and the free operad functor by $F^{\mathbb{T}}$. This construction generalizes the construction of the tensor algebra on a vector space $V$.

**Proposition 1.10.** *Given an operad $\mathbf{a}$ and a map $f : \mathbf{v} \to \mathbf{a}$ of $\mathbb{S}$-objects, there is a map of operads $h : \mathbb{T}\mathbf{v} \to \mathbf{a}$ such that the following diagram commutes:*

$$\begin{array}{ccc} \mathbf{v} & \xrightarrow{h} & \mathbb{T}\mathbf{v} \\ & \searrow{\eta\mathbf{v}} & \downarrow{f} \\ & & \mathbf{a} \end{array}$$

*Proof.* It suffices to show that for an operad $\mathbf{a}$, there is a natural map of operads $\mathbb{T}\mathbf{a} \to \mathbf{a}$, its structure map. Given a tree $\mathcal{S} \in \mathcal{T}(S)$, choose a total ordering of the vertices $\mathrm{v}(\mathcal{S})$. We define the map $\mathbb{T}\mathbf{a}(S) \to \mathbf{a}(S)$ by mapping an element

$$\bigotimes_{s \in \mathrm{v}(\mathcal{S})} a(s) \in \mathbf{a}(\mathcal{S}) \subset \mathbb{T}\mathbf{a}(S),$$

to the element of $\mathbf{a}(S)$ obtained by successively applying the product map of $\mathbf{a}$ to the elements $a(s)$: we start at the maximal vertices and work our way down the tree. It is easily seen that this map is equivariant, and sends the result of the grafting operation in $\mathbb{T}\mathbf{a}$ to the product in $\mathbf{a}$; hence, it does define a map of operads. □

**Corollary 1.11.** *The free operad functor $F^{\mathbb{T}} : \mathrm{Cat}(\mathbb{S}, \mathcal{C}) \to \mathrm{Op}(\mathcal{C})$ is the right adjoint of the underlying $\mathbb{S}$-object functor $U^{\mathbb{T}} : \mathrm{Op}(\mathcal{C}) \to \mathrm{Cat}(\mathbb{S}, \mathcal{C})$.*

It follows from this corollary that the functor

$$\mathbb{T} = U^{\mathbb{T}} F^{\mathbb{T}} : \text{Cat}(\mathbb{S}, \mathcal{C}) \to \text{Cat}(\mathbb{S}, \mathcal{C})$$

has the natural structure of a triple.

**Proposition 1.12.** *The category of algebras for the triple $\mathbb{T}$ is naturally isomorphic to the category of operads.*

*Proof.* The result follows from the Crude Tripleablity Theorem (MacLane [32], Exercise VI.7.7), which says that any functor $U : \text{Alg} \to \mathcal{C}$ satisfying the following three conditions is tripleable:

(1) $U$ has a left adjoint;
(2) $U$ reflects isomorphisms;
(3) Alg has coequalizers of reflexive pairs $(f, g) : X \to Y$ (pairs with a common right inverse) such that $(Uf, Ug)$ has a coequalizer, and $U$ preserves such coequalizers.

We have shown that the functor $U^{\mathbb{T}} : \text{Op}(\mathcal{C}) \to \text{Cat}(\mathbb{S}, \mathcal{C})$ has a left adjoint. The other two conditions are obvious. □

**1.5. Limits and colimits of operads.** In this section, we will show that the category of operads has all limits and colimits. This follows from the following theorem, which is a rather special case of the results of Section 9.3 of Barr and Wells [6].

**Theorem 1.13.** *Let $\mathcal{C}$ be a category with all limits and colimits, and let $\mathsf{T} : \mathcal{C} \to \mathcal{C}$ be a triple which preserves filtered colimits. Then the category of $\mathsf{T}$-algebras $\mathcal{C}^{\mathsf{T}}$ has all limits and colimits.*

The difficult part of this theorem is the construction of pushouts in $\mathcal{C}^{\mathsf{T}}$; however, for the category of operads, this may be done explicitly, as follows.

First, recall that if $A$ and $B$ are associative algebras, their coproduct may be explicitly realized as the direct sum

$$A \coprod B = \Big(\bigoplus_{n=1}^{\infty} (A \otimes B)^{(n)}\Big) \oplus \Big(\bigoplus_{n=0}^{\infty} (A \otimes B)^{(n)} \otimes A\Big) \oplus \Big(\bigoplus_{n=0}^{\infty} B \otimes (A \otimes B)^{(n)}\Big)$$
$$\oplus \Big(\bigoplus_{n=0}^{\infty} B \otimes (A \otimes B)^{(n)} \otimes A\Big).$$

Similarly, if **a** and **b** are operads, we may realize their coproduct as follows. We take a sum over trees with vertices coloured either black or white: to a black vertex of valence $k$ we attach the space $\mathbf{a}(k)$, while to a white vertex of valence $k$ we attach the space $\mathbf{b}(k)$. Taking a tensor product over the vertices and a sum over trees, we obtain the free operad $\mathbb{T}(\mathbf{a} \oplus \mathbf{b})$. Suppose we have a black (white) vertex $s$ all of whose parents are coloured black (white): we may form a new tree in which $s$ and its parents are combined into a single black (white) vertex, and the corresponding elements of **a** (respectively **b**) are multiplied using the operad structure of **a** (resp. **b**). The coproduct of **a** and **b** is this quotient of $\mathbb{T}(\mathbf{a} \oplus \mathbf{b})$: thus $\mathbf{a} \coprod \mathbf{b}$ may be identified with the subspace spanned by those trees such that no black (white) vertex has all of its parents black (white).

Now suppose we are given an operad **c** and maps $\Phi : \mathbf{c} \to \mathbf{a}$ and $\Psi : \mathbf{c} \to \mathbf{b}$. To construct the pushout $\mathbf{a} \coprod_{\mathbf{c}} \mathbf{b}$, we further quotient $\mathbf{a} \coprod \mathbf{b}$ by the subspace which identifies trees in which one vertex of valence $k$ is changed from black to white, and the associated label is changed from $\Phi(c)$ to $\Psi(c)$, where $c \in \mathbf{c}(k)$. It may be easily shown that the resulting quotient is an operad in a natural way, and that it has the universal property of the coproduct $\mathbf{a} \coprod_{\mathbf{c}} \mathbf{b}$.

By Theorem 8.1 of MacLane [32], a category with filtered colimits and pushouts has all small colimits. To show that the free operad functor $\mathbb{T}$ preserves filtered colimits, we apply the following lemma.

**Lemma 1.14.** *In a symmetric monoidal category such that $-\otimes-$ preserves colimits, the functor $V \mapsto V^{(k)}$ preserves filtered colimits.*

*Proof.* Let $I$ be a filtered category, and let $\phi : I \to \mathcal{C}$ be an $I$-diagram in $\mathcal{C}$. We may factor the functor $V \mapsto V^{(k)}$ into the diagonal functor
$$V \mapsto V \times \cdots \times V : \mathcal{C} \to \mathcal{C}^k,$$
followed by the tensor product functor
$$V_1 \times \cdots \times V_k \mapsto V_1 \otimes \ldots \otimes V_k : \mathcal{C}^k \to \mathcal{C}.$$
Since the colimit over $I^k$ of $\phi(i_1) \otimes \ldots \otimes \phi(i_k)$ can be expressed as an iterated colimit, and $X \otimes -$ preserves colimits, we see that
$$\left(\operatorname*{colim}_{i \in I} \phi(i)\right)^{(k)} = \operatorname*{colim}_{(i_1,\ldots,i_k) \in I^k} \phi(i_1) \otimes \ldots \otimes \phi(i_k).$$
But since the category $I$ is filtered, its diagonal is cofinal, so that
$$\operatorname*{colim}_{(i_1,\ldots,i_k) \in I^k} \phi(i_1) \otimes \ldots \otimes \phi(i_k) \cong \operatorname*{colim}_{i \in I} \phi(i)^{(k)}. \quad \square$$

Using this lemma, we may show that the category of operads $\operatorname{Op}(\mathcal{C})$ has all filtered colimits. Suppose that $\phi : I \to \operatorname{Op}(\mathcal{C})$ is a filtered diagram of operads, and let $\mathbf{a}$ be the colimit in $\operatorname{Cat}(\mathbb{S},\mathcal{C})$ of the underlying diagram of $\mathbb{S}$-objects. There is a family of compatible maps $\mathbb{T}\phi(i) \to \mathbf{a}$ in $\operatorname{Cat}(\mathbb{S},\mathcal{C})$ induced by the structure maps $\mathbb{T}\phi(i) \to \phi(i)$, which induce a map from the colimit $\mathbb{T}\mathbf{a} = \operatorname{colim}_I \mathbb{T}\phi(i)$ to $\mathbf{a}$. It is easily checked that this map gives $\mathbf{a}$ an operad structure.

Finally, the category of operads has all limits by general considerations: if $\mathsf{T}$ is a triple, the functor $U^\mathsf{T} : \mathcal{C}^\mathsf{T} \to \mathcal{C}$ creates limits. (This is Theorem 3.4.1 of Barr and Wells [6].) To see this, let $\phi : I \to \mathcal{C}^\mathsf{T}$ be a diagram of $\mathsf{T}$-algebras, and let $X$ be the limit in $\mathcal{C}$ of the diagram $U^\mathsf{T}\phi : I \to \mathcal{C}$. There is a unique $\mathsf{T}$-algebra structure induced on the limit $\lim U^\mathsf{T}\phi$ in $\mathcal{C}$ making it the limit of the diagram $\phi$ in $\mathcal{C}^\mathsf{T}$. Indeed, for each $i \in I$, there is a map of $\mathsf{T}$-algebras $\mathsf{T}X \to \phi(i)$ induced by the map $X \to U^\mathsf{T}\phi(i)$ in $\mathcal{C}$ of the limit cone. The limit of the maps $\mathsf{T}X \to \phi(i)$ induces a map $\mathsf{T}X \to X$ in $\mathcal{C}$, which gives $X$ a $\mathsf{T}$-algebra structure.

**1.6. Limits and colimits of algebras.** In this section, we will show that the category of algebras over an operad has all limits and colimits. Although this follows from Thorem 1.13, we prefer to imitate our proof of the corresponding result for operads. The construction of limits and filtered colimits is identical to the case of operads, and the only difficult part of the proof is the construction of pushouts. Their existence is a consequence of the construction of a direct image, or base change, functor.

A morphism of operads $\Phi : \mathbf{a} \to \mathbf{b}$ induces a functor $\Phi^* : \mathcal{C}^{\mathbf{b}} \to \mathcal{C}^{\mathbf{a}}$ on the categories of algebras: if $B$ is a $\mathbf{b}$-algebra, the object of $\mathcal{C}$ underlying $\Phi^* B$ is $B$ itself, and the $\mathbf{a}$-algebra structure on $B$ is given by the composition
$$\mathsf{T}(\mathbf{a},B) \xrightarrow{\mathsf{T}(\Phi,B)} \mathsf{T}(\mathbf{b},B) \xrightarrow{\rho} B.$$
The functor $\Phi^*$ has a left adjoint $\Phi_*$, the direct-image: if $A$ is an $\mathbf{a}$-algebra, the object of $\mathcal{C}$ underlying $\Phi_* A$ is the coequalizer
$$\mathsf{T}(\mathbf{b},\mathsf{T}(\mathbf{a},A)) \xrightarrow[\mu(A) \cdot \mathsf{T}(\mathbf{b},\mathsf{T}(\Phi,A))]{\mathsf{T}(\mathbf{b},\rho)} \mathsf{T}(\mathbf{b},A) \to \Phi_* A.$$
This generalizes the definition of the direct image $\Phi_* M$ of an $A$-module $M$ along a morphism of algebras $f : A \to B$, often denoted by $B \otimes_A M$, as the coequalizer
$$B \otimes A \otimes M \rightrightarrows B \otimes M \to \Phi_* M.$$

**Lemma 1.15.** *The object $\Phi_* A$ of $\mathcal{C}$ has a natural $\mathbf{b}$-algebra structure.*

*Proof.* Observe that $\mathsf{T}(\mathbf{b}, \Phi_* A)$ is the coequalizer in $\mathcal{C}$

$$(*) \quad \mathsf{T}(\mathbf{b}, \mathsf{T}(\mathbf{b}, \mathsf{T}(\mathbf{a}, A))) \oplus \mathsf{T}(\mathbf{b}, A)) \underset{\mathsf{T}(\mathbf{b},(\mu A)\cdot\mathsf{T}(\mathbf{b},\mathsf{T}(\Phi,A))\oplus\mathsf{T}(\mathbf{b},A))}{\overset{\mathsf{T}(\mathbf{b},\mathsf{T}(\mathbf{b},\rho)\oplus\mathsf{T}(\mathbf{b},A))}{\rightrightarrows}} \mathsf{T}(\mathbf{b}, \mathsf{T}(\mathbf{b}, A)) \to \mathsf{T}(\mathbf{b}, \Phi_* A).$$

Indeed, if $X \underset{g}{\overset{f}{\rightrightarrows}} Y \to Z$ is a coequalizer in $\mathcal{C}$, then so is

$$\mathsf{T}(\mathbf{b}, X \oplus Y) \underset{\mathsf{T}(\mathbf{b}, g \oplus Y)}{\overset{\mathsf{T}(\mathbf{b}, f \oplus Y)}{\rightrightarrows}} \mathsf{T}(\mathbf{b}, Y) \to \mathsf{T}(\mathbf{b}, Z) \quad ,$$

since the functor $\mathbf{b}(k) \otimes_{\mathbb{S}_k} -$ is right-exact. It now suffices to apply this general observation to the coequalizer diagram defining $\Phi_* A$.

But the composite morphism

$$\mathsf{T}(\mathbf{b}, \mathsf{T}(\mathbf{b}, A)) \xrightarrow{\mu A} \mathsf{T}(\mathbf{b}, A) \to \Phi_* A$$

coequalizes the parallel pair of arrows in $(*)$, so there is induced an arrow $\mathsf{T}(\mathbf{b}, \Phi_* A) \to \Phi_* A$. This may be checked to give $\Phi_* A$ a $\mathbf{b}$-algebra structure. $\square$

Note that since both of the arrows $\mathsf{T}(\mathbf{b}, \mathsf{T}(\mathbf{a}, A)) \rightrightarrows \mathsf{T}(\mathbf{b}, A)$ are morphisms of $\mathbf{b}$-algebras, $\Phi_* A$ is actually a coequalizer in the category of $\mathbf{b}$-algebras as well as in $\mathcal{C}$.

**Proposition 1.16.** *The direct image functor $\Phi_* : \mathcal{C}^{\mathbf{a}} \to \mathcal{C}^{\mathbf{b}}$ is the left adjoint of the functor $\Phi^* : \mathcal{C}^{\mathbf{b}} \to \mathcal{C}^{\mathbf{a}}$.*

*Proof.* If $A$ is an $\mathbf{a}$-algebra and $B$ is a $\mathbf{b}$-algebra, we may associate to a linear map $f : A \to B$ the square

$$\begin{array}{ccc} \mathsf{T}(\mathbf{a}, A) & \xrightarrow{\mathsf{T}(\Phi, f)} & \mathsf{T}(\mathbf{b}, B) \\ \rho_A \downarrow & & \downarrow \rho_B \\ A & \xrightarrow{f} & B \end{array}$$

This induces two linear maps from $\mathcal{C}(A, B)$ to $\mathcal{C}(\mathsf{T}(\mathbf{a}, A), B)$, and we may identify $\mathrm{Hom}_{\mathbf{a}}(A, \Phi^* B)$ with the equalizer

$$\mathrm{Hom}_{\mathbf{a}}(A, \Phi^* B) \to \mathcal{C}(A, B) \rightrightarrows \mathcal{C}(\mathsf{T}(\mathbf{a}, A), B).$$

Since $\mathrm{Hom}_{\mathbf{b}}(-, B)$ maps coequalizers in the category of $\mathbf{b}$-algebras to equalizers in the category of sets, we also have the equalizer diagram

$$\mathrm{Hom}_{\mathbf{b}}(\Phi_* A, B) \to \mathrm{Hom}_{\mathbf{b}}(\mathsf{T}(\mathbf{b}, A), B) \rightrightarrows \mathrm{Hom}_{\mathbf{b}}(\mathsf{T}(\mathbf{b}, \mathsf{T}(\mathbf{a}, A)), B).$$

These equalizers are identified under the isomorphism $\mathrm{Hom}_{\mathbf{b}}(\mathsf{T}(\mathbf{b}, -), B) \cong \mathcal{C}(-, B)$, showing that $\mathrm{Hom}_{\mathbf{b}}(\Phi_* A, B) \cong \mathrm{Hom}_{\mathbf{a}}(A, \Phi^* B)$. $\square$

**Corollary 1.17.** *The composite functor $\Phi_* F^{\mathbf{a}} : \mathcal{C} \to \mathcal{C}^{\mathbf{b}}$ is isomorphic to the functor $F^{\mathbf{b}}$.*

*Proof.* The composition of functors $\mathcal{C}^{\mathbf{b}} \xrightarrow{\Phi^*} \mathcal{C}^{\mathbf{a}} \xrightarrow{U^{\mathbf{a}}} \mathcal{C}$ is equal to $U^{\mathbf{b}}$. Taking adjoints, we see that $\Phi_* F^{\mathbf{a}} \cong F^{\mathbf{b}} : \mathcal{C} \to \mathcal{C}^{\mathbf{b}}$. $\square$

If $\mathbf{a}$ is an augmented operad with augmentation map $\varepsilon : \mathbf{a} \to \mathbb{1}$, the direct image functor $\varepsilon_* : \mathcal{C}^{\mathbf{a}} \to \mathcal{C}$ is the indecomposables, or linearization, functor. The composition $\varepsilon_* F^{\mathbf{a}}$ is the identity functor of $\mathcal{C}$: the indecomposables of a free object are its generators.

We will use the direct image construction to prove the existence of pushouts in $\mathcal{C}^{\mathbf{a}}$. Recall that if $\mathcal{C}$ is a category and $X$ is an object in $\mathcal{C}$, the category $X \setminus \mathcal{C}$ of objects under $X$ is the category of morphisms $f : X \to Y$ with source $X$.

**Lemma 1.18.** *Let $\mathbf{a}$ be an operad and let $X$ be an $\mathbf{a}$-algebra. There is an operad $\mathbf{a}[X]$ such that the category of $\mathbf{a}[X]$-algebras is naturally isomorphic to the category $X \setminus \mathcal{C}^{\mathbf{a}}$ of $\mathbf{a}$-algebras under $X$.*

*Proof.* In the special case that $\mathbf{a} = \mathbb{1}$, the $\mathbb{S}$-object underlying the operad $\mathbb{1}[X]$ is

$$\mathbb{1}[X](S) = \begin{cases} X, & |S| = 0, \\ \mathbb{1}, & |S| = 1, \\ 0, & |S| > 1. \end{cases}$$

More genereally, the $\mathbb{S}$-object underlying the operad $\mathbf{a}[X]$ equals $\mathbf{a} \circ \mathbb{1}[X]$ on nonempty finite sets $S$, while on the empty set it equals $X$: explicitly,

$$\mathbf{a}[X](S) = \begin{cases} X, & |S| = 0, \\ \bigoplus_{k=0}^{\infty} \mathbf{a}(S \coprod k) \otimes_{\mathbb{S}_k} X^{(k)}, & |S| > 0. \end{cases}$$

We define the product of $\mathbf{a}[X]$ as follows: if $|S| > 0$,

$$(\mathbf{a}[X] \circ \mathbf{a}[X])(S) \cong (\mathbf{a} \circ \mathbf{a})[X](S),$$

and the product $(\mathbf{a}[X] \circ \mathbf{a}[X])(S) \to \mathbf{a}[X](S)$ is defined using the product of $\mathbf{a}$. On the other hand,

$$(\mathbf{a}[X] \circ \mathbf{a}[X])(0) \subset \mathsf{T}(\mathbf{a} \circ \mathbf{a}, X),$$

and the product is defined by first mapping $\mathsf{T}(\mathbf{a} \circ \mathbf{a}, X)$ to $\mathsf{T}(\mathbf{a}, X)$ and then applying the structure map $\mathsf{T}(\mathbf{a}, X) \to X$. The unit of $\mathbf{a}[X]$ is the composition of the unit of $\mathbf{a}$ with the natural inclusion $\mathbf{a}(1) \to \mathbf{a}[X](1)$. These maps give $\mathbf{a}[X]$ an operad structure.

An algebra $Y$ over $\mathbf{a}[X]$ is defined by a structure map $\mathsf{T}(\mathbf{a}[X], Y) \to Y$. Composing with the inclusion $\mathsf{T}(\mathbf{a}, Y) \oplus X \to \mathsf{T}(\mathbf{a}[X], Y)$, we see that this determines maps $\mathsf{T}(\mathbf{a}, Y) \to Y$ and $X \to Y$. The operad structure of $\mathbf{a}[X]$ show that two such maps determine an $\mathbf{a}[X]$-algebra structure on $Y$ if and only if they give $Y$ the structure of an $\mathbf{a}$-algebra under $X$. □

If $f : X \to Y$ is a morphism of $\mathbf{a}$-algebras, and $\mathbf{a}[f] : \mathbf{a}[X] \to \mathbf{a}[Y]$ is the induced morphism of operads, then $\mathbf{a}[f]^*$ is the natural map from $\mathbf{a}$-algebras under $Y$ to $\mathbf{a}$-algebras under $X$ induced by $f$. We see that the pushout of the diagram

$$\begin{array}{ccc} X & \xrightarrow{f} & Y \\ \downarrow & & \\ Z & & \end{array}$$

is the direct image $\mathbf{a}[f]_* Z$. This gives an explicit construction of the pushout in $\mathcal{C}$. In the special case where $X$ is the initial object $\mathbb{1}$, we obtain the coproduct in $\mathcal{C}^{\mathbf{a}}$, which we denote by $Y \coprod Z$.

**1.7. Cooperads and Coalgebras.** The theory of cooperads and their coalgebras is almost dual to the theory of operads and their algebras. This duality is not perfect, since the Schur functor $\mathsf{S}(\mathbf{a})$ does not commute with limits without an extra hypothesis on $\mathbf{a}$. This leads us to impose the additional assumption that the $\mathbb{S}$-object underlying the cooperad is exact.

**Definition 1.19.** *An $\mathbb{S}$-object $\mathbf{v}$ is exact if for all $k \geq 0$, the functor $\mathbf{v}(k) \otimes_{\mathbb{S}_k} -$ preserves limits.*

For example, in the category of chain complexes over a field of characteristic 0, every $\mathbb{S}$-object is exact.

A cotriple on a category is a comonoid in $\mathrm{End}(\mathcal{C})$, that is, a functor $\mathsf{C} : \mathcal{C} \to \mathcal{C}$ together with a coproduct $\Delta : \mathsf{C} \to \mathsf{C}\mathsf{C}$ and counit $\varepsilon : \mathsf{C} \to \mathbb{1}$ satisfying axioms dual to those of a triple. Cotriples arise when one has an adjoint pair of functors $U : \mathcal{C} \rightleftarrows \mathrm{Alg} : F$. The composition $UF : \mathrm{Alg} \to \mathrm{Alg}$ is a cotriple, with coproduct $\Delta = U\eta F : UFUF \to UF$, where $\eta : \mathbb{1} \to FU$ is the unit of the adjunction, and the counit $\varepsilon : UF \to \mathbb{1}$ of the adjunction is the counit of the triple.

Dualizing algebras over a triple, we obtain coalgebras over a cotriple: a coalgebra is an object $X$ of $\mathcal{C}$ together with a morphism $\rho : X \to \mathsf{C}X$, satisfying axioms dual to those for algebras over a triple. In particular, if $X$ is an object of $\mathcal{C}$, then $\mathsf{C}X$ is the underlying object of a coalgebra over $\mathsf{C}$, called the cofree $\mathsf{C}$-coalgebra on $X$. The category of $\mathsf{C}$-coalgebras is denoted by $\mathcal{C}_\mathsf{C}$.

A cooperad $\mathbf{z}$ in a symmetric monoidal category $\mathcal{C}$ is a comonoid in the monoidal category $(\mathrm{Hom}(\mathbb{S}, \mathcal{M}), \circ, \mathbb{1})$, that is, an $\mathbb{S}$-object $\mathbf{z}$ together with morphisms $\Delta : \mathbf{z} \to \mathbf{z} \circ \mathbf{z}$ and $\varepsilon : \mathbf{z} \to \mathbb{1}$ satisfying the coassociativity and counit axioms. We will denote the category of cooperads by $\mathrm{Coop}(\mathcal{C})$.

If $\mathbf{z}$ is a cooperad, let $\mathsf{C}(\mathbf{z})$ be the associated cotriple, with underlying functor $\mathsf{S}(\mathbf{z})$; we denote the category of $\mathsf{C}(\mathbf{z})$-coalgebras by $\mathcal{C}_\mathbf{z}$, and call an object of $\mathcal{C}_\mathbf{z}$ a $\mathbf{z}$-coalgebra.

The following proposition is an example of how the hypothesis of exactness is used in the theory of cooperads.

**Proposition 1.20.** *If $\mathbf{z}$ is an exact cooperad, the category of $\mathbf{z}$-coalgebras has all limits and colimits.*

*Proof.* This follows from the dual of Theorem 1.13. The hypothesis of exactness is required in order that $\mathsf{C}(\mathbf{z})$ preserve filtered limits. □

One may dualize the construction of direct images of Section 1.6: if $\Phi : \mathbf{w} \to \mathbf{z}$ is a morphism of cooperads and $\mathbf{w}$ is exact, the functor $\Phi_*$ has a right adjoint $\Phi^*$, the inverse-image.

One may also dualize the explicit construction of pushouts, to obtain a construction of pullbacks in the category of $\mathbf{z}$-colagebras, for $\mathbf{z}$ an exact operad. The category $\mathcal{C}/Z$ of $\mathbf{z}$-coalgebras over a $\mathbf{z}$-coalgebra $Z$ may be identified with the category of coalgebras over a cooperad $\mathbf{z}[Z]$, whose underlying $\mathbb{S}$-object is

$$\mathbf{z}[Z](S) = \begin{cases} \bigoplus_{k=0}^{\infty} \mathbf{z}(S \coprod k) \otimes_{\mathbb{S}_k} Z^{(k)}, & |S| > 0, \\ Z, & |S| = 0. \end{cases}$$

The pullback of a diagram

$$\begin{array}{ccc} & & X \\ & & \downarrow \\ Y & \xrightarrow{f} & Z \end{array}$$

in $\mathcal{C}_\mathbf{z}$ is given by the explicit formula $\mathbf{z}[f]^*X$.

There is a cotriple $\mathbb{C}$ on the category of $\mathbb{S}$-objects whose category of coalgebras is the category of cooperads. The functor underlying $\mathbb{C}$ is the same endofunctor $\mathbb{T} : \mathrm{Cat}(\mathbb{S}, \mathcal{C}) \to \mathrm{Cat}(\mathbb{S}, \mathcal{C})$ as underlies the free operad triple $\mathbb{T}$. The coproduct $\Delta : \mathbb{C}\mathbf{v} \to (\mathbb{C}\mathbf{v}) \circ (\mathbb{C}\mathbf{v})$ of the cooperad $\mathbb{C}\mathbf{v}$ is defined by grafting of trees. The summands of $((\mathbb{C}\mathbf{v}) \circ (\mathbb{C}\mathbf{v}))(S)$ are labelled by a partition $\pi \in \Pi(S, k)$ of $S$, a tree $\mathcal{S}_0 \in \mathcal{T}(k)$ with $k$ incoming external edges, and $k$ trees $\mathcal{S}_i \in \mathcal{T}(\pi^{-1}(i))$, $1 \leq i \leq k$, labelled by the sets $\pi^{-1}(i) \subset S$. Such a summand is isomorphic to

$$\mathbf{v}(\mathcal{S}_0) \otimes \mathbf{v}(\mathcal{S}_1) \otimes \ldots \otimes \mathbf{v}(\mathcal{S}_k),$$

which in turn is naturally isomorphic to the summand $\mathbf{v}(\mathcal{S}_0[\mathcal{S}_1, \ldots, \mathcal{S}_k])$ of $\mathbb{C}\mathbf{v}$: the coproduct $\Delta : \mathbb{C}\mathbf{v} \to (\mathbb{C}\mathbf{v}) \circ (\mathbb{C}\mathbf{v})$ is defined by means of this identification. The counit $\varepsilon : \mathbb{C}\mathbf{v} \to \mathbb{1}$ of the cooperad $\mathbb{C}\mathbf{v}$ is the projection to the summand $\mathbb{1}$ of $\mathbb{C}\mathbf{v}(1)$.

If $\mathbf{v}$ is an $\mathbb{S}$-object, the cooperad $\mathbb{C}\mathbf{v}$ is the cofree cooperad of $\mathbf{v}$: any map of $\mathbb{S}$-objects $\mathbf{z} \to \mathbf{v}$, where $\mathbf{z}$ is a cooperad, factors as $\mathbf{z} \to \mathbb{C}\mathbf{v} \to \mathbf{v}$, where $\mathbb{C}\mathbf{v} \to \mathbf{v}$ is the canonical projection of $\mathbb{C}\mathbf{v}$ onto its summand $\mathbf{v}$, and $\mathbf{z} \to \mathbb{C}\mathbf{v}$ is a map of cooperads. It follows that the functor $G_{\mathbb{C}}$ such that $G_{\mathbb{C}}(\mathbf{v}) = \mathbb{C}\mathbf{v}$ is the right adjoint of the underlying $\mathbb{S}$-object functor $U_{\mathbb{C}} : \operatorname{Coop}(\mathcal{C}) \to \operatorname{Hom}(\mathbb{S}, \mathcal{C})$, and that the functor $U_{\mathbb{C}}$ is cotripleable.

**Proposition 1.21.** *The category of coalgebras for the cotriple $\mathbb{C} = U_{\mathbb{C}} G_{\mathbb{C}}$ is isomorphic to the category of cooperads $\operatorname{Coop}(\mathcal{C})$.*

## 2. The bar construction for operads and algebras

We now begin the study of operads and cooperads in the category of chain complexes $\mathcal{M}$. In Section 2.1, we define the bar cooperad $\mathcal{B}\mathbf{a}$ of an operad $\mathbf{a}$: this construction is based on Ginzburg and Kapranov's bar operad of an operad [22], which is its linear dual, and generalizes the bar construction of Eilenberg and MacLane for associative algebras.

As will be explained in Chapter 4, Quillen's homotopical algebra for algebras over an operad is the correct generalization of homological algebra for modules over an algebra. In Section 2.2, we study the algebras which play the role of the free complexes in the category of modules: the almost free $\mathbf{a}$-algebras. In homotopical algebra, the role of projective complexes will be taken by cofibrant $\mathbf{a}$-algebras, the retracts of almost free $\mathbf{a}$-algebras.

In Section 2.3, we use the bar cooperad to construct a homotopy equivalence between the categories of $\mathbf{a}$-algebras and $\mathcal{B}\mathbf{a}$-coalgebras. By this, we mean that there is an adjoint pair of functors

$$\Omega(\mathbf{a}) : \mathcal{M}_{\mathcal{B}\mathbf{a}} \rightleftarrows \mathcal{M}^{\mathbf{a}} : \mathbb{B}(\mathbf{a})$$

such that the unit $\Omega(\mathbf{a}, \mathbb{B}(\mathbf{a}, A)) \to A$ and counit $C \to \mathbb{B}(\mathbf{a}, \Omega(\mathbf{a}, C))$ of the adjunction are weak equivalences.

In Section 2.4, we recall the definition of a Koszul operad, due to Ginzburg and Kapranov: this generalizes Priddy's notion of a Koszul algebra. One of the main results of this paper, proved in Chapter 3, is that the $n$-algebra operad $\mathbf{e}_n$ of Section 1.3 is Koszul, for $1 \leq n \leq \infty$: for $n = 1$ and $n = \infty$, this is equivalent to the well known theorems which identify the Hochschild homology of a free associative algebra and the Lie algebra homology of a free Lie algebra.

**2.1. The bar construction for operads.** The bar construction of Eilenberg and MacLane is a functor from augmented dg-algebras to coaugmented dg-coalgebras. If $A$ is an augmented algebra, let $\bar{A}$ be the kernel of the augmentation map $\varepsilon : A \to \mathbb{1}$, and let $\Sigma \bar{A}$ be its suspension. The bar construction is obtained by twisting the differential of the cofree coalgebra generated by $\Sigma \bar{A}$ in a way which reflects the algebra structure of $A$. In this section, following Ginzburg and Kapranov [22], we define the bar cooperad $\mathcal{B}\mathbf{a}$ of a dg-operad $\mathbf{a}$; this construction specializes to the bar construction of Eilenberg-MacLane if $\mathbf{a}$ is an algebra.

If $\mathbf{v}$ is an $\mathbb{S}$-module (an $\mathbb{S}$-object in the category of chain complexes $\mathcal{M}$), denote by $\mathbf{v}^{\#}$ the $\mathbb{S}$-module with zero differential underlying $\mathbf{v}$ (this is Moore's notation [36]).

**Definition 2.1.** *An operad $\mathbf{a}$ is almost free if the operad $\mathbf{a}^{\#}$ is free. A cooperad $\mathbf{z}$ is almost cofree if the cooperad $\mathbf{z}^{\#}$ is cofree.*

If **a** is an augmented operad, let $\bar{\mathbf{a}}$ be the kernel of the augmentation map $\varepsilon : \mathbf{a} \to \mathbb{1}$. The bar cooperad $\mathcal{B}\mathbf{a}$ is defined by twisting the differential of the cofree cooperad $\mathbb{C}(\Sigma\bar{\mathbf{a}})$ in a way which refects the operad structure of **a**; thus $\mathcal{B}\mathbf{a}$ is an almost cofree cooperad.

Let $\mathcal{S}$ be a tree, and let $e$ be an internal edge of $\mathcal{S}$, starting at the vertex $s$ and ending at the vertex $t$. Define a new tree $\mathcal{S} \setminus e$ by contracting the edge $e$ and merging the vertices $s$ and $t$. Denote this merged vertex of $\mathcal{S} \setminus e$ by $t \cdot s$.

In defining the contribution of the edge $e$ to the differential of $\mathcal{B}\mathbf{a}$, it is convenient to suppose that the vertices of $\mathcal{S}$ are ordered in such a way that $s$ comes first among the parents of $t$. The multiplication map

$$\mathbf{a}(\mathrm{in}(t)) \otimes \mathbf{a}(\mathrm{in}(s)) \otimes \mathbb{1}^{\mathrm{in}(t)\setminus\{e\}} \xrightarrow{1 \otimes 1 \otimes \eta^{\mathrm{in}(t)\setminus\{e\}}} \mathbf{a}(\mathrm{in}(t)) \otimes \mathbf{a}(\mathrm{in}(s)) \otimes \mathbf{a}(1)^{\mathrm{in}(t)\setminus\{e\}} \to \mathbf{a}(\mathrm{in}(t \cdot s)),$$

induces by suspension a map $\partial_e : (\Sigma\bar{\mathbf{a}})(\mathcal{S}) \to (\Sigma\bar{\mathbf{a}})(\mathcal{S} \setminus e)$ of degree $-1$. The following picture illustrates the situation.

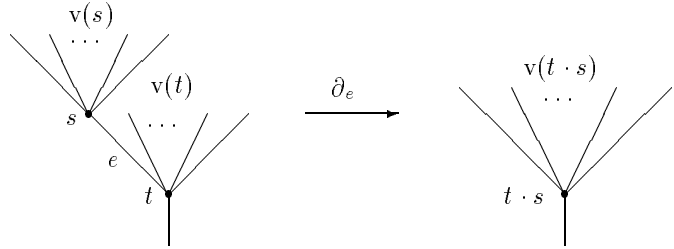

Denote by $\partial : \mathbb{C}(\Sigma\bar{\mathbf{a}}) \to \mathbb{C}(\Sigma\bar{\mathbf{a}})$ the sum of the maps $\partial_e : (\Sigma\bar{\mathbf{a}})(\mathcal{S}) \to (\Sigma\bar{\mathbf{a}})(\mathcal{S} \setminus e)$ over all trees $\mathcal{S}$ and internal edges $e \in \mathcal{S}$, and by $\delta_{\mathbf{a}}$ the internal differential of $\mathbb{C}(\Sigma\bar{\mathbf{a}})$ induced by that of **a**.

**Proposition 2.2.** *The operator $\delta_{\mathbf{a}} + \partial$ is a differential, and is compatible with the cooperad structure of $\mathbb{C}(\Sigma\bar{\mathbf{a}})$.*

*Proof.* We first show that $(\delta_{\mathbf{a}} + \partial)^2 = 0$. The internal differential $\delta_{\mathbf{a}}$ of **a** commutes with $\partial_e$ for all internal edges $e$, showing that $[\delta_{\mathbf{a}}, \partial] = 0$. If $\mathcal{S}$ is a tree and $e_1$ and $e_2$ are two internal edges of $\mathcal{S}$, it is easy to verify that $\partial_{e_1}\partial_{e_2} + \partial_{e_2}\partial_{e_1} = 0$, showing that $\partial^2 = 0$. It follows that $(\delta_{\mathbf{a}} + \partial)^2 = 0$.

Since it is clear that $\delta_{\mathbf{a}}$ is compatible with the coproduct of $\mathbb{C}(\Sigma\bar{\mathbf{a}})$, it remains to show that $\partial$ is. Denote by $\Delta_{\mathcal{S}_0[\mathcal{S}_1,\ldots,\mathcal{S}_k]}$ the component of the coproduct $\Delta : \mathbb{C}(\Sigma\bar{\mathbf{a}}) \to \mathbb{C}(\Sigma\bar{\mathbf{a}}) \circ \mathbb{C}(\Sigma\bar{\mathbf{a}})$ which maps the summand $(\Sigma\bar{\mathbf{a}})(\mathcal{S}_0[\mathcal{S}_1,\ldots,\mathcal{S}_k])$ of $\mathbb{C}(\Sigma\bar{\mathbf{a}})$ to the image of

$$(\Sigma\bar{\mathbf{a}})(\mathcal{S}_0) \otimes (\Sigma\bar{\mathbf{a}})(\mathcal{S}_1) \otimes \ldots \otimes (\Sigma\bar{\mathbf{a}})(\mathcal{S}_k)$$

in $\mathbb{C}(\Sigma\bar{\mathbf{a}}) \circ \mathbb{C}(\Sigma\bar{\mathbf{a}})$. If $e$ is an internal edge of a subtree $\mathcal{S}_i$, we see that

$$\Delta_{\mathcal{S}_0[\mathcal{S}_1,\ldots,\mathcal{S}_k]}\partial_e = (1 \otimes \ldots \otimes 1 \otimes \partial_e \otimes 1 \otimes \ldots \otimes 1)\Delta_{\mathcal{S}_0[\mathcal{S}_1,\ldots,\mathcal{S}_k]},$$

where on the right-hand side, $\partial_e$ acts on the factor $(\Sigma\bar{\mathbf{a}})(\mathcal{S}_i)$. Since $\Delta$ is the sum of $\Delta_{\mathcal{S}_0[\mathcal{S}_1,\ldots,\mathcal{S}_k]}$ over all $\mathcal{S}_i$, and $\partial$ is the sum of $\partial_e$ over all trees $\mathcal{S}$ and internal edges $e$, this shows that $\partial$ is compatible with the coproduct of $\mathbb{C}(\Sigma\bar{\mathbf{a}})$: $\mathbb{C}(\Sigma\bar{\mathbf{a}})$ with differential $\delta_{\mathbf{a}} + \partial$ is a cooperad. $\square$

**Definition 2.3.** The bar cooperad $\mathcal{B}\mathbf{a}$ of an augmented operad **a** is the cooperad $\mathbb{C}(\Sigma\bar{\mathbf{a}})$ with differential $\delta_{\mathcal{B}\mathbf{a}} = \delta_{\mathbf{a}} + \partial$. We denote the summand $(\Sigma\bar{\mathbf{a}})(\mathcal{S})$ of $\mathcal{B}\mathbf{a}$ by $\mathcal{B}\mathbf{a}(\mathcal{S})$.

A connected cooperad is a cooperad such that $\bar{\mathbf{z}}$ is concentrated in strictly positive degree: for example, the cooperad $\mathcal{B}\mathbf{a}$ is connected. There is a dual version of the bar construction for connected coaugmented cooperads **z**: the cobar operad $\mathcal{B}^*\mathbf{z}$ is the almost free operad $\mathbb{T}(\Sigma^{-1}\bar{\mathbf{z}})$ with differential $\delta_{\mathcal{B}^*\mathbf{z}} = \delta_{\mathbf{z}} + \partial^*$, where $\delta_{\mathbf{z}}$ is the internal differential of $\mathbb{T}(\Sigma^{-1}\bar{\mathbf{z}})$ induced by that of **z**, and $\partial^*$ is the differential defined by reversing all of the arrows in the definition of $\partial$ on $\mathbb{C}(\Sigma\bar{\mathbf{a}})$. The condition that the cooperad $\mathbf{z}$ is connected is needed in order for the operad $\mathcal{B}^*\mathbf{z}$ to not have any elements of negative degree. Note that $\mathcal{B}\Lambda\mathbf{a} \cong \Lambda\mathcal{B}\mathbf{a}$ and $\mathcal{B}^*\Lambda\mathbf{z} \cong \Lambda\mathcal{B}^*\mathbf{z}$.

In Section 2.3, we will prove that the functors $\mathcal{B}^*$ and $\mathcal{B}$ form an adjoint pair. In Theorem 3.2.16 of [22], Ginzburg and Kapranov show that the counit $\mathcal{B}^*\mathcal{B}\mathbf{a} \to \mathbf{a}$ and unit $\mathbf{z} \to \mathcal{B}\mathcal{B}^*\mathbf{z}$ of this adjunction are weak equivalences. Thus, the operad $\mathcal{B}^*\mathcal{B}\mathbf{a}$ may be thought of as a canonical almost free resolution of the operad $\mathbf{a}$. The analogous construction for algebras is well-known; see for example Moore [37].

**2.2. Almost-free algebras and differentials.** The following definition may be compared to that of an almost free operad of the last section.

**Definition 2.4.** If $\mathbf{a}$ is an operad, an $\mathbf{a}$-algebra $A$ is almost free if the underlying $\mathbf{a}^\#$-algebra $A^\#$ is free.

An almost free $\mathbf{a}$-algebra has an underlying free $\mathbf{a}$-algebra $\mathsf{T}(\mathbf{a}, V)$ with differential $\delta$ induced by a differential on $V$, and its differential may be decomposed as $\delta + d$, where $d$ is an endomorphism of $\mathsf{T}(\mathbf{a}, V)$ of degree $-1$. In order to classify all deformations of the differential on a free $\mathbf{a}$-algebra compatible with the $\mathbf{a}$-algebra structure, we will study the more general question of finding all operators $d$ on a general $\mathbf{a}$-algebra $A$ compatible with the $\mathbf{a}$-algebra structure of $A$, in a sense that we now explain.

Let $\mathbf{a}$, $\mathbf{b}$ and $\mathbf{c}$ be $\mathbb{S}$-modules in the category of chain complexes over a field $K$. The derivative $\partial(\Phi) : \mathbf{a} \circ \mathbf{b} \to \mathbf{a} \circ \mathbf{c}$ of a linear map $\Phi : \mathbf{b} \to \mathbf{c}$ is the unique map for which the following diagram commutes:

$$\begin{array}{ccc} \bigoplus_{k=0}^\infty \mathbf{a}(k) \otimes \left(\bigoplus_{\pi \in \Pi(S,k)} \mathbf{b}(\pi)\right) & \xrightarrow{\bigoplus_{i=1}^k 1 \otimes 1^{(i-1)} \otimes \Phi \otimes 1^{(k-i)}} & \bigoplus_{k=0}^\infty \mathbf{a}(k) \otimes \left(\bigoplus_{\pi \in \Pi(S,k)} \mathbf{c}(\pi)\right) \\ \downarrow & & \downarrow \\ \bigoplus_{k=0}^\infty \mathbf{a}(k) \otimes_{\mathbb{S}_k} \left(\bigoplus_{\pi \in \Pi(S,k)} \mathbf{b}(\pi)\right) & \xrightarrow{\partial(\Phi)} & \bigoplus_{k=0}^\infty \mathbf{a}(k) \otimes_{\mathbb{S}_k} \left(\bigoplus_{\pi \in \Pi(S,k)} \mathbf{c}(\pi)\right) \end{array}$$

In particular, if $D : V \to W$ where $V$ and $W$ are chain complexes, then $\partial(D) : \mathsf{T}(\mathbf{a}, V) \to \mathsf{T}(\mathbf{a}, W)$.

Let $\mathbf{a}$ be an operad with differential $\delta_\mathbf{a}$, and let $A$ be an $\mathbf{a}$-algebra, with differential $\delta$. Note that the following diagram commutes:

$$\begin{array}{ccc} \mathsf{T}(\mathbf{a}, A) & \xrightarrow{\delta_\mathbf{a} + \partial(\delta)} & \mathsf{T}(\mathbf{a}, A) \\ \rho \downarrow & & \downarrow \rho \\ A & \xrightarrow{\delta} & A \end{array}$$

**Definition 2.5.** An $\mathbf{a}$-derivation of an $\mathbf{a}$-algebra $A$ is a linear map $d : A \to A$ such that the following diagram commutes:

$$\begin{array}{ccc} \mathsf{T}(\mathbf{a}, A) & \xrightarrow{\partial(d)} & \mathsf{T}(\mathbf{a}, A) \\ \rho \downarrow & & \downarrow \rho \\ A & \xrightarrow{d} & A \end{array}$$

We write $\mathrm{Der}(\mathbf{a}, A)$ for the graded vector space of $\mathbf{a}$-derivations of an $\mathbf{a}$-algebra $A$.

If $\delta_1$ and $\delta_2$ are two differentials on $A$ compatible with its $\mathbf{a}$-algebra structure, then $d = \delta_1 - \delta_2$ is an $\mathbf{a}$-derivation of $A$ of degree $-1$. Conversely, if $A$ is an $\mathbf{a}$-algebra with differential $\delta$, and $d$ is a derivation of degree $-1$ such that $(\delta + d)^2 = 0$, the chain complex $(A, \delta + d)$ is an $\mathbf{a}$-algebra with the same structure map $\mathsf{T}(\mathbf{a}, A) \to A$ as $A$; we call this procedure twisting. An $\mathbf{a}$-derivation of degree $-1$ on an **a**-algebra $A$ such that $(\delta+d)^2 = 0$ is called an **a**-differential; the set of all **a**-differentials on $A$ is denoted $\mathrm{Diff}(\mathbf{a}, A)$.

**Proposition 2.6.** *If $A$ is an **a**-algebra, $\mathrm{Der}(\mathbf{a}, A)$ is a Lie dg-algebra, with bracket the graded commutator and differential the graded commutator with the differential $\delta$ of $A$.*

*Proof.* If $d_1$ and $d_2$ are endomorphisms of $A$, then the maps $\partial(d_i) : \mathsf{T}(\mathbf{a}, A) \to \mathsf{T}(\mathbf{a}, A)$ satisfy the formula
$$[\partial(d_1), \partial(d_2)] = \partial([d_1, d_2]).$$
It follows immediately from this formula that $\mathrm{Der}(\mathbf{a}, A)$ is closed under the graded commutator, and hence is a graded Lie algebra. Furthermore, $\mathrm{Der}(\mathbf{a}, A)$ is closed under the differential $d \mapsto [\delta, d]$; indeed,
$$\partial([\delta, d]) = [\delta_{\mathbf{a}} + \partial(\delta), \partial(d)] = [\partial(\delta), \partial(d)],$$
since $[\delta_{\mathbf{a}}, \partial(d)] = 0$. $\square$

The following proposition gives a more precise description of **a**-derivations on a free **a**-algebra.

**Proposition 2.7.** *The chain complex underlying the Lie algebra $\mathrm{Der}(\mathbf{a}, \mathsf{T}(\mathbf{a}, V))$ is isomorphic to the chain complex $\mathrm{Hom}(V, \mathsf{T}(\mathbf{a}, V))$, under the map which identifies an **a**-derivation $d : \mathsf{T}(\mathbf{a}, V) \to \mathsf{T}(\mathbf{a}, V)$ with the map*
$$V \xrightarrow{\eta V} \mathsf{T}(\mathbf{a}, V) \xrightarrow{d} \mathsf{T}(\mathbf{a}, V)$$
*in $\mathrm{Hom}(V, \mathsf{T}(\mathbf{a}, V))$.*

*Proof.* The inverse isomorphism is defined by sending a map $p : V \to \mathsf{T}(\mathbf{a}, V)$ to the **a**-derivation
$$\mathsf{T}(\mathbf{a}, V) \xrightarrow{\mathsf{T}(\mathbf{a}, \eta)} \mathsf{T}(\mathbf{a}, \mathsf{T}(\mathbf{a}, V)) \xrightarrow{\partial(p\varepsilon)} \mathsf{T}(\mathbf{a}, \mathsf{T}(\mathbf{a}, V)) \xrightarrow{\mu V} \mathsf{T}(\mathbf{a}, V) \quad \square$$

If $\mathbf{a}$ is an augmented operad, it is natural to restrict attention to those **a**-derivations of $\mathsf{T}(\mathbf{a}, V)$ such that the composition
$$V \xrightarrow{\eta V} \mathsf{T}(\mathbf{a}, V) \xrightarrow{d} \mathsf{T}(\mathbf{a}, V) \xrightarrow{\varepsilon V} V$$
vanishes. The set of such **a**-derivations forms an ideal of $\mathrm{Der}(\mathbf{a}, \mathsf{T}(\mathbf{a}, V))$, which we denote by $\mathrm{Der}_*(\mathbf{a}, \mathsf{T}(\mathbf{a}, V))$.

If $d_1$ and $d_2$ are two elements of $\mathrm{Hom}(V, \mathsf{T}(\mathbf{a}, V))$, we may define their product $d_1 \circ d_2$ by the composition
$$V \xrightarrow{d_1} \mathsf{T}(\mathbf{a}, V) \xrightarrow{\mathsf{T}(\mathbf{a}, d_2)} \mathsf{T}(\mathbf{a}, \mathsf{T}(\mathbf{a}, V)) \xrightarrow{\mu(V)} \mathsf{T}(\mathbf{a}, V).$$
It is easily verified that
$$[d_1, d_2] = d_1 \circ d_2 - (-1)^{|d_1| |d_2|} d_2 \circ d_1,$$
and that $\delta + d$ is a differential on $\mathsf{T}(\mathbf{a}, V)$ if and only if $\delta d + d \circ d = 0$.

It is now clear that an almost free **a**-algebra is determined by the following data: a chain complex $V$ together with an **a**-differential $d \in \mathrm{Diff}(\mathbf{a}, A)$; we denote the resulting almost free **a**-algebra by $\mathsf{T}(\mathbf{a}, V, d)$. The **a**-differential $d$ can change the differential on the chain complex $V$; however, if $d$ lies in $\mathrm{Diff}_*(\mathbf{a}, \mathsf{T}(\mathbf{a}, V))$, the intersection of $\mathrm{Diff}(\mathbf{a}, \mathsf{T}(\mathbf{a}, V))$ and $\mathrm{Der}_*(\mathbf{a}, \mathsf{T}(\mathbf{a}, V))$, this possibility is excluded, and the pair $(V, d)$ provides a pair of invariants specifying uniquely the almost free algebra.

**Definition 2.8.** An almost free **a**-algebra $A$ over an augmented operad **a** is minimal if the induced differential on the space of indecomposables $\varepsilon_* A$ vanishes.

Equivalently, an almost free algebra $\mathsf{T}(\mathbf{a}, V, d)$ is minimal if and only if the underlying differential on the chain complex $V$ vanishes.

**Proposition 2.9.** *A weak equivalence between two minimal algebras is an isomorphism.*

*Proof.* If $f : A \to B$ is such a weak equivalence, then so is $\varepsilon_* f : \varepsilon_* A \to \varepsilon_* B$. But by the definition of minimal algebras, the differentials on $\varepsilon_* A$ and $\varepsilon_* B$ vanish, so $\varepsilon_* A \cong \varepsilon_* B$. Since $A$ and $B$ are almost free on generators $\varepsilon_* A$ and $\varepsilon_* B$, this shows that $f$ is a bijection, and hence an isomorphism. □

The following alternative description of **a**-differentials on $\mathsf{T}(\mathbf{a}, V)$ is due to Ginzburg and Kapranov.

**Proposition 2.10.** ([22], 4.2.14) *Let $\mathbf{a}$ be an augmented dg-operad, and let $V$ be a chain complex. There is a natural bijection between $\mathbf{a}$-differentials $d \in \mathrm{Diff}_*(\mathbf{a}, \mathsf{T}(\mathbf{a}, V))$ and $\mathcal{B}\mathbf{a}$-coalgebra structures on $V$.*

*Proof.* Since $(\mathcal{B}\mathbf{a})^\# \cong \mathbb{C}(\Sigma \bar{\mathbf{a}}^\#)$, a $(\mathcal{B}\mathbf{a})^\#$-coalgebra structure on $V^\#$ corresponds to a linear map $d : V \to \mathsf{T}(\mathbf{a}, V)$ of degree $-1$ whose composition with the augmentation $\varepsilon : \mathsf{T}(\mathbf{a}, V) \to V$ vanishes. This identifies $(\mathcal{B}\mathbf{a})^\#$-coalgebra structures on $V^\#$ with elements of $\mathrm{Der}_*(\mathbf{a}, \mathsf{T}(\mathbf{a}, V))$ of degree $-1$.

In order for such a $(\mathcal{B}\mathbf{a})^\#$-coalgebra to underly a $\mathcal{B}\mathbf{a}$-coalgebra, it is necessary and sufficient that the coproduct be compatible with the differentials on $V$ and $\mathsf{C}(\mathcal{B}\mathbf{a}, V)$, that is, that the following diagram commute:

$$\begin{array}{ccc} V & \xrightarrow{\Delta} & \mathsf{C}(\mathcal{B}\mathbf{a}, V) \\ \delta \downarrow & & \downarrow \delta_{\mathcal{B}\mathbf{a}} + \delta + d \\ V & \xrightarrow{\Delta} & \mathsf{C}(\mathcal{B}\mathbf{a}, V) \end{array}$$

The map $\Delta \delta$ equals $d \cdot \delta : V \to \mathsf{T}(\mathbf{a}, V)$ followed by the inclusion, of degree $1$, of $\mathsf{T}(\Sigma \mathbf{a}, V)$ in $\mathsf{C}(\mathcal{B}\mathbf{a}, V)$. On the other hand, $(\delta_{\mathcal{B}\mathbf{a}} + \delta + d)\Delta$ equals the map

$$V \xrightarrow{(\delta_{\mathbf{a}} + \delta) \cdot d + d \circ d} \mathsf{T}(\mathbf{a}, V),$$

again followed by the inclusion of $\mathsf{T}(\Sigma \mathbf{a}, V)$ in $\mathsf{C}(\mathcal{B}\mathbf{a}, V)$. For these to be equal is precisely the equation $\delta d + d \circ d = 0$ for $d$ to be an **a**-differential. □

The above theory of almost free algebras, **a**-derivations and **a**-differentials has a parallel for coalgebras over a cooperad **z**.

**Definition 2.11.** *If $\mathbf{z}$ is a cooperad, and $C$ is a $\mathbf{z}$-coalgebra, then $C$ is almost cofree if the underlying $\mathbf{z}^\#$-coalgebra $C^\#$ is cofree.*

Suppose that the cooperad $\mathbf{z}$ is coaugmented. An almost cofree $\mathbf{z}$-coalgebra $C$ is minimal if the space of primitives $\eta^* C$ has vanishing differential.

Just as for minimal algebras, a weak equivalence between two minimal coalgebras is an isomorphism.

**Definition 2.12.** *A $\mathbf{z}$-coderivation of a $\mathbf{z}$-coalgebra $C$ is a linear map $d : C \to C$ such that the following diagram commutes:*

$$\begin{array}{ccc} C & \xrightarrow{d} & C \\ \rho \downarrow & & \downarrow \rho \\ \mathsf{C}(\mathbf{z}, C) & \xrightarrow{\partial(d)} & \mathsf{C}(\mathbf{z}, C) \end{array}$$

We write $\mathrm{Coder}(\mathbf{z}, C)$ for the graded vector space of $\mathbf{z}$-coderivations on a $\mathbf{z}$-coalgebra $C$.

A $\mathbf{z}$-codifferential on an $\mathbf{z}$-coalgebra $(C, \delta)$ is a $\mathbf{z}$-coderivation $d : C \to C$ of degree $-1$ such that $(\delta + d)^2 = 0$; the set of $\mathbf{z}$-codifferentials is denoted $\mathrm{Codiff}(\mathbf{z}, C)$. The chain complex $(C, \delta + d)$ is then a $\mathbf{z}$-coalgebra with the same structure map $C \to \mathsf{C}(\mathbf{z}, C)$ as $C$.

**Proposition 2.13.** *If $C$ is a $\mathbf{z}$-coalgebra, $\mathrm{Coder}(\mathbf{z}, C)$ is a Lie dg-algebras, with bracket the graded commutator and differential the graded commutator with the internal differential $\delta$ of $C$.*

If $\mathbf{z}$ is a coaugmented operad, it is natural to restrict attention to those $\mathbf{z}$-coderivations of $\mathsf{C}(\mathbf{z}, V)$ such that the composition
$$V \xrightarrow{\eta V} \mathsf{C}(\mathbf{z}, V) \xrightarrow{d} \mathsf{C}(\mathbf{z}, V) \xrightarrow{\varepsilon V} V$$
vanishes. The set of such $\mathbf{z}$-coderivations forms an ideal of $\mathrm{Coder}(\mathbf{z}, \mathsf{C}(\mathbf{z}, V))$, which we denote by $\mathrm{Coder}_*(\mathbf{z}, \mathsf{C}(\mathbf{z}, V))$.

An almost cofree $\mathbf{z}$-coalgebra is determined by a cofree $\mathbf{z}$-coalgebra $\mathsf{C}(\mathbf{z}, V)$ together with a $\mathbf{z}$-codifferential $d$; we denote the resulting almost cofree $\mathbf{z}$-coalgebra by $\mathsf{C}(\mathbf{z}, V, d)$. If $\mathbf{z}$ is coaugmented, there is a bijection between almost cofree coalgebras $\mathsf{C}(\mathbf{z}, C, d)$ and pairs $(V, d)$ where $V$ is a chain complex and $d \in \mathrm{Codiff}_*(\mathbf{z}, \mathsf{C}(\mathbf{z}, V)) = \mathrm{Codiff}(\mathbf{z}, \mathsf{C}(z, V)) \cap \mathrm{Coder}_*(\mathbf{z}, \mathsf{C}(\mathbf{z}, V))$. Such data determine a minimal $\mathbf{z}$-coalgebra if and only if the chain complex $V$ has vanishing differential.

**Proposition 2.14.** *The chain complex underlying the Lie algebra $\mathrm{Coder}(\mathbf{z}, \mathsf{C}(\mathbf{z}, V))$ is isomorphic to the chain complex $\mathrm{Hom}(\mathsf{C}(\mathbf{z}, V), V)$, under the map which identifies a coderivation $d : \mathsf{C}(\mathbf{z}, V) \to \mathsf{C}(\mathbf{z}, V)$ with the map*
$$\mathsf{C}(\mathbf{z}, V) \xrightarrow{d} \mathsf{C}(\mathbf{z}, V) \xrightarrow{\varepsilon V} V$$
*in $\mathrm{Hom}(\mathsf{C}(\mathbf{z}, V), V)$.*

*Proof.* The inverse isomorphism is given by sending a map $q : \mathsf{C}(\mathbf{z}, V) \to V$ to the $\mathbf{z}$-coderivation
$$\mathsf{C}(\mathbf{z}, V) \xrightarrow{\Delta V} \mathsf{C}(\mathbf{z}, \mathsf{C}(\mathbf{z}, V)) \xrightarrow{\mathsf{C}(\mathbf{z}, \partial(\eta q))} \mathsf{C}(\mathbf{z}, \mathsf{C}(\mathbf{z}, V)) \xrightarrow{\mathsf{C}(\mathbf{z}, \varepsilon)} \mathsf{C}(\mathbf{z}, V) \qquad \square$$

The following proposition is proved in the same way as Proposition 2.10.

**Proposition 2.15.** *Let $\mathbf{z}$ be a coaugmented dg-cooperad, and let $V$ be a chain complex. There is a natural bijection between $\mathbf{z}$-codifferentials $d \in \mathrm{Codiff}_*(\mathbf{z}, \mathsf{C}(\mathbf{z}, V))$ and $\mathcal{B}^*\mathbf{z}$-algebra structures on $V$.*

**2.3. Twisting cochains.** If $\mathbf{a}$ is an augmented operad, let $\bar{\mathbf{a}}$ be the kernel of the augmentation map $\varepsilon : \mathbf{a} \to \mathbb{1}$. Dually, if $\mathbf{z}$ is a coaugmented cooperad, let $\bar{\mathbf{z}}$ be the cokernel of the coaugmentation map $\eta : \mathbb{1} \to \mathbf{z}$. If $\Phi$ and $\Psi$ are linear maps from $\bar{\mathbf{z}}$ to $\bar{\mathbf{a}}$, define the cup product $\Phi \cup \Psi : \bar{\mathbf{z}} \to \bar{\mathbf{a}}$ to be the composition
$$\bar{\mathbf{z}} \to \bar{\mathbf{z}} \circ \bar{\mathbf{z}} \xrightarrow{\Phi \circ \partial(\Psi)} \bar{\mathbf{a}} \circ \bar{\mathbf{a}} \to \bar{\mathbf{a}}$$
of the map $\Phi \circ \partial(\Psi)$ with the coproduct of $\mathbf{z}$ and the product of $\mathbf{a}$.

**Definition 2.16.** A twisting cochain $\Phi \in \mathcal{T}(\mathbf{z}, \mathbf{a})$ is a linear map of degree $-1$ from $\bar{\mathbf{z}}$ to $\bar{\mathbf{a}}$ such that $\delta \Phi + \Phi \cup \Phi = 0$.

The analogous notion of a twisting cochain from a coassociative coalgebra to an associative algebra is well-known. The following theorem motivates the introduction of twisting cochains.

**Theorem 2.17.** *The functors $\mathcal{B}^*$ and $\mathcal{B}$ form an adjoint pair between the categories of connected coaugmented cooperads and augmented operads: there are natural isomorphisms*
$$\mathrm{Hom}_{\mathrm{Op}(\mathcal{M})}(\mathcal{B}^*\mathbf{z}, \mathbf{a}) \cong \mathcal{T}(\mathbf{z}, \mathbf{a}) \cong \mathrm{Hom}_{\mathrm{Coop}(\mathcal{M})}(\mathbf{z}, \mathcal{B}\mathbf{a}).$$
*The counit $\mathcal{B}^*\mathcal{B}\mathbf{a} \to \mathbf{a}$ and unit $\mathbf{z} \to \mathcal{B}\mathcal{B}^*\mathbf{z}$ of this adjunction are weak equivalences.*

*Proof.* The set $\mathrm{Hom}_{\mathrm{Op}(\mathcal{M})}(\mathcal{B}^*\mathbf{z}, \mathbf{a})$ may be identified with a subset of

$$\mathrm{Hom}_{\mathrm{Op}(\mathcal{M})}(\mathbb{T}\Sigma^{-1}\bar{\mathbf{z}}, \mathbf{a}) = \mathrm{Hom}_{\mathrm{Cat}(\mathbb{S}, \mathcal{M})}(\Sigma^{-1}\bar{\mathbf{z}}, \bar{\mathbf{a}}).$$

For such a map $\Phi : \Sigma^{-1}\bar{\mathbf{z}} \to \bar{\mathbf{a}}$ of $\mathbb{S}$-modules to give rise to a map in $\mathrm{Hom}_{\mathrm{Op}(\mathcal{M})}(\mathbf{z}, \mathcal{B}\mathbf{a})$, it must be compatible with the differentials $\delta_\mathbf{z} + \partial^*$ on $\mathcal{B}^*\mathbf{z}$ and $\delta_\mathbf{a}$ on $\mathbf{a}$. This condition coincides with the condition for $\Phi$ to be a twisting cochain $\Phi \in \mathcal{T}(\mathbf{z}, \mathbf{a})$. The proof that $\mathrm{Hom}_{\mathrm{Coop}(\mathcal{M})}(\mathbf{z}, \mathcal{B}\mathbf{a}) \cong \mathcal{T}(\mathbf{z}, \mathbf{a})$ is similar.

For the proof that the counit $\mathcal{B}^*\mathcal{B}\mathbf{a} \to \mathbf{a}$ and unit $\mathbf{z} \to \mathcal{B}\mathcal{B}^*\mathbf{z}$ are weak equivalences, we refer to Theorem 3.2.16 of Ginzburg and Kapranov [22]. □

There is a universal twisting cochain $\Phi \in \mathcal{T}(\mathcal{B}\mathbf{z}, \mathbf{a})$, obtained by composing the projection $\mathcal{B}\mathbf{a} \to \Sigma\bar{\mathbf{a}}$ onto those summands corresponding to trees with one internal vertex with the desuspension map from $\Sigma\bar{\mathbf{a}}$ to $\bar{\mathbf{a}} \subset \mathbf{a}$. This twisting cochain corresponds under the above proposition to the identity map of the cooperad $\mathcal{B}\mathbf{a}$.

To any twisting cochain $\Phi \in \mathcal{T}(\mathbf{z}, \mathbf{a})$, we will now associate a pair of adjoint functors

$$\Omega(\Phi) : \mathcal{M}_\mathbf{z} \rightleftarrows \mathcal{M}^\mathbf{a} : \mathbb{B}(\Phi).$$

A linear map $\Phi : \bar{\mathbf{z}} \to \bar{\mathbf{a}}$ induces a natural transformation from the functor underlying the cotriple $\mathbb{C}(\mathbf{z})$ to the functor underlying the triple $\mathsf{T}(\mathbf{a})$. If $A$ is an $\mathbf{a}$-algebra, let $d_{A,\Phi}$ be the $\mathbf{z}$-coderivation of $\mathsf{C}(\mathbf{z}, A)$ associated to the element

$$\mathsf{C}(\mathbf{z}, A) \xrightarrow{\Phi(A)} \mathsf{T}(\mathbf{a}, A) \xrightarrow{\rho} A$$

of $\mathrm{Hom}(\mathsf{C}(\mathbf{z}, A), A)$ by Proposition 2.14. Since $d_{A,\Phi} \cdot d_{A,\Psi} = d_{A,\Phi \cup \Psi}$, we see that $d_{A,\Phi}$ is a $\mathbf{z}$-codifferential precisely if $\Phi$ is a twisting cochain. Denote by $\mathbb{B}(\Phi) : \mathcal{M}^\mathbf{a} \to \mathcal{M}_\mathbf{z}$ the functor which associates to an $\mathbf{a}$-algebra $A$ the almost cofree $\mathbf{z}$-coalgebra $\mathsf{C}(\mathbf{z}, A, d_{A,\Phi})$

If $C$ is a $\mathbf{z}$-coalgebra, let $d_{C,\Phi}$ be the $\mathbf{a}$-derivation of $\mathsf{T}(\mathbf{a}, C)$ associated to the element

$$C \xrightarrow{\rho} \mathsf{C}(\mathbf{z}, C) \xrightarrow{\Phi(C)} \mathsf{T}(\mathbf{a}, C)$$

of $\mathrm{Hom}(C, \mathsf{T}(\mathbf{a}, C))$ by Proposition 2.7. Again, $d_{C,\Phi} \cdot d_{C,\Psi} = d_{C,\Phi \cup \Psi}$, and $d_{C,\Phi}$ is an $\mathbf{a}$-differential precisely if $\Phi$ is a twisting cochain. Denote by $\Omega(\Phi) : \mathcal{M}_\mathbf{z} \to \mathcal{M}^\mathbf{a}$ the functor which associates to a $\mathbf{z}$-coalgebra $C$ the almost free $\mathbf{a}$-algebra $\mathsf{T}(\mathbf{a}, C, d_{C,\Phi})$.

If $A$ is an $\mathbf{a}$-algebra, $C$ is a $\mathbf{z}$-algebra, and $f : C \to A$ is a map of degree 0, denote by $\Phi(f) : C \to A$ the map

$$C \to \mathsf{C}(\mathbf{z}, C) \xrightarrow{\Phi(C)} \mathsf{T}(\mathbf{a}, C) \xrightarrow{\mathsf{T}(\mathbf{a}, f)} \mathsf{T}(\mathbf{a}, A) \to A$$

A map $f : C \to A$ is called a $\Phi$-twisting cochain if $\delta f + \Phi(f) = 0$; denote the set of $\Phi$-twisting cochains by $\mathcal{T}_\Phi(C, A)$.

**Proposition 2.18.** *There are natural bijections*

$$\mathrm{Hom}_\mathbf{a}(\Omega(\Phi, C), A) \cong \mathcal{T}_\Phi(C, A) \cong \mathrm{Hom}_\mathbf{z}(C, \mathbb{B}(\Phi, A)),$$

*inducing an adjunction* $\Omega(\Phi) : \mathcal{M}_\mathbf{z} \rightleftarrows \mathcal{M}^\mathbf{a} : \mathbb{B}(\Phi)$.

*Proof.* The set $\mathrm{Hom}_\mathbf{a}(\Omega(\Phi, C), A)$ may be identified with the subset of

$$\mathrm{Hom}_\mathbf{a}(\mathsf{T}(\mathbf{a}, C), A) = \mathrm{Hom}(C, A)$$

consisting of those linear maps $f : C \to A$ which are compatible with the differentials $\delta_\mathbf{a} + \delta_C + d_{C,\Phi}$ on $\Omega(\Phi, C)$ and $\delta_{A,\Phi}$ on $A$. This condition is easily seen to coincide with the condition defining the set of twisting cochains $f \in \mathcal{T}_\Phi(C, A)$. The proof that $\mathrm{Hom}_\mathbf{z}(C, \mathbb{B}(\Phi, A)) \cong \mathcal{T}_\Phi(C, A)$ is similar. □

If $\Phi \in \mathcal{T}(\mathcal{B}\mathbf{a}, \mathbf{a})$ is the universal twisting cochain, we denote the resulting adjunction by

$$\Omega(\mathbf{a}) : \mathcal{M}_{\mathcal{B}\mathbf{a}} \rightleftarrows \mathcal{M}^{\mathbf{a}} : \mathbb{B}(\mathbf{a}).$$

The following theorem shows that the almost free algebra $\Omega(\mathbf{a}, \mathbb{B}(\mathbf{a}, A))$ is a natural resolution of the $\mathbf{a}$-algebra $\mathbf{a}$.

**Theorem 2.19.** *The counit $\Omega(\mathbf{a}, \mathbb{B}(\mathbf{a}, A)) \to A$ and unit $C \to \mathbb{B}(\mathbf{a}, \Omega(\mathbf{a}, C))$ of the adjunction $\Omega(\mathbf{a}) : \mathcal{M}_{\mathcal{B}\mathbf{a}} \rightleftarrows \mathcal{M}^{\mathbf{a}} : \mathbb{B}(\mathbf{a})$ are weak equivalences.*

*Proof.* First, we prove that the counit $\Omega(\mathbf{a}, \mathbb{B}(\mathbf{a}, A)) \to A$ is a weak equivalence. Denote by $\bar{\mathbb{B}}(\mathbf{a}, A)$ the kernel of the projection from $\mathbb{B}(\mathbf{a}, A)$ onto $A$. Consider the short exact sequence

$$\begin{array}{ccc} \Sigma^{-1}\bar{\mathbb{B}}(\mathbf{a}, A) & & \Omega(\mathbf{a}, \mathbb{B}(\mathbf{a}, A)) \\ \| & & \| \end{array}$$

$$0 \longrightarrow \bigoplus_{k=0}^{\infty} \bar{\mathbf{a}}(k) \otimes_{\mathbb{S}_k} \mathbb{B}(\mathbf{a}, A)^{(k)} \longrightarrow \bigoplus_{k=0}^{\infty} \mathbf{a}(k) \otimes_{\mathbb{S}_k} \mathbb{B}(\mathbf{a}, A)^{(k)} \longrightarrow \mathbb{B}(\mathbf{a}, A) \longrightarrow 0$$

The unit $\eta : \mathbb{1} \to \mathbf{a}$ induces a natural splitting $\mathbb{B}(\mathbf{a}, A) \to \Omega(\mathbf{a}, \mathbb{B}(\mathbf{a}, A))$ of this sequence. Let $h : \Omega(\mathbf{a}, \mathbb{B}(\mathbf{a}, A)) \to \Omega(\mathbf{a}, \mathbb{B}(\mathbf{a}, A))$ be the map of degree $+1$ which maps the summand $\Sigma^{-1}\bar{\mathbb{B}}(\mathbf{a}, A)$ to $\mathbb{B}(\mathbf{a}, A)$. Then

$$\delta h + h\delta = 1 - \pi,$$

where $\pi$ is the projection on the summand $A$ of $\Omega(\mathbf{a}, \mathbb{B}(\mathbf{a}, A))$, and $\delta$ is the total differential of $\Omega(\mathbf{a}, \mathbb{B}(\mathbf{a}, A))$, showing that $\Omega(\mathbf{a}, \mathbb{B}(\mathbf{a}, A)) \to A$ is a weak equivalence.

We now prove that $C \to \mathbb{B}(\mathbf{a}, \Omega(\mathbf{a}, C))$ is a weak equivalence. Recall that a maximal vertex of a tree $\mathcal{S}$ is a vertex with no parents. In the proof, we will use the term marked tree for a pair $(\mathcal{S}, U)$ where $\mathcal{S}$ is a tree and $U$ is a subset of its set of maximal vertices, called the marked vertices.

The space $\mathbb{B}(\mathbf{a}, \Omega(\mathbf{a}, C))$ is a sum over trees, where a tree contributes the tensor product of factors $\Sigma\bar{\mathbf{a}}(k)$ corresponding to vertices of valence $k$ and factors of $\Omega(\mathbf{a}, C)$ corresponding to elements of $\mathrm{in}(\mathcal{S})$. Decomposing $\Omega(\mathbf{a}, C)$ into the direct sum $C \oplus \mathsf{T}(\bar{\mathbf{a}}, C)$, we see that $\mathbb{B}(\mathbf{a}, \Omega(\mathbf{a}, C))$ may be written as a sum over marked trees, where a tree contributes the tensor product of factors $\Sigma\bar{\mathbf{a}}(k)$ corresponding to unmarked vertices of valence $k$, factors of $\bar{\mathbf{a}}$ corresponding to marked vertices, and factors of $C$ corresponding to elements of $\mathrm{in}(\mathcal{S})$.

Filter $\mathbb{B}(\mathbf{a}, \Omega(\mathbf{a}, C))$ by the number of vertices:

$$F_i \mathbb{B}(\mathbf{a}, \Omega(\mathbf{a}, C)) = \text{sum over marked trees with at most } i \text{ vertices.}$$

The differential $d^0$ on $E^0 \mathbb{B}(\mathbf{a}, \Omega(\mathbf{a}, C))$ is the sum of three terms:

(1) the internal differential on $\mathbf{a}$;
(2) the internal differential on $C$;
(3) a sum, over maximal unmarked vertices, of the map which converts the unmarked vertex to a marked vertex.

For each isomorphism class of trees, choose a maximal vertex. The set of marked trees is then partitioned into two subsets: those in which the chosen maximal vertex is marked and unmarked, respectively. We may construct a contracting homotopy $h$ for the complex $E^0 \mathbb{B}(\mathbf{a}, \Omega(\mathbf{a}, C))$ by sending a term corresponding to a marked tree of the first type to the term corresponding to the same tree but in chosen vertex is unmarked: the second of these terms is the suspension of the first, so this map has degree 1. The internal differentials commute with $h$, and we see that $d^0 h + h d^0$ equals the projection onto trees with at least one vertex, and that $E^1 \mathbb{B}(\mathbf{a}\Omega(\mathbf{a}, C))$ may be identified with $C$. Thus, the spectral sequence collapses at $E^1$, and $C \to \mathbb{B}(\mathbf{a}, \Omega(\mathbf{a}, C))$ is a weak equivalence. $\square$

**2.4. Koszul operads.** The class of Koszul algebras was singled out by Priddy [39] as sharing some of the good homological properties of polynomial algebras. Examples include free associative algebras, universal enveloping algebras of Lie algebras, the Steenrod algebra, and the Dyer-Lashoff algebra. A Koszul algebra is quadratic: it is generated by a vector space $V$, and all relations lie in $V \oplus (V \otimes V)$. Ginzburg and Kapranov [22] have introduced a class of operads, called Koszul, which share many properties with Koszul algebras. Like those authors, we restrict attention to Koszul operads which are the analogue of homogeneous Koszul algebras.

An operad $\mathbf{a}$ is generated by an $\mathbb{S}$-module $\mathbf{v}$ with relations the $\mathbb{S}$-submodule $\mathbf{r} \subset \mathbb{T}\mathbf{v}$ if it is a pushout in the category of operads

$$\begin{array}{ccc} \mathbb{T}\mathbf{r} & \longrightarrow & \mathbb{T}\mathbf{v} \\ \downarrow & & \downarrow \\ \mathbb{1} & \longrightarrow & \mathbf{a} \end{array}$$

Informally, there is a surjective map of operads $\mathbb{T}\mathbf{v} \to \mathbf{a}$ whose kernel is the ideal of $\mathbb{T}\mathbf{v}$ generated by $\mathbf{r}$.

Similarly, a cooperad $\mathbf{z}$ is cogenerated by an $\mathbb{S}$-module $\mathbf{w}$ with co-relations the quotient $\mathbb{S}$-module $\mathbf{q}$ of $\mathbb{C}\mathbf{w}$ if it is a pullback in the category of cooperads

$$\begin{array}{ccc} \mathbf{z} & \longrightarrow & \mathbb{1} \\ \downarrow & & \downarrow \\ \mathbb{C}\mathbf{w} & \longrightarrow & \mathbb{C}\mathbf{q} \end{array}$$

**Definition 2.20.** ([22], 2.1.7) An operad $\mathbf{a}$ is quadratic if it has generators $\mathbf{v}$ with $\mathbf{v}(S) = 0$ for $|S| \neq 2$, and relations $\mathbf{r}$ with $\mathbf{r}(S) = 0$ for $|S| \neq 3$.

A cooperad $\mathbf{z}$ is quadratic if it is cogenerated by an $\mathbb{S}$-module $\mathbf{u}$ such that $\mathbf{u}(S) = 0$ for $|S| \neq 2$, with co-relations $\mathbf{q}$ such that $\mathbf{q}(S) = 0$ for $|S| \neq 3$.

Note that if $\mathbf{v}$ satisfies $\mathbf{v}(S) = 0$ for $|S| \neq 2$, then $\mathbb{T}\mathbf{v}(3) = (\mathbf{v} \circ \mathbf{v})(3)$ is the direct sum of three copies of $\mathbf{v}(2) \otimes \mathbf{v}(2)$, corresponding to the three binary trees with three leaves: as an $\mathbb{S}_3$-module, it is the induced representation $\mathrm{Ind}_{\mathbb{S}_2}^{\mathbb{S}_3}(\mathbf{v}(2) \otimes \mathbf{v}(2))$, where $\mathbb{S}_2$ acts trivially on the first factor $\mathbf{v}(2)$. Thus, a quadratic operad is determined by the chain complex $\mathbf{v}(2)$ and the relations $\mathbf{r}(3)$, which may be any $\mathbb{S}_3$-invariant subcomplex of $\mathbf{v}(2) \circ \mathbf{v}(2)$. Similarly, a quadratic cooperad is determined by the chain complex $\mathbf{u}(2)$ and the co-relations $\mathbf{q}(3)$, which may be any $\mathbb{S}_3$-covariant quotient of $\mathbf{u}(2) \circ \mathbf{u}(2)$.

**Definition 2.21.** The dual $\mathbf{a}^\perp$ of a quadratic operad $\mathbf{a}$ with generators $\mathbf{v}$ and relations $\mathbf{r}$ is the quadratic cooperad cogenerated by $\Sigma \mathbf{v}$, with co-relations $\mathbf{q}$ such that

$$\mathbf{q}(3) = (\Sigma \mathbf{v} \circ \Sigma \mathbf{v})(3)/\Sigma^2 \mathbf{r}(3).$$

The dual $\mathbf{z}^\perp$ of a quadratic cooperad $\mathbf{a}$ is the quadratic operad generated by $\Sigma^{-1}\mathbf{u}$, with relations $\mathbf{r}$ such that $\mathbf{r}(3) = \ker\bigl(\Sigma^{-2}\mathbf{q}(3) \to (\Sigma^{-1}\mathbf{u} \circ \Sigma^{-1}\mathbf{u})(3)\bigr)$.

It is clear that $(\mathbf{a}^\perp)^\perp \cong \mathbf{a}$. Ginzburg and Kapranov consider instead the operad $\mathbf{a}^! = \Lambda^{-1}(\mathbf{a}^\perp)^*$; we prefer to work with the cooperad $\mathbf{a}^\perp$, in order to avoid taking linear duals.

A Koszul operad is a quadratic operad $\mathbf{a}$ whose bar cooperad is weakly equivalent to its dual $\mathbf{a}^\perp$, in a sense which we now explain.

**Definition 2.22.** Let $\mathcal{T}(S, i) \subset \mathcal{T}(S)$, $1 \leq i < |S|$, be the set of trees on a finite set $S$ with $i$ vertices. A tree $\mathcal{S} \in \mathcal{T}(S, |S| - 1)$ is called a binary tree.

If $\mathcal{S}$ is a binary tree, then all of its vertices have valence 2. Similarly, if $\mathcal{S} \in \mathcal{T}(S, |S|-2)$, then one of the vertices of $\mathcal{S}$ has valence 3 and the remainder have valence 2.

The grading of $\mathcal{T}(S)$ induces a grading of the bar cooperad $\mathcal{B}\mathbf{a}$ of an operad $\mathbf{a}$:
$$(\mathcal{B}_i \mathbf{a})(S) = \bigoplus_{\mathcal{S} \in \mathcal{T}(S,i)} (\Sigma \bar{\mathbf{a}})(\mathcal{S}).$$

Furthermore, the differential $\partial$ of $\mathcal{B}\mathbf{a}$ maps $\mathcal{B}_i \mathbf{a}$ to $\mathcal{B}_{i-1}\mathbf{a}$:
$$\mathcal{B}_{|S|-1}\mathbf{a}(S) \xrightarrow{\partial} \mathcal{B}_{|S|-2}\mathbf{a}(S) \xrightarrow{\partial} \ldots \xrightarrow{\partial} \mathcal{B}_1 \mathbf{a}(S) \xrightarrow{\partial} \mathcal{B}_0 \mathbf{a}(S) \to 0.$$

If $\mathbf{a}$ is a quadratic operad with dual cooperad $\mathbf{a}^\perp$, the kernel of the differential
$$\partial : \mathcal{B}_{|S|-1}\mathbf{a}(S) \to \mathcal{B}_{|S|-2}\mathbf{a}(S)$$
may be naturally identified with the dual quadratic cooperad $\mathbf{a}^\perp(S)$.

**Definition 2.23.** ([22], 4.1.3) A quadratic operad $\mathbf{a}$ is Koszul if the inclusion $\mathbf{a}^\perp \hookrightarrow \mathcal{B}\mathbf{a}$ is a weak equivalence.

There is an analogous grading of the cobar operad $\mathcal{B}^*\mathbf{z}$ of a connected cooperad $\mathbf{z}$,
$$(\mathcal{B}_i^* \mathbf{z})(S) = \bigoplus_{\mathcal{S} \in \mathcal{T}(S,i)} (\Sigma^{-1} \bar{\mathbf{z}})(\mathcal{S}).$$

Furthermore, the differential $\partial^*$ of $\mathcal{B}^*\mathbf{z}$ maps $\mathcal{B}_i^*\mathbf{z}$ to $\mathcal{B}_{i+1}^*\mathbf{z}$:
$$0 \to \mathcal{B}_0^* \mathbf{z}(S) \xrightarrow{\partial^*} \mathcal{B}_1^*\mathbf{z}(S) \xrightarrow{\partial^*} \ldots \xrightarrow{\partial} \mathcal{B}_{|S|-2}^*\mathbf{z}(S) \xrightarrow{\partial^*} \mathcal{B}_{|S|-1}^*\mathbf{z}(S).$$

If $\mathbf{z}$ is a quadratic cooperad, the cokernel of the differential $\mathcal{B}_{|S|-2}^*\mathbf{z}(S) \xrightarrow{\partial^*} \mathcal{B}_{|S|-1}^*\mathbf{z}(S)$ may be naturally identified with the dual quadratic operad $\mathbf{z}^\perp$. The quadratic cooperad $\mathbf{z}$ is Koszul if the projection $\mathcal{B}^*\mathbf{z} \to \mathbf{z}^\perp$ is a weak equivalence.

**Proposition 2.24.** ([22], 4.1.4) *A quadratic operad* $\mathbf{a}$ *is Koszul if and only if its dual cooperad* $\mathbf{a}^\perp$ *is Koszul.*

*Proof.* The operad $\mathbf{a}$ is Koszul if and only if the inclusion $\mathbf{a}^\perp \hookrightarrow \mathcal{B}\mathbf{a}$ is a weak equivalence. Applying the cobar functor, we see that this is the case if and only if the inclusion $\mathcal{B}^*\mathbf{a}^\perp \hookrightarrow \mathcal{B}^*\mathcal{B}\mathbf{a}$ is a weak equivalence. But the counit $\mathcal{B}^*\mathcal{B}\mathbf{a} \to \mathbf{a} \cong (\mathbf{a}^\perp)^\perp$ is a weak equivalence, and the result follows. $\square$

The associative operad $\mathbf{e}_1$ is Koszul: its dual is the cooperad $\mathbf{z}_1 = \Lambda^{-1}\mathbf{e}_1^*$ whose coalgebras are the suspensions of coassociative coalgebras. In Belinson-Ginzburg [8], it is proved that the operad $\mathbf{e}_\infty$ in characteristic zero is also Koszul, with dual the cooperad $\mathbf{z}_\infty = \Lambda^{-1}\mathcal{L}^*$ whose coalgebras are the suspensions of Lie coalgebras. In Chapter 3, we will prove that in characteristic zero, the operads $\mathbf{e}_n$ introduced in Section 1.3 are Koszul for all $1 \leq n \leq \infty$, with dual $\Lambda^{-n}\mathbf{e}_n^*$.

If $\mathbf{a}$ is a quadratic operad, the inclusion $\mathbf{a}^\perp \hookrightarrow \mathcal{B}\mathbf{a}$ induces a twisting cochain $\Phi \in \mathcal{T}(\mathbf{a}^\perp, \mathbf{a})$; explicitly, this is the map which sends the cogenerators $\Sigma \mathbf{v}$ of $\mathbf{a}^\perp$ to the generators $\mathbf{v}$ of $\mathbf{a}$ and otherwise vanishes. We write $\mathsf{B}(\mathbf{a})$ for the associated functor $\mathbb{B}(\Phi) : \mathcal{M}^{\mathbf{a}} \to \mathcal{M}_{\mathbf{a}^\perp}$, and, slightly abusively, $\Omega(\mathbf{a})$ for the functor $\Omega(\Phi) : \mathcal{M}_{\mathbf{a}^\perp} \to \mathcal{M}^{\mathbf{a}}$. The following theorem is an evident corollary of Theorem 2.19.

**Theorem 2.25.** *If* $\mathbf{a}$ *is a quadratic operad, there is an adjunction*
$$\Omega(\mathbf{a}) : \mathcal{M}_{\mathbf{a}^\perp} \rightleftarrows \mathcal{M}^{\mathbf{a}} : \mathsf{B}(\mathbf{a}).$$

*If* $\mathbf{a}$ *is Koszul, the counit* $\Omega(\mathbf{a}, \mathsf{B}(\mathbf{a}, A)) \to A$ *and unit* $C \to \mathsf{B}(\mathbf{a}, \Omega(\mathbf{a}, C))$ *of this adjunction are weak equivalences.*

## 3. The $n$-algebra operad is Koszul

In his work on $A_\infty$-algebras, Stasheff introduced a series of $(k-2)$-dimensional polytopes $K(k)$ [49]. In this chapter, we prove that the operads $\mathsf{e}_n$ are Koszul, using a generalization $\mathsf{F}_n(k)$ of $K(k)$ due to Fulton and MacPherson [15]. The space $\mathsf{F}_n(k)$ is obtained by compactifying the quotient of $\mathbb{F}_n(k)$ by translations and dilatations: it is a manifold with corners, whose codimension $p$ faces correspond to trees $\mathcal{S} \in \mathcal{T}(k)$ with $p+1$ vertices.

Beilinson and Ginzburg [8] have shown that a similar approach may be used to prove that the commutative operad $\mathsf{e}_\infty$ is Koszul, with dual the cooperad $\Lambda^{-1}\mathcal{L}^*$. They prove this result by studying the spectral sequence for the stratified space $\overline{\mathcal{M}}_{0,k+1}$, the moduli space of $(k+1)$-pointed genus zero curves: this is a smooth $(k-2)$-dimensional projective variety. Our proof that $\mathsf{e}_n$ is Koszul is influenced by their work. One difference is that their spectral sequence collapses by Deligne's mixed Hodge theory. Our proof of collapse is more elementary, and relies on consideration of all $1 < n < \infty$ simultaneously.

For $1 \le n < \infty$, denote by $\mathbf{z}_n$ the cooperad $\mathbf{z}_n(k) = \Lambda^{-n}\mathsf{e}_n^*(k)$. Explicitly, $\mathbf{z}_n(k)$ is the graded vector space such that
$$(\mathbf{z}_n(k))_i = \begin{cases} H^{n(k-1)-i}(\mathbb{F}_n(k)), & k > 0, \\ 0, & k = 0. \end{cases}$$

It is clear that $\mathbf{z}_n$ is connected and coaugmented. We call a $\mathbf{z}_n$-coalgebra an $n$-coalgebra, and denote the cofree $n$-coalgebra $\mathsf{C}(\mathbf{z}_n, V)$ by $\mathsf{C}_n(V)$.

The following theorem is the main result of this chapter.

**Theorem 3.1.** *If $1 < n < \infty$, then the operad $\mathsf{e}_n$ is Koszul, and $\mathsf{e}_n^\perp \cong \mathbf{z}_n$.*

The operad $\mathsf{e}_1$ is also Koszul, but this result is a lot easier to prove: it is a consequence of the fact that the Hochschild homology of the free associative algebra $\mathsf{T}_1(V)$ is isomorphic to $V$ in degree 1, and vanishes in other degrees.

The proof proceeds by considering the cobar complex of $\mathbf{z}_n$:
$$0 \to \mathcal{B}_1^* \mathbf{z}_n(k) \xrightarrow{\partial^*} \mathcal{B}_2^* \mathbf{z}_n(k) \xrightarrow{\partial^*} \ldots \to \mathcal{B}_{k-2}^* \mathbf{z}_n(k) \xrightarrow{\partial^*} \mathcal{B}_{k-1}^* \mathbf{z}_n(k).$$

In Section 3.1, we calculate the cokernel of the map $\mathcal{B}_{k-2}^* \mathbf{z}_n(k) \xrightarrow{\partial^*} \mathcal{B}_{k-1}^* \mathbf{z}_n(k)$, showing that it equals the Poisson operad $\mathbf{p}_n(k)$, where $\mathbf{p}_n$ is the $n$-Poisson operad. The more difficult part of the proof is devoted to showing that elsewhere the cobar complex is exact, showing that $\mathcal{B}^* \mathbf{z}_n \simeq \mathbf{p}_n$. Since by Lemma 1.8, $\mathbf{p}_n$ and $\mathsf{e}_n$ have the same Poincaré series, we see that $\mathsf{e}_n$ must be the same operad as $\mathbf{p}_n$, which is Theorem 1.6. The proof of Theorem 3.1 follows easily from combining all of these partial results.

We identify the cobar complex of $\mathbf{z}_n$ with the $E^1$-term of the homology spectral sequence for the manifold with corners $\mathsf{F}_n(k)$: Lefschetz duality for this manifold with corners lies at the origin of the self-duality $\mathsf{e}_n^\perp \cong \Lambda^{-n}\mathsf{e}_n^*$. From collapse of the spectral sequence, we see that the cobar complex of $\mathbf{z}_n$ is exact. In the next chapter, we will prove a homotopy equivalence between the categories of $n$-algebras and $n$-coalgebras, generalizing the well-known cases $n = 1$, relating associative dg-algebras and connected coassociative dg-coalgebras, and $n = \infty$, relating commutative dg-algebras and connected co-Lie dg-coalgebras.

**3.1. The dual of $\mathbf{p}_n$.** This section is devoted to the proof of the following result.

**Lemma 3.2.** *The cokernel of the map $\mathcal{B}_{k-2}^* \mathbf{z}_n(k) \xrightarrow{\partial^*} \mathcal{B}_{k-1}^* \mathbf{z}_n(k)$ is naturally isomorphic to $\mathbf{p}_n(k)$, where $\mathbf{p}_n$ is the $n$-Poisson operad.*

*Proof.* It is convenient to rewrite the quadratic relations of the operad $\mathbf{p}_n$ in terms of the bracket $\{a,b\}$, related to the bracket $[a,b]$ in the statement of Theorem 1.6 by the formula

$$[a,b] = (-1)^{(n-1)|a|}\{a,b\}.$$

In terms of this bracket (which of course equals the old one if $n$ is odd), these relations may be written

$$\{v,w\} = (-1)^{|v||w|+n}\{w,v\},$$
$$\delta\{v,w\} = (-1)^{n-1}\bigl(\{\delta v,w\} + (-1)^{|v|}\{v,\delta w\}\bigr),$$
$$\{\{u,v\},w\} + (-1)^{|u|(|v|+|w|)}\{\{v,w\},u\} + (-1)^{|w|(|u|+|v|)}\{\{w,u\},v\} = 0,$$
$$\{uv,w\} = (-1)^{|u|(|v|+|w|)}\{v,w\}u + (-1)^{|w|(|u|+|v|)+n}\{w,u\}v.$$

This redefinition of the bracket has the effect of making $\{v,w\}$ behave, as far as the sign convention is concerned, as if it were written $\{-,-\} \otimes v \otimes w$, where $\{-,-\}$ is a symbol having degree $n-1$.

Let $\mathbf{v}$ be the $\mathbb{S}$-module

$$\mathbf{v}(S) = \begin{cases} \Sigma^{-1}\mathbf{z}_n(S), & |S| = 2, \\ 0, & |S| \neq 2. \end{cases}$$

We see that $\mathcal{B}^*_{k-1}\mathbf{z}_n(k) \cong \mathbb{T}\mathbf{v}(k)$ for all $k$. We wish to identify the cokernel of the differential

$$\mathcal{B}^*_{k-2}\mathbf{z}_n(k) \xrightarrow{\partial^*} \mathcal{B}^*_{k-1}\mathbf{z}_n(k) \cong \mathbb{T}\mathbf{v}(k)$$

with $\mathbf{p}_n(k)$. It is clear that this cokernel is a quadratic operad, with generators $\Sigma^{-1}\mathbf{z}_n(2)$, and to calculate the relations, it suffices to consider the above complex in the case $k=3$.

For $k=3$, the map $\partial^*$ has domain

$$\mathcal{B}^*_1\mathbf{z}_n(3) = \Sigma^{-1}\mathbf{z}_n(3) \cong \Sigma^{2n-1}\mathbf{e}^*_n(3),$$

since there is only one tree with one vertex and three incoming external edges. Proposition 1.5 shows that

$$\Sigma^{-1}\mathbf{z}_n(3) \text{ has basis } \begin{cases} \Sigma^{2n-1}(1), & \text{in degree } 2n-1, \\ \Sigma^{2n-1}(\omega_{12}), \Sigma^{2n-1}(\omega_{23}), \Sigma^{2n-1}(\omega_{31}), & \text{in degree } n, \\ \Sigma^{2n-1}(\omega_{12}\omega_{23}), \Sigma^{2n-1}(\omega_{23}\omega_{31}), & \text{in degree } 1, \end{cases}$$

and vanishes in other degrees.

The codomain $\mathcal{B}^*_2\mathbf{z}_n(3) \cong \mathbb{T}\mathbf{v}(3)$ of $\partial^*$ is the sum of three copies of

$$\Sigma^{-1}\mathbf{z}_n(2) \otimes \Sigma^{-1}\mathbf{z}_n(2) \cong \Sigma^{n-1}\mathbf{e}^*_n(2) \otimes \Sigma^{n-1}\mathbf{e}^*_n(2),$$

corresponding to the three labelled binary trees in $\mathcal{T}(3)$:

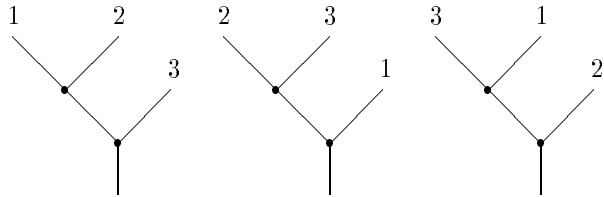

The space $\mathbf{v}(2) = \Sigma^{-1}\mathbf{z}_n(2) \cong \Sigma^{n-1}\mathbf{e}^*_n(2)$ has basis $\Sigma^{n-1}(\omega)$, in degree 0, and $\Sigma^{n-1}(1)$, in degree $n-1$. Thus, the free operad $\mathbb{T}\mathbf{v}$ is generated by two operations: a graded commutative operation of degree 0, denoted $ab$, and an operation of degree $n-1$, graded commutative if $n$ is even, and graded anticommutative if $n$ is odd, denoted $\{a,b\}$. A basis for $\mathbb{T}\mathbf{v}(3)$ is formed by ordered triples, each entry of which is a basis of the tensor product $\Sigma^{n-1}\mathbf{e}^*_n(2) \otimes \Sigma^{n-1}\mathbf{e}^*_n(2)$, or alternatively by the words that can be formed from three indeterminates $a,b,c$, representing even elements of a free $n$-algebra, by means of these two products.

(1) In degree $2(n-1)$, $\mathbb{T}\mathbf{v}(3)$ has basis
$$(\Sigma^{n-1}(1) \otimes \Sigma^{n-1}(1), 0, 0) \equiv \{\{a,b\},c\},$$
$$(0, \Sigma^{n-1}(1) \otimes \Sigma^{n-1}(1), 0) \equiv \{\{b,c\},a\},$$
$$(0, 0, \Sigma^{n-1}(1) \otimes \Sigma^{n-1}(1)) \equiv \{\{c,a\},b\}.$$

(2) In degree $n-1$, $\mathbb{T}\mathbf{v}(3)$ has basis
$$(\Sigma^{n-1}(\omega) \otimes \Sigma^{n-1}(1), 0, 0) \equiv \{a,b\}c,$$
$$(\Sigma^{n-1}(1) \otimes \Sigma^{n-1}(\omega), 0, 0) \equiv \{ab,c\},$$
$$(0, \Sigma^{n-1}(\omega) \otimes \Sigma^{n-1}(1), 0) \equiv \{b,c\}a,$$
$$(0, \Sigma^{n-1}(1) \otimes \Sigma^{n-1}(\omega), 0) \equiv \{bc,a\},$$
$$(0, 0, \Sigma^{n-1}(\omega) \otimes \Sigma^{n-1}(1)) \equiv \{c,a\}b,$$
$$(0, 0, \Sigma^{n-1}(1) \otimes \Sigma^{n-1}(\omega)) \equiv \{ca,b\}.$$

(3) In degree $0$, $\mathbb{T}\mathbf{v}(3)$ has basis
$$(\Sigma^{n-1}(\omega) \otimes \Sigma^{n-1}(\omega), 0, 0) \equiv (ab)c,$$
$$(0, \Sigma^{n-1}(\omega) \otimes \Sigma^{n-1}(\omega), 0) \equiv (bc)a,$$
$$(0, 0, \Sigma^{n-1}(\omega) \otimes \Sigma^{n-1}(\omega)) \equiv (ca)b.$$

In other degrees, $\mathbb{T}\mathbf{v}(3)$ vanishes. Note that, by the sign rule,
$$\Sigma^{n-1}(\omega) \otimes \Sigma^{n-1}(\omega) = (-1)^{n-1}\Sigma^{2n-2}(\omega \otimes \omega),$$
and similarly for the other vectors of the basis.

We must identify the relations in $\mathbf{p}_n(3)$ with the image of the map $\partial^* : \mathcal{B}_1^*\mathbf{z}_n(3) \to \mathcal{B}_2^*\mathbf{z}_n(3)$. In degree $2(n-1)$, there is just one relation,
$$\partial^* \Sigma^{2n-1}(1) = (\Sigma^{n-1}(1) \otimes \Sigma^{n-1}, \Sigma^{n-1}(1) \otimes \Sigma^{n-1}, \Sigma^{n-1}(1) \otimes \Sigma^{n-1})$$
$$= \{\{a,b\},c\} + \{\{b,c\},a\} + \{\{c,b\},a\},$$
which is the Jacobi relation for the bracket $\{-,-\}$.

To calculate the differential $\partial^*$ in degrees $n-1$ and $2(n-1)$, we use the explicit formulas for the three maps $\mathbf{e}_n^*(3) \to \mathbf{e}_n^*(2) \otimes \mathbf{e}_n^*(2)$ corresponding to the three binary trees $\mathcal{S} \in \mathcal{T}(3,2)$:
$$\omega_{12} \mapsto (1 \otimes \omega, (-1)^n \omega \otimes 1, \omega \otimes 1),$$
$$\omega_{23} \mapsto (\omega \otimes 1, 1 \otimes \omega, (-1)^n \omega \otimes 1),$$
$$\omega_{31} \mapsto ((-1)^n \omega \otimes 1, \omega \otimes 1, 1 \otimes \omega).$$

Applying $\Sigma^{2n-1}$ to the left-hand side and $\Sigma^{2n-2}$ to the right-hand side, we obtain the desired formula for $\partial^*$ in degree $2n-1$:
$$\partial^* \Sigma^{2n-1}(\omega_{12}) = (\Sigma^{n-1}(1) \otimes \Sigma^{n-1}(\omega), -\Sigma^{n-1}(\omega) \otimes \Sigma^{n-1}(1), (-1)^{n-1}\Sigma^{n-1}(\omega) \otimes \Sigma^{n-1}(1))$$
$$\equiv \{ab,c\} - \{b,c\}a + (-1)^{n-1}\{c,a\}b,$$
$$\partial^* \Sigma^{2n-1}(\omega_{23}) = ((-1)^{n-1}\Sigma^{n-1}(\omega) \otimes \Sigma^{n-1}(1), \Sigma^{n-1}(1) \otimes \Sigma^{n-1}(\omega), -\Sigma^{n-1}(\omega) \otimes \Sigma^{n-1}(1))$$
$$\equiv (-1)^{n-1}\{a,b\}c + \{bc,a\} - \{c,a\}b,$$
$$\partial^* \Sigma^{2n-1}(\omega_{31}) = (-\Sigma^{n-1}(\omega) \otimes \Sigma^{n-1}(1), (-1)^{n-1}\Sigma^{n-1}(\omega) \otimes \Sigma^{n-1}(1), \Sigma^{n-1}(1) \otimes \Sigma^{n-1}(\omega))$$
$$\equiv -\{a,b\}c + (-1)^{n-1}\{b,c\}a + \{ca,b\}.$$

Bearing in mind that $\{a,b\} = (-1)^n\{b,a\}$, we obtain the Poisson relations

$$\{ab,c\} = \{b,c\}a + \{a,c\}b,$$
$$\{bc,a\} = \{c,a\}b + \{b,a\}c,$$
$$\{ca,b\} = \{a,b\}c + \{c,b\}a.$$

To calculate the operator $\partial^*$ in degree 1, we use the formula

$$\omega_{12}\omega_{23} \mapsto ((-1)^{n-1}\omega \otimes \omega, (-1)^n\omega \otimes \omega, 0),$$
$$\omega_{23}\omega_{31} \mapsto (0, (-1)^{n-1}\omega \otimes \omega, (-1)^n\omega \otimes \omega),$$
$$\omega_{31}\omega_{12} \mapsto ((-1)^n\omega \otimes \omega, 0, (-1)^{n-1}\omega \otimes \omega).$$

These add up to zero, as they must by Arnold's relation. We see that $\partial^*$ in degree 1 is given by the formulas

$$\partial^* \Sigma^{2n-1}(\omega_{12}\omega_{23}) = (\Sigma^{n-1}(\omega) \otimes \Sigma^{n-1}(\omega), -\Sigma^{n-1}(\omega) \otimes \Sigma^{n-1}(\omega), 0) \equiv (ab)c - (bc)a,$$
$$\partial^* \Sigma^{2n-1}(\omega_{23}\omega_{31}) = (0, \Sigma^{n-1}(\omega) \otimes \Sigma^{n-1}(\omega), -\Sigma^{n-1}(\omega) \otimes \Sigma^{n-1}(\omega)) \equiv (bc)a - (ca)b.$$

This gives the associativity of the product $ab$. □

**3.2. The Fulton-MacPherson compactification of configuration space.** Denote by $G(n)$ the subgroup of affine transformations of $\mathbb{R}^n$, of dimension $n+1$, generated by translations and dilatations by a positive real number. This group acts freely on the configuration spaces $\mathbb{F}_n(S)$ if $|S| > 1$; denote by $\mathring{\mathsf{F}}_n(S)$ the quotient $\mathbb{F}_n(S)/G(n)$. Thus, $\mathring{\mathsf{F}}_n(S)$ is a manifold of dimension $n(|S|-1) - 1$, and the action of $\mathbb{S}(S)$ on $\mathring{\mathsf{F}}_n(S)$ is free. With the convention that $\mathring{\mathsf{F}}_n(1)$ and $\mathring{\mathsf{F}}_n(0)$ are empty, the spaces $\mathring{\mathsf{F}}_n(S)$ assemble to form an $\mathbb{S}$-space. Furthermore, the group $\mathrm{GL}(n)$ acts on $\mathring{\mathsf{F}}_n$, and this action commutes with the action of $\mathbb{S}$.

Fulton and MacPherson [15] have constructed a natural compactification $\mathsf{F}_n$ of the $\mathbb{S}$-space $\mathring{\mathsf{F}}_n$. The underlying point set of this compactification is the free operad $\mathbb{T}\mathring{\mathsf{F}}_n$ generated by $\mathring{\mathsf{F}}_n$. Observe that only the nests $\mathcal{S} \in \mathcal{N}(S)$ contribute to $\mathbb{T}\mathring{\mathsf{F}}_n(S)$, since $\mathring{\mathsf{F}}_V(1)$ and $\mathring{\mathsf{F}}_V(0)$ are empty.

Fulton and MacPherson show how to glue $\mathbb{T}\mathring{\mathsf{F}}_n(S)$ together naturally by a continuous bijective map $\mathbb{T}\mathring{\mathsf{F}}_n(S) \to \mathsf{F}_n(S)$, to form a manifold with corners. This construction has the following properties:

(1) The gluing map $\mathbb{T}\mathring{\mathsf{F}}_n(S) \to \mathsf{F}_n(S)$ is equivariant for the actions of $\mathbb{S}$ and $\mathrm{GL}(n)$.
(2) $\mathring{\mathsf{F}}_n$ is an equivariant deformation retract of $\mathsf{F}_n$, and hence $\mathsf{F}_n(S)$ is homotopy equivalent to $\mathbb{F}_n(S)$.
(3) The operad structure of $\mathbb{T}\mathring{\mathsf{F}}_n$ descends to a differentiable operad structure on $\mathsf{F}_n$.
(4) For $T \subset S$, let $D(T)$ be the union of the faces of $\mathsf{F}_n(S)$ which correspond to components of $\mathbb{T}\mathring{\mathsf{F}}_n(S)$ indexed by those trees $\mathcal{S} \in \mathcal{T}(S)$ such that $T$ is the set of leaves of a subtree of $\mathcal{S}$. Then $D(T)$ is a submanifold with corners of $\mathsf{F}_n(S)$ of codimension one, and $\mathsf{F}_n(\mathcal{S})$ is the transverse intersection of those $D(T)$ such that $\mathsf{F}_n(\mathcal{S}) \subset D(T)$.

The construction of $\mathsf{F}_n$ is made by a sequence of blowings up. Fulton and MacPherson work in the category of complex varieties: the blow up $\mathrm{Bl}_X(M)$ of smooth variety $M$ along a smooth subvariety $X \subset M$ is obtained by gluing together $M \setminus X$ and the projectivized normal bundle of $X$ in $M$. This blow-up is again smooth, and blowing up has the effect of replacing the original submanifold by a divisor (codimension one submanifold). By contrast, we work in the category of real manifolds with corners. In this category, the blow up along a submanifold $X \subset M$ transverse to all faces of $M$ is obtained by gluing together $M \setminus X$ and the normal sphere bundle of $X$: this is once more a manifold with corners. (For a more detailed description of the Fulton-MacPherson compactification in this context, see Axelrod and Singer [4]).

If $S$ is a finite set, let $\mathrm{Bl}_\Delta((\mathbb{R}^n)^S)$ be the blow-up of $(\mathbb{R}^n)^S$ along the diagonal $\Delta \hookrightarrow (\mathbb{R}^n)^S$. The group $G(n)$ acts freely on $\mathrm{Bl}_\Delta((\mathbb{R}^n)^S)$, and the quotient is a manifold with boundary. There is a natural inclusion $\mathring{\mathsf{F}}_n(S) \hookrightarrow \mathrm{Bl}_\Delta((\mathbb{R}^n)^S)/G(n)$, and we define $\mathsf{F}_n(S)$ to be the closure of the diagonal embedding of $\mathring{\mathsf{F}}_n(S)$ in the product

$$\prod_{T \subset S} \mathrm{Bl}_\Delta((\mathbb{R}^n)^T)/G(n).$$

Fulton and MacPherson prove that $\mathsf{F}_n(S)$ has all of the asserted properties.

The $\mathbb{S}$-spaces $\mathsf{F}_1$ and $\mathsf{F}_2$ are classical:

(1) The space $\mathsf{F}_1(k)$ is equivariantly diffeomorphic to $K(k) \times \mathbb{S}_k$, where $K(k)$ is the $(k-2)$-dimensional polytope, the associahedron, introduced by Stasheff [49] in the study of homotopy associative spaces.
(2) The space $\mathsf{F}_2(k)$ is a circle bundle over the moduli space $\overline{\mathcal{M}}_{0,k+1}$ of stable rational curves with $k+1$ marked points constructed by Knudsen [28].

We may also describe $\mathsf{F}_n(2)$ and $\mathsf{F}_n(3)$ for general $n$.

(1) The space $\mathsf{F}_n(2) = \mathring{\mathsf{F}}_n(2)$ is the sphere $S^{n-1}$, reflecting the fact that there is only one tree with two leaves:

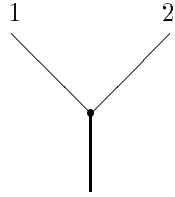

The group $\mathrm{GL}(n)$ acts by rotating $\mathbb{R}^n$, while the involution in $\mathbb{S}_2$ acts by the antipodal map.
(2) The space $\mathsf{F}_n(3)$ has four strata, the interior $\mathring{\mathsf{F}}_n(3)$ and the three boundary components $S^{n-1} \times S^{n-1}$, corresponding to the four trees with three leaves:

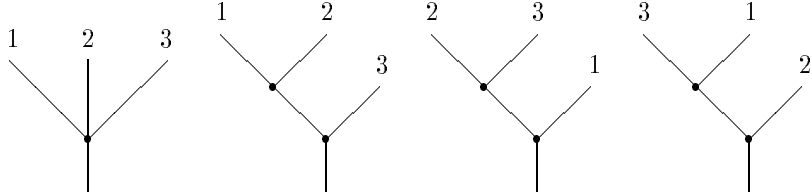

The compactification $\mathsf{F}_n(4)$ already has 26 strata, and we leave it to the curious reader to draw a picture of it.

Although the natural homotopy equivalence $\mathcal{F}_n(k) \to \mathsf{F}_n(k)$ obtained by the composition

$$\mathcal{F}_n(k) \to \mathbb{F}_n(k) \to \mathring{\mathsf{F}}_n(k) \hookrightarrow \mathsf{F}_n(k)$$

is not a map of operads, on taking homology it induces an isomorphism of dg-operads $\mathsf{e}_n = H_\bullet(\mathcal{F}_n) \cong H_\bullet(\mathring{\mathsf{F}}_n)$. The topological operad $\mathsf{F}_n$ expresses the structure of the dg-operad $\mathsf{e}_n$ in its geometry much better than the topological operad $\mathcal{F}_n$.

**3.3. $\mathsf{e}_n$ is Koszul.** This section is devoted to the proof of Theorem 3.1. In its proof, we employ the Fulton-MacPherson compactifications $\mathsf{F}_n(k)$ of Section 3.2.

If $M$ is a compact $n$-dimensional manifold with corners, denote by $M[p]$ the union of the faces of $M$ with codimension $p$, and by $F_p M$ its closure, the union of the faces of $M$ with codimension $i \geq p$. The Lefschetz duality theorem relates the homology of $M[p]$ to the cohomology of the pair $(F_p M, F_{p+1} M)$:

$$H_q(M[p]) \cong H^{n-p-q}(F_p M, F_{p+1} M).$$

Here, homology and cohomology are taken with respect to any abelian group of coefficients.

**Lemma 3.3.** *Let $M$ be a compact $n$-dimensional manifold with corners. There is a spectral sequence converging to $H^{n-\bullet}(M)$ with $E^1_{pq} = H_q(M[p])$, such that $d^i : E^{pq}_i \to E^i_{p-i,q+i-1}$. The differential $d^1 : E^1_{pq} \to E^1_{p-1,q}$ is identified, by the Lefschetz duality theorem, with the boundary map $\partial$ of the cohomology exact sequence for the triple $(F_{p-1}M, F_p M, F_{p+1}M)$, by the commutative diagram*

$$\begin{array}{ccc} H^{n-p-q}(F_p M, F_{p+1}M) & \xrightarrow{\partial} & H^{n-p-q+1}(F_{p-1}M, F_p M) \\ \| & & \| \\ H_q(M[p]) & \xrightarrow{d^1} & H_q(M[p-1]) \end{array}$$

*Proof.* Associated to the filtration of $M$ by the closed subspaces $F_p M$, there is an ascending filtration of the singular cochains of $M$ by subcomplexes $F_p C^{n-\bullet}(M) = C^{n-\bullet}(M, F_{p+1}M)$. The spectral sequence of the proposition is the spectral sequence of this filtered complex, which clearly converges to $H^{n-\bullet}(M)$, since the filtration has finite depth.

It is easy to identify $E^0_{pq}$ with

$$F_p C^{n-p-q}(M)/F_{p-1} C^{n-p-q}(M) = C^{n-p-q}(M, F_{p+1}M)/C^{n-p-q}(M, F_p M)$$
$$= C^{n-p-q}(F_p M, F_{p+1}M).$$

This shows that $E^1_{pq}$ is isomorphic to $H^{n-p-q}(F_p M, F_{p-1}M) = H_q(M[p])$. □

The following lemma is the main step in the proof of Theorem 3.1.

**Lemma 3.4.** *In the spectral sequence for the manifold with corners $\mathsf{F}_n(k)$, the complex $(E^1_{p,q}, d^1)$ is naturally isomorphic to the bar cooperad $\mathcal{B}\mathbf{e}_n$:*

$$\begin{array}{ccc} E^1_{pq} & \xrightarrow{d^1} & E^1_{p-1,q} \\ \| & & \| \\ (\mathcal{B}_{p+1}\mathbf{a}(k))_{p+q+1} & \xrightarrow{\partial} & (\mathcal{B}_p \mathbf{e}_n(k))_{p+q} \end{array}$$

*Proof.* The space $\mathsf{F}_n(k)[p]$ of codimension $p$ faces of $\mathsf{F}_n(k)$ may be identified with the disjoint union over trees $\mathcal{S} \in \mathcal{T}(k, p+1)$ of the products

$$\mathring{\mathsf{F}}_n(\mathcal{S}) = \prod_{s \in \mathcal{S}} \mathring{\mathsf{F}}_n(\mathrm{in}(s)).$$

By the Künneth theorem, we see that

$$E^1_{p,\bullet} \cong \bigoplus_{\mathcal{S} \in \mathcal{T}(k,p+1)} \bigotimes_{s \in \mathcal{S}} H_\bullet(\mathring{\mathsf{F}}_n(\mathrm{in}(s)))$$
$$\cong \bigoplus_{\mathcal{S} \in \mathcal{T}(k,p+1)} \bigotimes_{s \in \mathcal{S}} \mathbf{e}_n(\mathrm{in}(s)) \cong \Sigma^{-p-1} \mathcal{B}_{p+1} \mathbf{e}_n(\mathcal{S}).$$

It remains to identify the differential $d^1 : E^1_{p,q} \to E^1_{p-1,q}$ with the differential

$$\partial : (\mathcal{B}_{p+1}\mathbf{a}(k))_{p+q+1} \to (\mathcal{B}_p \mathbf{a}(k))_{p+q}.$$

Consider the map of manifolds with corners $\mathsf{F}_n(\mathcal{S}) \to \mathsf{F}_n(k)$, where $\mathcal{S} \in \mathcal{T}(k,p)$. This induces a map of spectral sequences: summing over $\mathcal{S} \in \mathcal{T}(k,p)$, we obtain a map of $E^1$-terms

$$\begin{array}{ccc}
\bigoplus_{\mathcal{S} \in \mathcal{T}(k,p)} H_q(\mathsf{F}_n(\mathcal{S})[1]) & \xrightarrow{d^1} & \bigoplus_{\mathcal{S} \in \mathcal{T}(k,p)} H_q(\mathsf{F}_n(\mathcal{S})[0]) \\
\downarrow & & \parallel \\
H_q(\mathsf{F}_n(k)[p]) & \xrightarrow{d^1} & H_q(\mathsf{F}_n(k)[p-1])
\end{array}$$

The left-hand vertical arrow is surjective, so it suffices to calculate the differential in the top row.

If $M$ and $N$ are compact manifolds with boundaries,

$$(M \times N)[p] = \bigcup_{i=0}^{p} M[i] \times N[p-i].$$

The differential $H_q((M \times N)[p]) \xrightarrow{d^1_{p,q}} H_q((M \times N)[p-1])$ in the spectral sequence for the manifold with corners $M \times N$ may be rewritten by means of this isomorphism:

$$H_q((M \times N)[p]) \cong \bigoplus_{i=0}^{p} H_q(M[i] \times N[p-i]) \cong \bigoplus_{i=0}^{p} \bigoplus_{j=0}^{q} H_j(M[i]) \otimes H_{q-j}(N[p-i])$$

$$\xrightarrow{\bigoplus d^1_{i,j} \otimes 1 + 1 \otimes d^1_{p-i,q-j}} \bigoplus_{i=0}^{p-1} \bigoplus_{j=0}^{q} H_j(M[i]) \otimes H_{q-j}(N[p-i-1])$$

$$\longrightarrow H_q((M \times N)[p-1]),$$

reducing the calculation of $d^1_{p,q}$ to the corresponding calculation for $M$ and $N$.

Applying this observation to the manifold with corners $\mathsf{F}_n(\mathcal{S})$, we see that the differential $d^1_{1,\bullet}$ in its spectral sequence is the sum of differentials $d^1_{1,\bullet}$ over the factors $\mathsf{F}_n(\mathrm{in}(s))$, indexed by the vertices $s$ of the tree $\mathcal{S}$. Of course, the differential $\partial$ on the summand $\mathcal{B}\mathbf{e}_n(\mathcal{S})$ of the bar complex $\mathcal{B}\mathbf{e}_n$ is also a sum of operators $\partial_s$ over internal vertices of the tree $\mathcal{S}$; thus, it suffices to examine the case where $\mathcal{S}$ is a tree with a single vertex.

If $S$ is a finite set, $\mathsf{F}_n(k)[1]$ is a union of spaces $\coprod_{T \subset S} \mathring{\mathsf{F}}_n((S \setminus T) \cup \{T\}) \times \mathring{\mathsf{F}}_n(T)$. On each of these faces, the differential $d^1_{1,\bullet}$ may be identified with the corresponding product in the operad $\mathbf{e}_n$, and thus equals the corresponding term in the differential of the bar complex. Assembling all of these facts, we obtain the identification between $d^1_{p,\bullet}$ and $\partial : \Sigma^{-p-1} \mathcal{B}_{p+1} \mathbf{e}_n(k) \to \Sigma^{-p} \mathcal{B}_p \mathbf{e}_n(k)$. □

**Corollary 3.5.** *The spectral sequence for the $n(k-1)-1$-dimensional manifold with corners $\mathsf{F}_n(k)$ collapses at the $E^2$-term; in other words,*

$$H_\bullet(\mathcal{B}\mathbf{e}_n(k)) \cong H^{n(k-1)-\bullet}(\mathsf{F}_n(k)) \cong \mathbf{z}_n(k).$$

*Proof.* If $n > 1$, $\mathbf{e}_n \cong \mathbf{e}_{n+2}$ as $\mathbb{Z}/2$-graded operads by Proposition 1.5, and hence $\mathcal{B}\mathbf{e}_n \cong \mathcal{B}\mathbf{e}_{n+2}$ as $\mathbb{Z}/2$-graded chain complexes. This shows that

$$\bigoplus_{pq} \dim E^2_{pq} = \bigoplus_i \dim H_i(\mathcal{B}\mathbf{e}_n(k))$$

only depends on the parity of $n$. On the other hand,

$$\sum_{pq} \dim E^\infty_{pq} = k!$$

is independent of $n$. But $E^1_{pq} = 0$ unless $0 \le p \le k-1$ and $(n-1)|q$, the condition on $p$ coming from the fact that any nest with $k$ leaves has at most $k-1$ vertices, the condition on $q$ coming from Proposition 1.5. It follows that $d^i$ is non-zero only if $i \le k-2$ and $(n-1)|(i-1)$. Thus, for $n \ge k$, the spectral sequence collapses at the $E^2$-term. The collapse of the spectral sequence at $E^2$ for all $n > 1$ follows. □

We now have sufficient information on $\mathcal{B}\mathbf{e}_n$ to prove Theorem 3.1. First, note that the linear dual of the bar cooperad $\mathcal{B}\mathbf{e}_n$ may be identified with the cobar operad $\mathcal{B}^*\mathbf{e}_n^* \cong \Lambda^n \mathcal{B}^* \mathbf{z}_n$. Thus, the above corollary implies that
$$H_\bullet(\mathcal{B}^* \mathbf{z}_n(k)) \cong \mathbf{e}_n(k).$$

We argue that following sequence is exact:
$$0 \to \mathcal{B}_1^* \mathbf{z}_n(k) \xrightarrow{\partial^*} \mathcal{B}_2^* \mathbf{z}_n(k) \xrightarrow{\partial^*} \ldots \to \mathcal{B}_{k-2}^* \mathbf{z}_n(k) \xrightarrow{\partial^*} \mathcal{B}_{k-1}^* \mathbf{z}_n(k) \to \mathbf{e}_n(k) \to 0.$$

In Section 3.1, we proved that the cokernel of the map $\partial^* : \mathcal{B}_{k-2}^* \mathbf{z}_n(k) \to \mathcal{B}_{k-1}^* \mathbf{z}_n(k)$ is naturally isomorphic to $\mathbf{p}_n(k)$. Lemma 1.8 shows that the vector space $\mathbf{p}_n(k)$ has the same dimension as $\mathbf{e}_n(k)$, namely $k!$, and hence that the homology of $\mathcal{B}^* \mathbf{z}_n(k)$ is concentrated in the cokernel of $\partial^* : \mathcal{B}_{k-2}^* \mathbf{z}_n(k) \to \mathcal{B}_{k-1}^* \mathbf{z}_n(k)$; in other words, $\mathbf{e}_n = \mathbf{p}_n$ as operads. This completes the proof of Theorem 1.6 in the non-unital case; in particular, the operad $\mathbf{e}_n$ is seen to be quadratic.

Returning to the bar complex $\mathcal{B}\mathbf{e}_n(k)$, we see that the homology of this complex, isomorphic to $\mathbf{z}_n(k)$, is the kernel of the differential $\partial : \mathcal{B}_{k-1}\mathbf{e}_n(k) \to \mathcal{B}_{k-2}\mathbf{e}_n(k)$; that is, $E_{pq}^2 = 0$ for $p \neq k-2$. Thus, $\mathbf{e}_n$ is Koszul, and $\mathbf{e}_n^\perp = \mathbf{z}_n$, completing the proof of Theorem 3.1. □

There is a natural embedding of operads $\mathsf{F}_n(k) \hookrightarrow \mathsf{F}_{n+1}(k)$ induced by the embedding $\mathbb{R}^n \to \mathbb{R}^{n+1}$. Taking homology, we obtain a map of operads $\mathbf{e}_n \to \mathbf{e}_{n+1}$ which identifies the degree 0 spaces and vanishes in other degrees. Thus, if $A$ is an $(n+1)$-algebra, it is an $n$-algebra in a natural way: this $n$-algebra has the same commutative product as $A$, but vanishing bracket.

The embedding $\mathsf{F}_n(k) \hookrightarrow \mathsf{F}_{n+1}(k)$ has codimension $k-1$, and taking its Gysin map gives a map of dg-operads $\Lambda \mathbf{e}_{n+1} \to \mathbf{e}_n$. Thus, if $A$ is an $n$-algebra, its suspension $\Sigma A$ is in a natural way an $(n+1)$-algebra. We will now show that this map is the one which suspends the Lie bracket on $A$ and sets the product to zero.

Taking the linear dual of the map $\Lambda \mathbf{e}_{n+1} \to \mathbf{e}_n$ and applying the functor $\Lambda^{-n}$, we obtain a map of cooperads $\mathbf{z}_n \to \mathbf{z}_{n+1}$, which is the natural map $\mathbf{e}_n^\perp \to \mathbf{e}_{n+1}^\perp$ induced by the map of operads $\mathbf{e}_n \to \mathbf{e}_{n+1}$. This map fits into a map of exact sequences

$$\begin{array}{ccccccc}
\mathcal{B}_{k-2}\mathbf{e}_n(k) & \xrightarrow{\partial} & \mathcal{B}_{k-1}\mathbf{e}_n(k) & \longrightarrow & \mathbf{z}_n(k) & \longrightarrow & 0 \\
\downarrow & & \downarrow & & \downarrow & & \\
\mathcal{B}_{k-2}\mathbf{e}_{n+1}(k) & \xrightarrow{\partial} & \mathcal{B}_{k-1}\mathbf{e}_{n+1}(k) & \longrightarrow & \mathbf{z}_{n+1}(k) & \longrightarrow & 0
\end{array}$$

The arrow $\mathcal{B}_i\mathbf{e}_n(k) \to \mathcal{B}_i\mathbf{e}_{n+1}(k)$ is zero except in degree $i$, where it is the identity map, proving the following result.

**Proposition 3.6.** *The map of graded vector spaces $\mathbf{z}_n(k) \to \mathbf{z}_{n+1}(k)$ is the identity in degree $k-1$, and zero in other degrees. The map of graded vector spaces $\Lambda \mathbf{e}_{n+1}(k) \to \mathbf{e}_n(k)$ is the identity in degree $(n-1)(k-1)$, and zero in other degrees.*

Denote by $\mathbf{z}_\infty$ the colimit
$$\mathbf{z}_\infty = \operatorname*{colim}_{n \to \infty} \mathbf{z}_n.$$

We see that $\mathbf{e}_\infty$ is Koszul, that $\mathbf{e}_\infty^\perp = \mathbf{z}_\infty$, and that $\Lambda^{-1} \mathbf{z}_\infty^*$ is the Lie operad:
$$\mathcal{L}(k) \cong H_{(n-1)(k-1)}(\mathbb{F}_n(k)) \otimes \operatorname{sgn}_k^{(n-1)}.$$

In particular, we obtain a proof that the Lie operad $\mathcal{L}$ is Koszul, and that $\mathcal{L}^\perp \cong \Lambda^{-1} \mathbf{e}_\infty^*$.

## 4. Homotopy theory for **a**-algebras

Closed model categories were developed by Quillen in order to extend the methods of homotopy theory to categories of algebras (Quillen [40], [42]). Examples of closed model categories abound: simplicial sets, topological spaces, spaces with a group action, small categories, simplicial algebras, spectra. In Section 4.2, we show that the categories of algebras over an exact dg-operad, and coalgebras over a connected exact dg-cooperad, are closed model categories; this generalizes the closed model categories of commutative dg-algebras over $\mathbb{Q}$, Lie dg-algebras over $\mathbb{Q}$ (Section II.5, Quillen [42]) and associative dg-algebras over any field (Munkholm [38]):

In Section 4.3, we prove that the adjoint pair of functors $\Omega(\mathbf{a})$ and $\mathbb{B}(\mathbf{a})$ defines an equivalence between the homotopy categories of **a**-operads and $\mathcal{B}\mathbf{a}$-cooperads. We also study the total left derived functor $\mathsf{L}\Phi_*$ of the direct image functor, where $\Phi : \mathbf{a} \to \mathbf{b}$ is a map of operads, proving that if $\Phi$ is a weak equivalence, $\mathsf{L}\Phi_*$ induces an equivalence of homotopy categories.

In Section 4.4, we define for a Koszul operad **a** the notion of a homotopy **a**-algebra $A$: this is an $\mathbf{a}^\perp$-codifferential on the cofree $\mathbf{a}^\perp$-coalgebra $\mathsf{C}(\mathbf{a}^\perp, A)$, or equivalently, the structure of an algebra on $A$ for the almost free resolution $\mathcal{B}^*\mathbf{a}^\perp$ of **a**. We show that for the Lie operad $\mathcal{L}$, we recover the notion of a homotopy Lie algebra of Drinfeld [12], Hinich and Schechtman [25] and Lada and Stasheff [30].

All operads in this section will be exact and augmented, and all maps of operads will respect the augmentation. All cooperads will be connected, exact and coaugmented.

**4.1. Closed model categories.** We say that the map $i$ has the left lifting property (LLP) with respect to the map $p$, and the map $p$ has the right lifting property (RLP) with respect to $i$, if given any commuting diagram of the form

$$\begin{array}{ccc} A & \longrightarrow & X \\ i \downarrow & \nearrow f & \downarrow p \\ B & \longrightarrow & Y \end{array}$$

there is a lift $f : B \to X$ making the upper and lower triangles commute.

A map $g : V_1 \to V_2$ is a retract of a map $f : W_1 \to W_2$ if there is a diagram

$$\begin{array}{ccccc} V_1 & \xrightarrow{i_1} & W_1 & \xrightarrow{j_1} & V_1 \\ g \downarrow & & f \downarrow & & g \downarrow \\ V_2 & \xrightarrow{i_2} & W_2 & \xrightarrow{j_2} & V_2 \end{array}$$

where the endomorphisms $j_1 i_1 : V_1 \to V_1$ and $j_2 i_2 : V_2 \to V_2$ are the identity.

**Definition 4.1.** Let $\mathcal{C}$ be a category with classes of morphisms called weak equivalences, cofibrations and fibrations. A map which is both a (co)fibration and a weak equivalence, is called an acyclic (co)fibration. The category $\mathcal{C}$ is a closed model category if it satisfies the following axioms:

**CM1:** $\mathcal{C}$ has finite limits and colimits.

**CM2:** If any two of $f$, $g$ and $fg$ are weak equivalences, then so is the third.

**CM3:** A retract $f$ of a weak equivalence (cofibration, fibration) $g$ is a weak equivalence (cofibration, fibration).

**CM4 (i):** A cofibration has the LLP with respect to all acyclic fibrations.

**CM4 (ii):** A fibration has the RLP with respect to all acyclic cofibrations.

**CM5 (i):** Any map $f : A \to B$ may be factored $f = pi$ where $i : A \to X$ is a cofibration and $p : X \to B$ is an acyclic fibration.

**CM5 (ii):** Any map $f : A \to B$ may be factored $f = pi$ where $i : A \to X$ is an acyclic cofibration and $i : X \to B$ is a fibration.

A proper closed model category is a closed model category such that

- **P (i):** the pushout of a weak equivalence by a cofibration is a weak equivalence;
- **P (ii):** the pullback of a weak equivalence by a fibration is a weak equivalence.

It follows from Axiom CM4 that a pullback of a fibration $p$ is again a fibration, which is acyclic if $f$ is, while a pushout of a cofibration $i$ is again a cofibration, and is acyclic if $i$ is. We will use these properties of closed model categories without further comment.

These axioms are self-dual: the opposite category $\mathcal{C}^{\mathrm{op}}$, with the same weak equivalences and the sets of fibrations and cofibrations exchanged, is a closed model category, and if $\mathcal{C}$ is proper then so is $\mathcal{C}^{\mathrm{op}}$. In a closed model category, the weak equivalences together with the fibrations determine the cofibrations, while the weak equivalences together with the cofibrations determine the fibrations.

An object $A$ of $\mathcal{C}$ is called cofibrant if the unique map $* \to A$ is a cofibration, where $*$ is the initial object of $\mathcal{C}$. There is a dual definition of fibrant, using the terminal object of $\mathcal{C}$.

The homotopy category $\operatorname{Ho} \mathcal{C}$ of a closed model category $\mathcal{C}$ is the localization of $\mathcal{C}$ with respect to all weak equivalences. Denote by $\gamma : \mathcal{C} \to \operatorname{Ho} \mathcal{C}$ the localization functor.

If $F : \mathcal{C}_1 \to \mathcal{C}_2$ is a functor between closed model categories, the total left-derived functor $\mathsf{L}F$ of $F$ is a colimit over all pairs $(\Phi, \varepsilon)$ where $\Phi : \operatorname{Ho}\mathcal{C}_1 \to \operatorname{Ho}\mathcal{C}_2$ is a functor and $\varepsilon : F \to \Phi$ is a natural transformation. Such a functor, if it exists, is clearly unique up to isomorphism. If $F : \mathcal{C}_1 \to \mathcal{C}_2$ carries weak equivalences between cofibrant objects into weak equivalences, then it possesses a total left-derived functor (Quillen [40], Section I.4): if $X$ is an object of $\mathcal{C}_1$, then $\mathsf{L}F(\gamma_1(X)) = \gamma_2(F(Q))$ where $Q$ is a cofibrant object of $\mathcal{C}_1$ and $p : Q \to X$ is an acyclic fibration. Similarly, if $G : \mathcal{C}_2 \to \mathcal{C}_1$ is a functor which carries weak equivalences between fibrant objects into weak equivalences, there is a total right-derived functor $\mathsf{R}F : \operatorname{Ho}\mathcal{C}_2 \to \operatorname{Ho}\mathcal{C}_1$.

The following theorem is due to Quillen (Section I.4 of [40]).

**Theorem 4.2.** *Let $\mathcal{C}_1$ and $\mathcal{C}_2$ be closed model categories, and let $F : \mathcal{C}_1 \rightleftarrows \mathcal{C}_2 : G$ be a pair of adjoint functors, such that*

(1) *$F$ carries cofibrations in $\mathcal{C}_1$ into cofibrations in $\mathcal{C}_2$, and $G$ carries fibrations in $\mathcal{C}_2$ into fibrations in $\mathcal{C}_1$;*
(2) *$F$ carries weak equivalences between cofibrant objects in $\mathcal{C}_1$ into weak equivalences in $\mathcal{C}_2$, and $G$ carries weak equivalences between fibrant objects in $\mathcal{C}_2$ into weak equivalences in $\mathcal{C}_1$.*

*Then the total derived functors $\mathsf{L}F : \operatorname{Ho}\mathcal{C}_1 \rightleftarrows \operatorname{Ho}\mathcal{C}_2 : \mathsf{R}G$ form an adjoint pair.*

*Suppose in addition that for $X$ a cofibrant object of $\mathcal{C}_1$ and $Y$ a fibrant object of $\mathcal{C}_2$, the weak equivalences correspond under the identification*

$$\mathcal{C}_1(FX, Y) \cong \mathcal{C}_2(X, GY).$$

*Then the unit $\operatorname{Id} \to (\mathsf{L}F)(\mathsf{R}G)$ and counit $(\mathsf{R}G)(\mathsf{L}G) \to \operatorname{Id}$ of this adjunction are natural equivalences of functors, and thus $\mathsf{L}F$ and $\mathsf{R}G$ define an equivalence between the categories $\operatorname{Ho}\mathcal{C}_1$ and $\operatorname{Ho}\mathcal{C}_2$.*

The following lemma is a special case of Lemma I.1.4 of Baues [7].

**Lemma 4.3.** *In a closed model category, if all objects are cofibrant, then Axiom P (i) for a proper closed model category holds. Dually, if all objects are fibrant, then Axiom P (ii) for a proper closed model category holds.*

*Proof.* By duality, it suffices to prove the first assertion. Given a map $f : A \to X$, factor the map $f \coprod X : A \coprod X \to X$ as in CM5 (i)

$$A \coprod X \xrightarrow{i \coprod j} Z_f \xrightarrow{p} X$$

where $i \coprod j$ and $p$ are respectively a cofibration and an acyclic fibration. The space $Z_f$ is known as the mapping cylinder of $f$: it is unique up to weak equivalence. Since $A$ and $X$ are cofibrant, it follows that $i : A \to Z_f$ and $j : X \to Z_f$ are cofibrations. Since $pj : X \to X$ is the identity and $p$ is a weak equivalence, it follows from CM1 that that $j$ is a weak equivalence.

Consider the following commutative diagram, in which each square is a pushout, and each vertical arrow is a cofibration:

$$\begin{array}{ccccccc} A & \xrightarrow{i} & Z_f & \xrightarrow{p} & X & \xrightarrow{j} & Z_f \\ \downarrow & & \downarrow & & \downarrow & & \downarrow \\ B & \xrightarrow{i'} & W & \xrightarrow{p'} & Y & \xrightarrow{j'} & W \end{array}$$

Since $i$ and $j$ are acyclic cofibrations, it follows that $i'$ and $j'$ are as well. It only remains to show that $p'$ is a weak equivalence: but $pj : X \to X$ is the identity, and hence so is $p'j' : Y \to Y$, and the result follows by a further application of CM1. □

**4.2. The closed model categories of a-algebras and z-coalgebras.** The category $\mathcal{M}$ of chain complexes over a field $K$ is a proper closed model category:

(1) a map is a weak equivalence if it induces an isomorphism on homology;
(2) a map is a fibration if it is surjective in degree $i > 0$ (this reflects the topological fact that a fibration need not be surjective on components);
(3) a map is a cofibration if it is injective.

In this category, all objects are cofibrant and fibrant.

We will now show that the category of **a**-algebras $\mathcal{M}^{\mathbf{a}}$ over an exact dg-operad **a** is a proper closed model category. In the category of algebras over an operad **a**, we define weak equivalences, fibrations and cofibrations in the following way:

(1) a map in $\mathcal{M}^{\mathbf{a}}$ is a weak equivalence (fibration) if it is a weak equivalence (fibration) in the category of chain complexes $\mathcal{M}$;
(2) a map $f : A \to B$ in $\mathcal{M}^{\mathbf{a}}$ is a cofibration if the $\mathbf{a}[A]$-algebra $B$ is a retract of an almost free **a**-algebra $\mathsf{T}(\mathbf{a}[A], V, d)$. (The $\mathbf{a}[A]$-algebra structure on $B$ defined by the map $f$ is described in Section 1.6.)

In particular, an **a**-algebra is cofibrant if it is a retract of an almost free **a**-algebra $\mathsf{T}(\mathbf{a}, V, d)$.

In the following proof, we will use the fact that if an operad **a** is exact, and $A$ is an **a**-algebra, then $\mathbf{a}[A]$ is exact.

**Theorem 4.4.** *Let **a** be an exact operad. With the above definitions of weak equivalences, fibrations and cofibrations, $\mathcal{M}^{\mathbf{a}}$ is a closed model category.*

*Proof.* The existence of limits and colimits in $\mathcal{M}^{\mathbf{a}}$ was established in Section 1.6. Axioms CM2, and CM3 for weak equivalences and fibrations, are inherited from the closed model category $\mathcal{M}$, while CM3 for cofibrations is immediate from the definition of a cofibration.

Next, we prove CM4 (i): a cofibration $i : A \to B$ of **a**-algebras has the LLP with respect to acyclic fibrations. Replacing the operad **a** by $\mathbf{a}[A]$, we see that it suffices to establish that a cofibrant **a**-algebra has the LLP with respect to acyclic fibrations. But the retract of a map with the LLP with respect to acyclic fibrations inherits this property. Thus, it suffices to show that an almost free **a**-algebra has the LLP with respect to acyclic fibrations.

If $\mathsf{T}(\mathbf{a}, V, d)$ is an almost free **a**-algebra, write the underlying free algebra $\mathsf{T}(\mathbf{a}, V)$ as a colimit

$$\mathsf{T}(\mathbf{a}, V) = \operatorname*{colim}_{n \to \infty} \mathsf{T}(\mathbf{a}, \tau_n V),$$

where $\tau_n V$ is the truncation of $V$ to degrees $i \leq n$. Since the differential $d$ has degree $-1$, it maps $\tau_n V \subset \mathsf{T}(\mathbf{a}, V)$ to $\mathsf{T}(\mathbf{a}, \tau_{n-1} V)$. Thus the almost free **a**-algebra $\mathsf{T}(\mathbf{a}, V, d)$ is also a colimit

$$\mathsf{T}(\mathbf{a}, V, d) = \operatorname*{colim}_{n \to \infty} \mathsf{T}(\mathbf{a}, \tau_n V, d).$$

Thus, to show that the almost free algebra $\mathsf{T}(\mathbf{a},V,d)$ has the LLP with respect to an acyclic fibration $p$, it suffices to show that the injection

$$\mathsf{T}(\mathbf{a},\tau_{n-1}V,d) \to \mathsf{T}(\mathbf{a},\tau_n V,d)$$

has the LLP with respect to $p$, for all $n > 0$. To see this, we write the injection as a pushout

$$\begin{array}{ccc} \mathsf{T}(\mathbf{a},U) & \xrightarrow{f} & \mathsf{T}(\mathbf{a},\tau_{n-1}V,d) \\ \mathsf{T}(\mathbf{a},i)\downarrow & & \downarrow \\ \mathsf{T}(\mathbf{a},W)) & \longrightarrow & \mathsf{T}(\mathbf{a},\tau_n V,d) \end{array}$$

Here, $U = \Sigma^{-1}\tau_n V/\tau_{n-1}V$ is the chain complex with the vector space $V_n$ concentrated in degree $n-1$ and vanishing differential, and $W$ is isomorphic as a graded vector space to $U \oplus \Sigma U$ and has as differential the identity map of $U$. The map $f$ is determined by its restriction to $U$, where it may be identified with the restriction of the differential $d : \tau_n V \to \mathsf{T}(\mathbf{a},\tau_{n-1}V)$ to $V_n \subset \tau_n V$, and $i$ is the inclusion of $U$ in $W$. The proof of CM4 (i) is completed by applying the following lemma.

**Lemma 4.5.** *Let $A$ be an $\mathbf{a}$-algebra with the LLP with respect to acyclic fibrations. Let $i : V \to W$ be a cofibration in $\mathcal{M}$, and let $B$ be the following pushout in $\mathcal{M}^{\mathbf{a}}$:*

$$\begin{array}{ccc} \mathsf{T}(\mathbf{a},V) & \longrightarrow & A \\ \mathsf{T}(\mathbf{a},i)\downarrow & & \downarrow \\ \mathsf{T}(\mathbf{a},W) & \longrightarrow & B \end{array}$$

*Then $B$ has the LLP with respect to acyclic fibrations.*

*Proof.* The pushout of a map with the LLP with respect to a map $p$ also has the LLP with respect to $p$. Thus, it suffices to show that if $i : V \to W$ is a cofibration in $\mathcal{M}$, then the map $\mathsf{T}(\mathbf{a},i) : \mathsf{T}(\mathbf{a},V) \to \mathsf{T}(\mathbf{a},W)$ has the LLP with respect to acyclic fibrations in $\mathcal{M}^{\mathbf{a}}$. This follows from the universality property of $\mathsf{T}(\mathbf{a},W)$: to define a lift

$$\begin{array}{ccccc} V & \xrightarrow{\eta V} & \mathsf{T}(\mathbf{a},V) & \longrightarrow & X \\ i\downarrow & & \mathsf{T}(\mathbf{a},i)\downarrow \;\;{}^{f} & & \downarrow p \\ W & \xrightarrow{\eta W} & \mathsf{T}(\mathbf{a},W) & \longrightarrow & Y \end{array}$$

where $p$ is an acyclic fibration of $\mathbf{a}$-algebras, it suffices to find a lift $W \to X$ in the category of chain complexes making the outer square commute, which exists since $i$ has the LLP with respect to $p$ in the closed model category $\mathcal{M}$. $\square$

CM5 (i) is proved using the bar construction of Section 2.3. Given a map $f : A \to B$ of $\mathbf{a}$-algebras, let $p : \Omega(\mathbf{a}[A], \mathbb{B}(\mathbf{a}[A], B)) \to B$ be the unit of the adjunction

$$\Omega(\mathbf{a}[A]) : \mathcal{M}_{\mathcal{B}\mathbf{a}[A]} \rightleftarrows \mathcal{M}^{\mathbf{a}[A]} : \mathbb{B}(\mathbf{a}[A]),$$

applied to the $\mathbf{a}[A]$-algebra $B$. By Theorem 2.19, it is an acyclic fibration. Clearly, $\Omega(\mathbf{a},\mathbb{B}(\mathbf{a},A))$ is an almost free $\mathbf{a}[A]$-algebra, and thus

$$i : A = \Omega(\mathbf{a}[A], 0) \to \Omega(\mathbf{a}[A], \mathbb{B}(\mathbf{a}[A], B))$$

is a cofibration of $\mathbf{a}$-algebras.

In the proof of CM5 (ii), we use the fact that if $\mathbf{a}$ is a exact $\mathbb{S}$-module, the functor $\mathsf{T}(\mathbf{a})$ is homotopy invariant, that is, if $f : V \to W$ is a weak equivalence of chain complexes, then $\mathsf{T}(\mathbf{a},f) :$

$\mathsf{T}(\mathbf{a}, V) \to \mathsf{T}(\mathbf{a}, W)$ is a weak equivalence. This is the only place in the proof of the theorem where the hypothesis that the operad $\mathbf{a}$ is exact is needed.

If $B$ is an $\mathbf{a}$-algebra, let $V$ be a contractible chain complex such that there is a fibration $V \to B$ in the closed model category of chain complexes. Given a map $f : A \to B$ of $\mathbf{a}$-algebras, let

$$X = A \coprod \mathsf{T}(\mathbf{a}, V) \cong \mathsf{T}(\mathbf{a}[A], V).$$

Then $f = pi$, where $p : X \to B$ is the natural surjection induced by the maps $f : A \to B$ and $V \to B$, and $i : A \to X$ is the natural inclusion. Obviously, $p$ is a fibration. The map $i$ is free, so is a cofibration. The operad $\mathbf{a}[A]$ is exact, since $\mathbf{a}$ is, so the functor $\mathsf{T}(\mathbf{a}[A])$ preserves weak equivalences; thus, it sends the weak equivalence $0 \to V$ to the weak equivalence

$$i : A = \mathsf{T}(\mathbf{a}[A], 0) \to \mathsf{T}(\mathbf{a}[A], V) = X.$$

It is now easy to prove CM4 (ii): given an acyclic cofibration $f : A \to B$, factor it as in the proof of CM5 (ii)

$$A \xrightarrow{i} A \coprod \mathsf{T}(\mathbf{a}, V) \xrightarrow{p} B.$$

By CM2, $p$ is a weak equivalence. Since $f$ is a cofibration, it has the LLP with respect to $p$, while $i$ inherits the LLP with respect to fibrations from the closed model category $\mathcal{M}$; in this way, we see that $f$ has the LLP with respect to any fibration. $\square$

**Corollary 4.6.** *A cofibrant $\mathbf{a}$-algebra $A$ is a deformation retract of an almost free $\mathbf{a}$-algebra $X$; in other words, there are weak equivalences $p : X \to A$ and $j : A \to X$ such that $pj : A \to A$ is the identity of $A$.*

*Proof.* Factor $f : * \to A$ as in the proof of CM5 (i) into a product $f = pi$, where $i : * \to X$ is the inclusion of the initial object into an almost free $\mathbf{a}$-algebra, and $p : X \to A$ is an acyclic fibration. Since $f$ is a cofibration, it has the LLP with respect to $p$, so that there is a map $j : A \to X$ such that $pj : A \to A$ is the identity of $A$. It follows by CM2 that $j$ is a weak equivalence. $\square$

There is also a closed model structure on the category of coalgebras over an exact connected cooperad $\mathbf{z}$. The proof will use the adjoint pair of functors

$$\mathbb{B}^*(\mathbf{z}) = \Omega(\Phi^*) : \mathcal{M}_{\mathbf{z}} \rightleftarrows \mathcal{M}^{\mathcal{B}^*\mathbf{z}} : \Omega^*(\mathbf{z}) = \mathbb{B}(\Phi^*)$$

associated to the universal twisting cochain $\Phi^* : \mathbf{z} \to \mathcal{B}^*\mathbf{z}$.

In the category of coalgebras over $\mathbf{z}$, we define weak equivalences, fibrations and cofibrations in the following way:
  (1) a map in $\mathcal{M}_{\mathbf{z}}$ is a weak equivalence (cofibration) if it is a weak equivalence (cofibration) in the category of chain complexes $\mathcal{M}$;
  (2) a map $f : C \to D$ in $\mathcal{M}_{\mathbf{z}}$ is a fibration if the $\mathbf{z}[D]$-algebra $C$ is a retract of $\Omega^*(\mathbf{z}[D], A)$ for some $\mathcal{B}^*\mathbf{z}[D]$-algebra $A$.

Note that if $D$ is a $\mathbf{z}$-coalgebra, then the cooperad $\mathbf{z}[D]$ is connected if $\mathbf{z}$ is.

**Theorem 4.7.** *Let $\mathbf{z}$ be an exact connected cooperad. With the above definitions of weak equivalences, fibrations and cofibrations, the category $\mathcal{M}_{\mathbf{z}}$ of $\mathbf{z}$-coalgebras is a closed model category.*

*Proof.* The proofs of all of the axioms of a closed model category other than Axiom CM4 (ii) are the precise dual of the proofs of the corresponding axioms for $\mathcal{M}^{\mathbf{a}}$. For example, given a map $f : C \to D$ of $\mathbf{z}$-coalgebras, CM5 (ii) is proved by the factorization

$$C \to \Omega^*(\mathbf{z}[D], \mathbb{B}^*(\mathbf{z}[D], C)) \to \Omega^*(\mathbf{z}[D], \mathbb{B}^*(\mathbf{z}[D], 0)) = D,$$

while CM5 (i) is proved by the factorization

$$C \to D \coprod \mathsf{C}(\mathbf{z}, V) \cong \Omega^*(\mathbf{z}[D], V) \to D,$$

where $V$ is a contractible chain complex such that there is a fibration $V \to D$.

To prove CM4 (ii), it suffices to show that if $A$ is a $\mathcal{B}^*\mathbf{z}$-algebra, the $\mathbf{z}$-coalgebra $\Omega^*(\mathbf{z}, A)$ has the RLP with respect to acyclic cofibrations $i : C \to D$. By adjointness, liftings in the diagram

$$\begin{array}{ccc} C & \longrightarrow & \Omega^*(\mathbf{z}, A) \\ {\scriptstyle i}\downarrow & \nearrow & \\ D & & \end{array}$$

are in bijective correspondence with liftings in the diagram

$$\begin{array}{ccc} \mathbb{B}^*(\mathbf{z}, C) & \longrightarrow & A \\ {\scriptstyle \mathbb{B}^*(\mathbf{z}, i)}\downarrow & \nearrow & \\ \mathbb{B}^*(\mathbf{z}, D) & & \end{array}$$

But $A$, like all $\mathcal{B}^*\mathbf{z}$-algebras, is fibrant, and thus has the RLP with respect to the acyclic cofibration of $\mathcal{B}^*\mathbf{z}$-coalgebras $\mathbb{B}^*(\mathbf{z}, f) : \mathbb{B}^*(\mathbf{z}, C) \to \mathbb{B}^*(\mathbf{z}, D)$. $\square$

If $\mathbf{a}$ is an operad, then $\mathcal{B}\mathbf{a}$ is a connected cooperad, which is exact if $\mathbf{a}$ is. The total derived functors

$$\mathsf{L}\Omega(\mathbf{a}) : \mathcal{M}_{\mathcal{B}\mathbf{a}} \rightleftarrows \mathcal{M}^{\mathbf{a}} : \mathsf{R}\mathbb{B}(\mathbf{a})$$

form an adjoint pair, and define an equivalence of homotopy theories.

If $\mathbf{a}$ is a Koszul operad, then $\mathbf{a}^\perp$ is a connected cooperad, which is exact if $\mathbf{a}$ is. From Theorem 2.25, we see the total derived functors

$$\mathsf{L}\Omega(\mathbf{a}) : \mathcal{M}_{\mathbf{a}^\perp} \rightleftarrows \mathcal{M}^{\mathbf{a}} : \mathsf{R}\mathbb{B}(\mathbf{a})$$

form an adjoint pair, and define an equivalence of homotopy theories. Taking for $\mathbf{a}$ the Lie operad $\mathcal{L}$, we obtain an equivalence between the homotopy category of Lie dg-algebras over $\mathbb{Q}$, and the homotopy category of connected cocommutative dg-coalgebras, also over $\mathbb{Q}$, a result due to Quillen ([42], Section II.5). Actually, Quillen realizes the equivalence by means of the adjoint pair $P\Sigma^{-1} : \mathcal{M}_{\mathcal{L}^\perp} \rightleftarrows \mathcal{M}^{\mathcal{L}} : \Sigma U$, where $P$ is the primitive functor and $U$ is the universal enveloping algebra functor. However, if $C$ is an almost cofree cocommutative coalgebra, there is a weak equivalence between $PC$ and $\Omega(\mathcal{L}, \Sigma C)$: this shows that our equivalence of homotopy categories is the same as Quillen's.

We now turn to the theory of cochain algebras and coalgebras. Let $\mathcal{N}$ be the category of cochain complexes: that is, the category of complexes $V_\bullet$ such that $V_i = 0$ for $i > 0$. The name cochain complex is motivated by the alternative notation $V^i = V_{-i}$; the differential is then a map $\delta : V^i \to V^{i+1}$, and $V^i = 0$ for $i < 0$. Since $\mathcal{N}$ is a symmetric monoidal category with respect to the tensor product, we obtain corresponding categories of operads, cooperads, and their categories of algebras and coalgebras.

A cochain complex is called $n$-connected if $A_i = 0$ for $i \geq -n$; we say connected instead of 0-connected.

The category $\mathcal{N}$ of cochain complexes over a field $K$ is a closed model category:

(1) a map is a weak equivalence if it induces an isomorphism on homology;
(2) a map is a fibration if it is surjective;
(3) a map is a cofibration if is injective.

In this category, all objects are cofibrant and fibrant.

We have seen that there is an asymmetry in the construction of a closed model structure between categories of algebras and coalgebras. For example, it is only if a cooperad is connected that its category of coalgebras is a closed model category. In the category of cochain complexes, the situation is similar, but the roles of operads and cooperads is reversed: for example, the category of **a**-algebras over a cochain operad is a closed model category only if **a** is connected. This is reflected in the fact that the bar cooperad $\mathbb{B}\mathbf{a}$ is a cochain cooperad if and only if **a** is a connected cochain operad.

We will content ourselves with defining the closed model structures on categories of algebras and coalgebras: the proofs that these are indeed closed model categories may be obtained by dualizing the above proofs for chain algebras and coalgebras, exchanging operads and cooperads, algebras and coalgebras, and reversing the arrows.

If **a** is an exact, connected cochain operad, define weak equivalences, fibrations and cofibrations in the category of **a**-algebras in the following way:

(1) a map in $\mathcal{N}^{\mathbf{a}}$ is a weak equivalence (fibration) if it is a weak equivalence (fibration) in the category of cochain complexes $\mathcal{N}$;
(2) a map $f : A \to B$ in $\mathcal{N}^{\mathbf{a}}$ is a cofibration if the $\mathbf{a}[A]$-algebra $B$ is a retract of $\Omega(\mathbf{a}[A], C)$ for some $\mathcal{B}\mathbf{a}[A]$-coalgebra $C$.

If **z** is an exact cochain cooperad, define weak equivalences, fibrations and cofibrations in the category of **z**-coalgebras in the following way:

(1) a map in $\mathcal{N}_{\mathbf{z}}$ is a weak equivalence (cofibration) if it is a weak equivalence (cofibration) in the category of cochain complexes $\mathcal{N}$;
(2) a map $f : C \to D$ in $\mathcal{N}_{\mathbf{z}}$ is a fibration if the $\mathbf{z}[D]$-coalgebra $C$ is a retract of an almost cofree $\mathbf{z}[D]$-coalgebra $\mathsf{C}(\mathbf{z}[D], V, d)$.

If **z** is a connected dg-cooperad, its linear dual $\mathbf{z}^*$ is a connected cochain operad. The closed model structure on the category of **z**-coalgebras is closely related to the closed model structure on the category of cochain $\mathbf{z}^*$-coalgebras: indeed, the contravariant functor $C \mapsto C^*$ between these categories carries weak equivalences to weak equivalences, cofibrations to fibrations and fibrations to cofibrations. If **a** is a dg-operad, there is a similar duality between **a**-algebras and $\mathbf{a}^*$-coalgebras.

Let $\mathbf{e}_n$ be the $n$-algebra operad. The operad $\Lambda^n \mathbf{e}_n$ is a cochain operad with zero differential, and since it is connected, the category of cochain $\Lambda^n \mathbf{e}_n$-algebras is a closed model category. But a $\Lambda^n \mathbf{e}_n$-algebra structure on a cochain complex $A$ is the same thing as an $n$-algebra structure on the cochain complex $\Sigma^{-n} A$. In this way, we obtain the following result.

**Proposition 4.8.** *The category of $(n-1)$-connected cochain $n$-algebras is a closed model category.*

Similarly, there is a closed model category structure on the category of connected commutative cochain algebras, since this category is isomorphic to the category of cochain $\Lambda \mathbf{e}_\infty$-algebras.

**4.3. André-Quillen homology.** In this section, we study the total derived functors of the adjoint pair $\Phi_* : \mathcal{M}^{\mathbf{a}} \rightleftarrows \mathcal{M}^{\mathbf{b}} : \Phi^*$, where $\Phi : \mathbf{a} \to \mathbf{b}$ is a map of operads. In order to show that the functor $\Phi_*$ has a total left derived functor, we must study more carefully its effect on an almost free **a**-algebra. We will say that $\Phi : \mathbf{a} \to \mathbf{b}$ is a weak equivalence of operads if it is a weak equivalence of $\mathbb{S}$-modules, that is, if $\Phi(S) : \mathbf{a}(S) \to \mathbf{b}(S)$ is a weak equivalence of chain complexes for all finite sets $S$.

**Definition 4.9.** A contractible pair in a category is a diagram

$$A \xrightarrow[f_1]{f_0} B \xrightarrow{t} A$$

such that $f_0 t$ is the identity of $B$ and $f_1 t f_0 = f_1 t f_1$. The coequalizer of a contractible pair is called a split coequalizer.

A split coequalizer is absolute: that is, it remains a coequalizer after the application of any functor (MacLane [32], Section VI.6).

**Proposition 4.10.** *(1) If $A$ is a cofibrant $\mathbf{a}$-algebra and $\Phi : \mathbf{a} \to \mathbf{b}$ is a map of operads, then*

$$\mathsf{T}(\mathbf{b}, \mathsf{T}(\mathbf{a}, A)) \xrightarrow[\mu(A)\cdot\mathsf{T}(\mathbf{b},\mathsf{T}(\Phi,A))]{\mathsf{T}(\mathbf{b},\rho)} \mathsf{T}(\mathbf{b}, A) \longrightarrow \Phi_* A$$

*is a split coequalizer in $\mathcal{M}^{\mathbf{b}}$.*

*(2) If $f : A \to B$ is a weak equivalence of cofibrant $\mathbf{a}$-algebras and $\Phi : \mathbf{a} \to \mathbf{b}$ is a map of operads, then $\Phi_* f : \Phi_* A \to \Phi_* B$ is a weak equivalence.*

*(3) If $\Phi : \mathbf{a} \to \mathbf{b}$ is a weak equivalence of operads and $A$ is a cofibrant $\mathbf{a}$-algebra, then $A \to \Phi^* \Phi_* A$ is a weak equivalence.*

*Proof.* Since any cofibrant $\mathbf{a}$-algebra is a retract of an almost free $\mathbf{a}$-algebra, we may suppose that $A = \mathsf{T}(\mathbf{a}, V, d)$ is an almost free $\mathbf{a}$-algebra. Let $t$ be the map of $\mathbf{b}$-algebras

$$\mathsf{T}(\mathbf{b}, \mathsf{T}(\mathbf{a}, V)) \xrightarrow{\mathsf{T}(\mathbf{b}, \mathsf{T}(\mathbf{a}, \eta V))} \mathsf{T}(\mathbf{b}, \mathsf{T}(\mathbf{a}, \mathsf{T}(\mathbf{a}, V))).$$

It is easily seen that $t$ is a contraction for the above parallel pair, and that it is compatible with the differential on $A$, proving (1).

To prove (2), suppose that $A$ and $B$ are cofibrant $\mathbf{a}$-algebras. Applying the homology functor to the split coequalizers which define $\Phi_* A$ and $\Phi_* B$, we see that the rows of the diagram of graded vector spaces

$$\begin{array}{ccccc}
H_\bullet(\mathsf{T}(\mathbf{b}, \mathsf{T}(\mathbf{a}, A))) & \rightrightarrows & H_\bullet(\mathsf{T}(\mathbf{b}, A)) & \longrightarrow & H_\bullet(\Phi_* A) \\
& & \| & & \| \\
H_\bullet(\mathsf{T}(\mathbf{b}, \mathsf{T}(\mathbf{a}, B))) & \rightrightarrows & H_\bullet(\mathsf{T}(\mathbf{b}, B)) & \longrightarrow & H_\bullet(\Phi_* B)
\end{array}$$

are split coequalizers. Since $\mathbf{a}$ and $\mathbf{b}$ are exact, the functors $\mathsf{T}(\mathbf{a})$ and $\mathsf{T}(\mathbf{b})$ preserve weak equivalences, hence the vertical arrows in this diagram are isomorphisms. From this, it follows that $H_\bullet(\Phi_* A) \cong H_\bullet(\Phi_* B)$.

To prove (3), suppose that $A$ is a cofibrant $\mathbf{a}$-algebra. Again, we see that the rows of the diagram of graded vector spaces

$$\begin{array}{ccccc}
H_\bullet(\mathsf{T}(\mathbf{a}, \mathsf{T}(\mathbf{a}, A))) & \rightrightarrows & H_\bullet(\mathsf{T}(\mathbf{a}, A)) & \longrightarrow & H_\bullet(A) \\
& & \| & & \| \\
H_\bullet(\mathsf{T}(\mathbf{b}, \mathsf{T}(\mathbf{a}, A))) & \rightrightarrows & H_\bullet(\mathsf{T}(\mathbf{b}, A)) & \longrightarrow & H_\bullet(\Phi^* \Phi_* A)
\end{array}$$

are split coequalizers and the vertical arrows are isomorphisms, and hence $H_\bullet(A) \cong H_\bullet(\Phi^* \Phi_* A)$. □

**Corollary 4.11.** *Let $\Phi : \mathbf{a} \to \mathbf{b}$ be a map of operads. The total derived functors*

$$\mathsf{L}\Phi_* : \operatorname{Ho} \mathcal{M}^{\mathbf{a}} \rightleftarrows \operatorname{Ho} \mathcal{M}^{\mathbf{b}} : \mathsf{R}\Phi^*$$

*form an adjoint pair.*

*If $\Phi$ is a weak equivalence of operads, then the functors $\mathsf{L}\Phi_*$ and $\mathsf{R}\Phi^*$ induce an equivalence of homotopy categories.*

*The category of $\mathbf{a}$-algebras $\mathcal{M}^{\mathbf{a}}$ is a proper closed model category.*

*Proof.* We will verify the hypotheses of Theorem 4.2 for the adjoint pair of functors $\Phi_*$ and $\Phi^*$. It is evident that $\Phi^*$ carries fibrations and weak equivalences in $\mathcal{M}^{\mathbf{b}}$ to fibrations and weak equivalences in $\mathcal{M}^{\mathbf{a}}$. Let $i : A \to B$ be a cofibration of $\mathbf{a}$-algebras. If $p : X \to Y$ is an acyclic fibration of $\mathbf{b}$-algebras, then $i$ has the LLP with respect to $\Phi^* p$. But since $\Phi_*$ is left adjoint to $\Phi^*$, $\Phi_* i$ has the LLP with respect to $p$, showing that $\Phi_* i$ is a cofibration. Finally, if $f : A \to B$ is a weak equivalence of cofibrant objects, then $\Phi_* f$ is a weak equivalence by part (2) of Proposition 4.10.

Now, suppose that $\Phi : \mathbf{a} \to \mathbf{b}$ is a weak equivalence of operads. We must show that if $A$ is a cofibrant $\mathbf{a}$-algebra and $B$ is a (necessarily fibrant) $\mathbf{b}$-algebra, and if $f : \Phi_* A \to B$ is a map, then the composition

$$A \to \Phi^* \Phi_* A \xrightarrow{\Phi^* f} \Phi^* B$$

is a weak equivalence if and only if the map $f$ is a weak equivalence; here $A \to \Phi^* \Phi_* A$ is the unit of the adjunction $\Phi_* : \mathcal{M}^{\mathbf{a}} \rightleftarrows \mathcal{M}^{\mathbf{b}} : \Phi^*$ applied to the $\mathbf{a}$-algebra $A$. Since $\Phi^* f$ is a weak equivalence if and only if $f$ is, this will follow by CM2 if we can show that $A \to \Phi^* \Phi_* A$ is a weak equivalence for cofibrant $\mathbf{a}$-algebras $A$. But this is part (3) of Proposition 4.10.

Let us show that $\mathcal{M}^{\mathbf{a}}$ is a proper closed model category. Axiom P (ii) follows by Lemma 4.3 from the fact that all $\mathbf{a}$-algebras are fibrant. We have just verified Axiom P (i): given a pushout

$$\begin{array}{ccc} A & \xrightarrow{f} & X \\ i \downarrow & & \downarrow \\ B & \longrightarrow & Y \end{array}$$

with $i : A \to B$ a cofibration and $f : A \to X$ a weak equivalence, we may think of $Y$ as the $\mathbf{a}[X]$-algebra $\mathbf{a}[f]_* B$. Since $\mathbf{a}[f] : \mathbf{a}[A] \to \mathbf{a}[X]$ is a weak equivalence of operads, it follows that $B \to \mathbf{a}[f]^* \mathbf{a}[f]_* B = \mathbf{a}[f]^* Y$ is a weak equivalence. $\square$

The proof of CM5 (ii) provides us with an explicit definition of the functor $\mathsf{L}\Phi_*$ applied to an $\mathbf{a}$-algebra $A$, as a certain almost free $\mathbf{b}$-algebra with underlying free $\mathbf{b}$-algebra $\mathsf{T}(\mathbf{b}, \mathbb{B}(\mathbf{a}, A))$. The special case where $\Phi$ is the augmentation $\varepsilon : \mathbf{a} \to \mathbb{1}$ of $\mathbf{a}$ is especially important.

**Definition 4.12.** The homology of $\mathbf{a}$-algebras is the total left derived functor of the indecomposables map

$$\mathsf{L}\varepsilon_* : \mathrm{Ho}\,\mathcal{M}^{\mathbf{a}} \to \mathrm{Ho}\,\mathcal{M}.$$

It is realized by the functor $A \mapsto \mathbb{B}(\mathbf{a}, A)$.

The homology functor $\mathsf{L}\varepsilon_*$ provides us with a general notion of homology for $\mathbf{a}$-algebras, generalizing the Hochschild homology of associative algebras and the André-Quillen homology of commutative algebras. If $\mathbf{a}$ is Koszul, then $\mathsf{B}(\mathbf{a}, A)$ is weakly equivalent to $\mathbb{B}(\mathbf{a}, A)$, and thus it provides a small complex calculating the André-Quillen homology in the category of $\mathbf{a}$-algebras.

We also see that the André-Quillen homology in the category of Lie algebras $\mathsf{L}\varepsilon_* L$ is represented by the $\mathcal{L}^{\perp}$-coalgebra $\mathsf{B}(\mathcal{L}, L)$, whose underlying graded vector space is $\Sigma^{-1} \mathsf{T}_{\infty}(\Sigma L)$, the desuspension of the Chevalley-Eilenberg complex of $L$. Note that $\mathsf{T}_{\infty}(\Sigma L)$ is just the exterior algebra of $L$ if $L$ is concentrated in degree 0.

If $\Phi : \mathbf{w} \to \mathbf{z}$ is a map of connected cooperads, there is an adjoint pair of total derived functors

$$\mathsf{L}\Phi_* : \mathrm{Ho}\,\mathcal{M}_{\mathbf{w}} \rightleftarrows \mathrm{Ho}\,\mathcal{M}_{\mathbf{z}} : \mathsf{R}\Phi^*,$$

which define an equivalence of homotopy theories if $\Phi$ is a weak equivalence. If $\mathbf{z}$ is a coaugmented operad, the total derived functor $\mathsf{R}\eta^* : \mathrm{Ho}\,\mathcal{M}_{\mathbf{z}} \to \mathrm{Ho}\,\mathcal{M}$ is called the André-Quillen cohomology, where $\eta : \mathbb{1} \to \mathbf{z}$ is the coaugmentation of $\mathbf{z}$.

**4.4. Algebras over an almost free operad.** In this section, we study the category of $\mathcal{A}$-algebras, where $\mathcal{A}$ is an almost free operad.

**Theorem 4.13.** *Let $\mathcal{A}$ be an almost free operad, let $B$ be an $\mathcal{A}$-algebra let $f : A \to B$ be an acyclic fibration of chain complexes. Then there exists an $\mathcal{A}$-algebra structure on $A$ such that the map $f : A \to B$ is a map of $\mathcal{A}$-algebras.*

*Proof.* Write $\mathcal{A}$ as a colimit $\mathcal{A} = \operatorname*{colim}_{n \to \infty} \mathcal{A}_n$, where $\mathcal{A}_{n+1}$ is a pushout

$$\begin{array}{ccc} \mathbb{T}\mathbf{v}_n & \xrightarrow{f} & \mathcal{A}_n \\ \mathbb{T}i_n \downarrow & & \downarrow \\ \mathbb{T}\mathbf{w}_n & \longrightarrow & \mathcal{A}_{n+1} \end{array}$$

and $i_n : \mathbf{v}_n \to \mathbf{w}_n$ is a pushout of $\mathbb{S}$-modules. By induction, we suppose that $A$ has been made into an $\mathcal{A}_n$-algebra. To extend this structure to that of an $\mathcal{A}_{n+1}$-algebra on $A$, we use the LLP in the category of chain complexes of the cofibration $\mathsf{S}(i, A) : \mathsf{S}(\mathbf{v}_n, A) \to \mathsf{S}(\mathbf{w}_n, A)$, in the diagram

$$\begin{array}{ccc} \mathsf{S}(\mathbf{v}_n, A) & \longrightarrow & A \\ \mathsf{S}(i,A) \downarrow & & f \downarrow \\ \mathsf{S}(\mathbf{w}_n, A) & \longrightarrow & B \end{array}$$

Here, the map $\mathsf{S}(\mathbf{v}_n, A) \to A$ is induced by the structure map $\mathsf{T}(\mathcal{A}_n, A) \to A$ of the $\mathcal{A}_n$-algebra structure on $A$, while the map $\mathsf{S}(\mathbf{w}_n, A) \to B$ is the composition of the map $\mathsf{S}(\mathbf{w}_n, f) : \mathsf{S}(\mathbf{w}_n, A) \to \mathsf{S}(\mathbf{w}_n, B)$ with the map $\mathsf{S}(\mathbf{w}_n, B) \to B$ induced by the $\mathcal{A}_{n+1}$-algebra structure on $B$. □

The interest of this result comes from the fact that if $\mathbf{a}$ is an exact dg-operad and $\mathcal{A} \to \mathbf{a}$ is an almost free resolution of $\mathbf{a}$, for example $\mathcal{A} = \mathcal{B}^*\mathcal{B}\mathbf{a}$, the results of the last section show that the homotopy theories of $\mathbf{a}$-algebras and $\mathcal{A}$-algebras are the same.

Let us give an example of the use of Theorem 4.13. Suppose that $A$ is a chain complex over a field $K$ and that there is given an $\mathbf{a}$-algebra structure on the homology $H_\bullet(A)$ of $A$. Choose a basis of $H_\bullet(A)$, lift each element of this basis to $A$, and extend this to a basis of all of $A$. This allows us to define an acyclic fibration $A \to H_\bullet(A)$. If $\mathcal{A} \to \mathbf{a}$ is an almost free resolution of the operad $\mathbf{a}$, then we conclude that $A$ has a $\mathcal{A}$-algebra structure, compatible with the $\mathbf{a}$-algebra structure on $H_\bullet(A)$.

If the almost free operad $\mathcal{A}$ has the form $\mathcal{B}^*\mathbf{z}$ for some operad $\mathbf{z}$, we see from Proposition 2.15 that a $\mathcal{A}$-algebra structure on $A$ is the same as a differential on the cofree $\mathbf{z}$-coalgebra $\mathsf{C}(\mathbf{z}, A)$. Applying this result to the operad $\mathcal{B}^*\mathbf{a}^\perp$, where $\mathbf{a}$ is a Koszul operad, we are led to the following definition.

**Definition 4.14.** If $\mathbf{a}$ is a Koszul operad, a homotopy $\mathbf{a}$-algebra structure $A$ is an $\mathbf{a}^\perp$-codifferential on the cofree $\mathbf{a}^\perp$-coalgebra $\mathsf{C}(\mathbf{a}^\perp, A)$, or equivalently, the structure of a $\mathcal{B}^*\mathbf{a}^\perp$-algebra structure on $A$, where $\mathcal{B}^*\mathbf{a}^\perp$ is the almost free resolution of $\mathbf{a}$.

There is a faithful embedding of the category of $\mathbf{a}$-algebras in the category of homotopy $\mathbf{a}$-algebras, induced by the functor which sends an $\mathbf{a}$-algebra $A$ to $\mathsf{B}(\mathbf{a}, A)$.

Since the $n$-algebra operad $\mathbf{e}_n$ is Koszul, we may define a homotopy $n$-algebra to be a graded vector space $A$ together with an $n$-codifferential on $\mathsf{C}_n(A)$. A morphism of homotopy $n$-algebras $A \to B$ is a homomorphism of $\mathbf{z}_n$-coalgebras $\mathsf{C}_n(A) \to \mathsf{C}_n(B)$ which respects the $n$-codifferentials.

Similarly, a homotopy Lie algebra is a graded vector space $L$ together with an $\mathcal{L}^\perp$-codifferential on

$$\mathsf{C}(\mathcal{L}^\perp, L) = \mathsf{C}(\Lambda^{-1}\mathbf{e}_\infty^*, L) \cong \Sigma^{-1}\mathsf{S}(\mathbf{e}_\infty^*, \Sigma L),$$

or equivalently a codifferential on the graded commutative coalgebra $S(e_\infty^*, \Sigma L)$. This definition of a homotopy Lie algebra coincides with that of Drinfeld [12], Hinich and Schechtman [25] and Lada and Stasheff [30].

## 5. Homotopy $n$-algebras

In the first three sections of this chapter, we study three operads: $A_\infty$, introduced by Stasheff [49], $B_\infty$, which we abstract from the work of Baues [7], and $C_\infty$, considered by Kadeishvili [27] and Kontsevich [29]. We define these operads in a uniform way: an algebra over one of these operads is a chain complex $A$ together with additional algebraic structure on the free coalgebra generated by $\Sigma A$: for $A_\infty$-algebras, the structure of a dg-coalgebra, for $B_\infty$-algebras that of a dg-bialgebra, and for $C_\infty$-algebras that of a dg-bialgebra with product equal to the shuffle product.

The operads $A_\infty$ and $C_\infty$ are almost free, isomorphic to $\mathcal{B}^* z_1$ and $\mathcal{B}^* z_\infty$, and thus describe homotopy associative and homotopy commutative algebras respectively. The case of $B_\infty$ is a little different, since it is not almost free. However, as we explain below, there is an almost free operad $\mathcal{E}_2$, resolving $e_2$, of which $B_\infty$ is a quotient.

Baues has shown that the space of singular cochains on a topological space is a $B_\infty$-algebra [7]. In Section 5.2, we show that if $A$ is an $A_\infty$-algebra, then the space of Hochschild cochains $C^\bullet(A, A)$ is a $B_\infty$-algebra. Interpreting the complex $C^\bullet(A, A)$ as the "homotopy centre" of the $A_\infty$-algebra $A$, this result takes on the following familiar form: the homotopy centre of a homotopy associative algebra is a homotopy commutative algebra.

In Section 5.4, we construct almost free resolutions $\mathcal{E}_n \to e_n$; in fact, the operad $\mathcal{E}_n$ is defined over the integers, and is generated by a free $\mathbb{Z}[\mathbb{S}]$-module. The chain complex $\mathcal{E}_n(k)$ is the cellular chain complex associated to an equivariant regular cell decomposition of the space $F_n(k)$. We call an algebra over $\mathcal{E}_n$ a homotopy $n$-algebra. The operads of the first three sections are related to these operads in the following way: the operads $A_\infty$ and $\mathcal{E}_1$ are isomorphic, $B_\infty$ is a quotient of $\mathcal{E}_2$, and $C_\infty$ is a quotient of $\mathcal{E}_\infty$. Finally, we show that the singular cochain functor $S^\bullet(X)$ takes values in the category of $\mathcal{E}_\infty$-algebras: in another paper, we will apply this to obtain an explicit model for the chains on $\Omega^n X$.

**5.1. $A_\infty$-algebras.** Let $V$ be a chain complex. The bar coalgebra of $V$ is the direct sum

$$BV = \bigoplus_{k=0}^{\infty} (\Sigma V)^{(k)},$$

and following the usual practice, we denote the element $(\Sigma a_1) \otimes \ldots \otimes (\Sigma a_n) \in BV$ by $[a_1|\ldots|a_n]$. The coproduct of $BV$ is given by the formula

$$\Delta[a_1|\ldots|a_k] = \sum_{i=0}^{k} [a_1|\ldots|a_i] \otimes [a_{i+1}|\ldots|a_k].$$

Denote by $\eta : \mathbb{1} \to BV$ the coaugmentation map given by the inclusion of the 0-chain $[\;]$ in $BV$, and by $\bar{B}V$ its cokernel, the non-counital coalgebra

$$\bar{B}V = \bigoplus_{k=1}^{\infty} (\Sigma V)^{(k)}.$$

There is a bijective correspondence between coassociative coalgebras $C$ and connected 1-coalgebras $\Sigma C$, under which the bar coalgebra $\bar{B}V$ is identified with the cofree 1-coalgebra $C_1(V) \cong \Sigma^{-1}\bar{B}V$.

A Hochschild cochain is an element of $C^\bullet(V, V) = \text{Hom}(BV, V)$. Proposition 2.14 gives a bijection from Hochschild cochains $c \in C^\bullet(V, V)$ to coderivations $\delta(c)$ of $BV$, given by the formula

$$\delta(c)[a_1|\ldots|a_k] = \sum_{0 \leq i < j \leq k} (-1)^{(|c|+1)(|a_1|+\cdots+|a_i|+i)} [a_1|\ldots|a_i|c[a_{i+1}|\ldots|a_j]|a_{j+1}|\ldots|a_k].$$

Note that $|\delta(c)| = |c| + 1$.

Gerstenhaber [16] has introduced an operation $c_1 \circ c_2$ of degree 1 on $C^\bullet(V, V)$:

$$(c_1 \circ c_2)[a_1| \ldots |a_k] = \sum_{0 \leq i < j \leq k} (-1)^{(|c_2|+1)(|a_1|+\cdots+|a_i|+i)} c_1[a_1|\ldots|a_i|c_2[a_{i+1}|\ldots|a_j]|a_{j+1}|\ldots|a_k].$$

The commutator $[c_1, c_2] = c_1 \circ c_2 - (-1)^{(|c_1|+1)(|c_2|+1)} c_2 \circ c_1$ of this operation is a Lie bracket of degree 1, as is shown by the formula

$$\delta([c_1, c_2]) = \delta(c_1)\delta(c_2) - (-1)^{(|c_1|+1)(|c_2|+1)} \delta(c_2)\delta(c_1).$$

A coderivation of $\mathsf{B}V$ of degree $-1$ is a codifferential if and only if the corresponding Hochschild cochain $m$ (of degree $-2$) satisfies the formula $m \circ m = 0$. The following definition is due to Stasheff [49] (see also [20]).

**Definition 5.1.** An $\mathrm{A}_\infty$-algebra structure on a graded vector space $A$ is one of the two equivalent data:
  (1) a Hochschild cochain $m \in \mathrm{Hom}(\mathsf{B}A, A)$ of degree $-2$ such that $m \circ m = 0$;
  (2) a codifferential on $\mathsf{B}A$ such that $\delta[\,] = 0$, or equivalently, an element of $\mathrm{Coder}_*(\mathbf{z}_1, \mathsf{C}_1(A))$.

The Hochschild cochain $m \in \mathrm{Hom}(\bar{\mathsf{B}}A, A)$ defining an $\mathrm{A}_\infty$-structure may be viewed as a sequence of multilinear maps $m_k : A^{(k)} \to A$, $k \geq 1$, of degree $k - 2$, defined by the formula

$$m_k(a_1, \ldots, a_k) = (-1)^{(k-1)|a_1|+(k-2)|a_2|+\cdots+|a_{k-1}|} m[a_1|\ldots|a_k].$$

The equation $m \circ m = 0$ translates to a sequence of identities generalizing those satisfied by a dg-algebra: in fact, $\mathrm{A}_\infty$ is isomorphic to the operad $\mathcal{B}^* \mathbf{z}_1$ whose algebras are homotopy associative algebras in the sense of Section 4.4.

In the next section, we will need a sequence of operations $c_0\{c_1, \ldots, c_k\}$ on the space of Hochschild cochains of a chain complex, constructed in [18], generalizing Gerstenhaber's operation $c_0\{c_1\} = c_0 \circ c_1$, given by the formula

$$(1) \quad c\{c_1, \ldots, c_k\}[a_1|\ldots|a_\ell] = \sum_{0 \leq i_1 \leq j_1 \leq \cdots \leq i_k \leq j_k \leq \ell} (-1)^{\eta_{i_1}|c_1|+\cdots+\eta_{i_k}|c_k|}$$
$$c[a_1|\ldots|a_{i_1}|c_1[a_{i_1+1}|\ldots|a_{j_1}]|a_{j_1+1}|\ldots|a_{i_k}|c_k[a_{i_k+1}|\ldots|a_{j_k}]|a_{j_k+1}|\ldots|a_\ell],$$

where $\eta_i = |a_1| + \cdots + |a_i| + i$. If $A$ is an $\mathrm{A}_\infty$-algebra, its space of Hochschild cochains $C^\bullet(A, A)$ is an $\mathrm{A}_\infty$-algebra, with differential $[m, -]$ and higher products $M_k$, $k > 1$, given by the formula

$$M_k(c_1, \ldots, c_k) = m\{c_1, \ldots, c_k\}.$$

If $A$ is a dg-algebra, this $\mathrm{A}_\infty$-structure on $C^\bullet(A, A)$ reduces to a dg-algebra structure, with product

$$(c_1 \cup c_2)[a_1|\ldots|a_k] = \sum_{i=0}^{k} c_1[a_1|\ldots|a_i] c_2[a_{i+1}|\ldots|a_k].$$

**5.2. $\mathrm{B}_\infty$-algebras.** A dg-bialgebra (or Hopf algebra) is a coalgebra $A$ which also carries the structure of a dg-algebra, in such a way that the product $A \otimes A \to A$ is a morphism of coalgebras and the differential $\delta : A \to A$ is a coderivation.

**Definition 5.2.** A $\mathrm{B}_\infty$-algebra is a chain complex $A$ together with the structure of a dg-bialgebra on $\mathsf{B}A$.

A $B_\infty$-algebra $A$ has an underlying $A_\infty$-structure, obtained by discarding the product on $\mathsf{B}A$, and retaining only the dg-coalgebra structure. As in the last section, this $A_\infty$-structure may be described by a sequence of multilinear maps $m_k : A^{(k)} \to A$, $k > 0$, of degree $k - 2$, satisfying certain identities generalizing those of a dg-algebra. Similarly, we will see that the product $\cup$ on $\mathsf{B}A$, where $A$ is a $B_\infty$-algebra, may be described by a sequence of multilinear maps $m_{k,\ell} : A^{(k+\ell)} \to A$, $k, \ell > 0$, of degree $k + \ell - 1$, satisfying certain identities which are equivalent to the associativity of this product on $\mathsf{B}A$ and its compatibility with the codifferential on $\mathsf{B}A$.

In order to do this, we must analyse the structure of bilinear maps $\Phi : \mathsf{B}A \otimes \mathsf{B}A \to \mathsf{B}A$ compatible with the coproduct on $\mathsf{B}A$. Such maps are determined by the composition of $\Phi$ with the projection $\mathrm{pr} : \mathsf{B}A \to \Sigma A$,
$$\mathsf{B}A \otimes \mathsf{B}A \xrightarrow{\Phi} \mathsf{B}A \xrightarrow{\mathrm{pr}} \Sigma A.$$
Indeed, the composition of the bilinear map $\Phi$ with the projection $\mathsf{B}A \to (\Sigma A)^{(k)}$ is given in terms of $\mathrm{pr} \cdot \Phi$ by the formula
$$\mathsf{B}A \otimes \mathsf{B}A \to (\mathsf{B}A)^{(k)} \otimes (\mathsf{B}A)^{(k)} \cong (\mathsf{B}A \otimes \mathsf{B}A)^{(k)} \xrightarrow{(\mathrm{pr} \cdot \Phi)^{(k)}} (\Sigma A)^{(k)}.$$
Thus, the bilinear map $\Phi$ is determined by a sequence of maps $m_{k,\ell} : A^{(k+\ell)} \to A$, $k + \ell > 0$, of degree $k + \ell - 1$, such that
$$m_{k,\ell}(a_1, \ldots, a_k; b_1, \ldots, b_\ell) =$$
$$(-1)^{(k-1)|a_1|+(k-2)|a_2|+\cdots+|a_{k-1}|}(-1)^{(\ell-1)|b_1|+(\ell-2)|b_2|+\cdots+|b_{\ell-1}|}(\mathrm{pr} \cdot \Phi)([a_1|\ldots|a_k],[b_1|\ldots|b_\ell]).$$

We may apply this analysis to the product $\cup$ on $\mathsf{B}A$, where $A$ is a $B_\infty$-algebra. In order for the zero-chain $[\ ] \in \mathsf{B}A$ to be a unit for $\cup$, it is necessary and sufficient that $m_{1,0}$ and $m_{0,1}$ be the identity map of $A$ and $m_{k,0} = m_{0,k} = 0$ for $k > 1$. The equations among the products $m_{k,\ell}$ which are equivalent to the associativity of $\cup$, and among the products $m_{k,\ell}$ and $m_k$ which are equivalent to the differential of $\mathsf{B}A$ being a derivation of $\cup$, are rather complicated, but it is clear that they could be written down if necessary.

In order to make the structure of $B_\infty$-algebras a little more familiar, we will now show explicitly that the homology of a $B_\infty$-algebra is a 2-algebra. This also follows from the fact, proved in Section 5.4, that $B_\infty$ is a quotient over an almost free resolution of the operad $\mathbf{e}_2$. We adopt the notations $\delta a$ for $m_1(a)$, $ab$ for $m_2(a,b)$, and $a \circ b$ for $m_{1,1}(a;b)$.

The product of two one-chains in $\mathsf{B}A$ is given by the formula
$$[a] \cup [b] = [a|b] + (-1)^{(|a|+1)(|b|+1)}[b|a] + [a \circ b].$$
Applying the differential of $\mathsf{B}A$ to both sides, we see that
$$ab - (-1)^{|a||b|}ba = (-1)^{|a|}(\delta a) \circ b - a \circ (\delta b) - (-1)^{|a|}\delta(a \circ b),$$
showing that $ab$ descends to a graded commutative product on $H_\bullet(A)$.

Define a bilinear operation of degree 1 on $A$ by the formula
$$[a,b] = a \circ b - (-1)^{(|a|+1)(|b|+1)}b \circ a.$$
It is automatic that this operation is a Lie bracket of degree 1, since it is related to the commutator of two one-chains by the formula
$$[a] \cup [b] - (-1)^{(|a|+1)(|b|+1)}[b] \cup [a] = [[a,b]].$$
Taking the differential of both sides of this formula, we see that
$$\delta[a,b] = [\delta a, b] + (-1)^{|a|+1}[a, \delta b],$$
so that the bracket descends to $H_\bullet(A)$.

It remains to prove the Poisson formula relating the products $ab$ and $[a,b]$. Using the fact that $\mathsf{B}A$ is a bialgebra, we see that the product $[a|b] \cup [c]$ is given by the formula

$$[a|b] \cup [c] = [a|b|c] + (-1)^{(|b|+1)(|c|+1)}[a|c|b] + (-1)^{(|c|+1)(|a|+|b|)}[c|a|b]$$
$$+ (-1)^{(|b|+1)(|c|+1)}[a \circ c|b] + [a|b \circ c] + (-1)^{|a|}[m_{2,1}(a,b;c)].$$

Similarly, the product $[c] \cup [a|b]$ is given by the formula

$$[c] \cup [a|b] = [c|a|b] + (-1)^{(|c|+1)(|a|+1)}[a|c|b] + (-1)^{(|c|+1)(|a|+|b|)}[a|b|c]$$
$$+ [c \circ a|b] + (-1)^{(|c|+1)(|a|+1)}[a|c \circ b] + (-1)^{|a|}[m_{1,2}(c;a,b)].$$

If we subtract $(-1)^{(|a|+|b|)(|c|+1)}$ times the second equation from the first, we see that

$$[a|b] \cup [c] - (-1)^{(|a|+|b|)(|c|+1)}[c] \cup [a|b] = (-1)^{(|b|+1)(|c|+1)}[[a,c]|b] + [a|[b,c]] + (-1)^{|a|}[n(a,b;c)],$$

where $n(a,b;c) = m_{2,1}(a,b;c) - (-1)^{(|a|+|b|)(|c|+1)}m_{1,2}(c;a,b)$. Since the differential of $\mathsf{B}A$ is a derivation with respect to the product $\cup$, we see that

$$[ab,c] - a[b,c] - (-1)^{|b|(|c|+1)}[a,c]b$$
$$= \delta(n(a,b;c)) + n(\delta a,b;c) + (-1)^{|a|}n(a,\delta b;c) + (-1)^{|a|+|b|+1}n(a,b;\delta c).$$

This proves the Poisson relation for the product $ab$ and the odd bracket $[a,b]$.

In practice, a restricted class of $\mathsf{B}_\infty$-algebras appears to arise most frequently, in which the operations $m_{k,\ell}$ vanish if $k > 1$. This condition leads to a great simplification in the equations for the cup product on $\mathsf{B}A$. Denoting $m_{1,k}(a;a_1,\ldots,a_k)$ by $a\{a_1,\ldots,a_k\}$, the formula for the cup product on $\mathsf{B}A$ is

$$[a_1|\ldots|a_k] \cup [b_1|\ldots|b_\ell] = \sum_{0 \le i_1 \le j_1 \le \cdots \le i_k \le j_k \le \ell} (-1)^{\eta_{i_1}|a_1|+\cdots+\eta_{i_k}|a_k|+\alpha_1+\cdots+\alpha_k}$$
$$[b_1|\ldots|b_{i_1}|a_1\{b_{i_1+1},\ldots,b_{j_1}\}|b_{j_1+1}|\ldots|b_{i_k}|a_k\{b_{i_k+1},\ldots,b_{j_k}\}|b_{j_k+1}|\ldots|b_\ell],$$

where $\eta_i = |b_1| + \cdots + |b_i| + i$ and $\alpha_i = (-1)^{(j_i-i_i-1)|b_{i_i+1}|+(j_i-i_i-2)|b_{i_i+2}|+\cdots+|b_{j_1-1}|}$. It is easy to see that the product is associative precisely when, in the same notation,

$$(c\{a_1,\ldots,a_k\})\{b_1,\ldots,b_\ell\} = \sum_{1 \le i_1 \le j_1 \le \cdots \le i_k \le j_k \le \ell} (-1)^{\eta_{i_1}|a_1|+\cdots+\eta_{i_k}|a_k|+\alpha_1+\cdots+\alpha_k}$$
$$c\{b_1,\ldots,b_{i_1},a_1\{b_{i_1+1},\ldots,b_{j_1}\},b_{j_1+1},\ldots,b_{i_k},a_k\{b_{i_k+1},\ldots,b_{j_k}\},b_{j_k+1},\ldots,b_\ell\},$$

and that it is compatible with the differential on $\mathsf{B}A$ when the map from $A$ to $C^\bullet(A,A)$ given by sending $a \in A$ to the inhomogeneous cochain $\sum_{k=0}^\infty a\{b_1,\ldots,b_k\}$ is a map of $\mathsf{A}_\infty$-algebras.

Baues [7] has shown that the complex of singular cochains $S^\bullet(X)$ on a topological space $X$ (with coefficients in any ring) is a $\mathsf{B}_\infty$-algebra of this type: the differential is the usual one, the product is the cup product, and the operations $m_{1,k}(a;b_1,\ldots,b_k)$ are multilinear analogues of Steenrod's operation $m_{1,1}(a;b) = a \cup_1 b$.

Another example of a $\mathsf{B}_\infty$-algebra is given by the space of Hochschild cochains $C^\bullet(A,A)$ on an $\mathsf{A}_\infty$-algebra. As we saw in Section 5.1, this is an $\mathsf{A}_\infty$-algebra; the products $m_{k,\ell}$ vanish if $k > 1$, while $m_{1,k}(c;c_1,\ldots,c_k) = c\{c_1,\ldots,c_k\}$ is given by (1). As we mentioned in the introduction to this chapter, this result may be interpreted as saying that the centre of an $\mathsf{A}_\infty$-algebra is a $\mathsf{B}_\infty$-algebra.

**5.3. $C_\infty$-algebras.** A $C_\infty$-algebra is a $B_\infty$-algebra such that the products $m_{k,\ell}$ vanish if $k+\ell > 1$; equivalently, it is an $A_\infty$-algebra such that the codifferential $\delta$ is a derivation with respect to the shuffle product on $BA$. We will show in this section that this is the same thing as an $A_\infty$-algebra whose multiplication cochain $m \in C^\bullet(A,A)$ is a Harrison cochain. It will follow that the operad $C_\infty$ is isomorphic to $\mathcal{B}^* \mathbf{z}_\infty$, whose algebras are homotopy commutative algebras in the sense of Section 4.4.

If $k, \ell \geq 0$, let $S(k,\ell)$ be the set of all $(k,\ell)$-shuffles, that is, permutations of $\{1, \ldots, k+\ell\}$ such that $\sigma(1) < \cdots < \sigma(k)$ and $\sigma(k+1) < \cdots < \sigma(k+\ell)$. The shuffle product $(\Sigma V)^{(k)} \otimes (\Sigma V)^{(\ell)} \to (\Sigma V)^{(k+\ell)}$ is defined by the formula

$$[a_1|\ldots|a_k] * [a_{k+1}|\ldots|a_{k+\ell}] = \sum_{\sigma \in S(k,\ell)} \pm [a_{\sigma(1)}|\ldots|a_{\sigma(k+\ell)}],$$

where the sign is chosen according to the sign convention for permuting the elements $\Sigma a_i \in \Sigma V$. This product is associative and graded commutative, and gives $BV$ the structure of a (unital, counital) bialgebra. The following result is clear.

**Proposition 5.3.** *The shuffle product of $BV$ is the unique associative product making the coalgebra $BV$ into a bialgebra for which $m_{1,0}$ and $m_{0,1}$ are the identity map of $V$, and $m_{k,\ell} = 0$ for $k+\ell > 0$.*

If $A$ is a bialgebra, denote by $\bar{A}$ the kernel of the counit $\varepsilon : A \to \mathbb{1}$, or the cokernel of the unit map $\eta : \mathbb{1} \to A$: these two spaces are naturally isomorphic, since $\varepsilon\eta : \mathbb{1} \to \mathbb{1}$ is the identity. Denote by $\mu$ the product and by $\Delta$ the coproduct of $A$. There are two filtrations on any bialgebra $A$ (see Milnor and Moore [35] and Quillen [42]).

(1) The descending filtration $F^i A$, where $F^0 A = A$ and $F^i A$ is the $i$-th power of $\bar{A}$ if $i \geq 1$.
(2) The ascending filtration $F_i A$, where $F_0 A = 0$ and $F_i A$ is the kernel of the iterated coproduct map $A \xrightarrow{\Delta^{(i-1)}} A^{(i)} \to \bar{A}^{(i)}$ for $i \geq 1$.

If $A$ is a bialgebra which is finite dimensional in each degree, then its dual $A^*$ is again a bialgebra. The two filtrations on $A$ are dual, in the sense that $(F^i A^*)^\perp = F_i A$.

If $A$ is a bialgebra, the quotient $QA = F^1 A/F^2 A$ is the space of indecomposables of $A$, while the quotient $PA = F_2 A/F_1 A$ is the space of primitives; clearly, $P(A^*) = (QA)^*$.

**Definition 5.4.** *If $A$ is a graded vector space, a Harrison cochain is a Hochschild cochain which vanishes on the image of the shuffle product and on the counit $[\,]$: the space of Harrison cochains may be identified with $\mathrm{Hom}(QBA, A) \subset \mathrm{Hom}(BA, A) \cong C^\bullet(A, A)$.*

By the associativity of the shuffle product, we see that the operation $c_1 \circ c_2$ on $C^\bullet(A, A)$ maps Harrison cochains into Harrison cochains. Thus, $\mathrm{Hom}(QBA, A)$ is a graded Lie subalgebra of $\mathrm{Coder}(BA)$.

**Proposition 5.5.** *The coderivation $\delta(c)$ of the coalgebra $BA$ corresponding to a Hochschild cochain $c$ is a derivation of the bialgebra $BA$ if and only if $c$ is a Harrison cochain.*

*Proof.* A derivation $\delta$ of the bialgebra $BA$ induces a linear map from $F^p BA$ to itself, for each $p > 0$. Since $F^2 BA \cap \Sigma A = 0$, the corresponding cochain $D \in \mathrm{Hom}(BA, A)$, which is obtained by composing $\delta$ with the projection from $BA$ onto $\Sigma A$, must vanish on $F^2 BA$; this shows that it lies in $\mathrm{Hom}(QBA, A)$.

Conversely, if $c$ is a Harrison cochain, it is easily seen that the associated coderivation $\delta(c)$ is a derivation with respect to the shuffle map. $\square$

Ree has studied the bialgebra $U$ dual to ours [43]. Its underlying algebra is the tensor algebra $\mathsf{T}_1^+(V)$ of a connected vector space $V$, with coproduct the sum over unshuffles

$$\Delta(a_1 \otimes \ldots \otimes a_n) = \sum_{k=1}^{n-1} \sum_{\sigma \in S(k,n-k)} \pm \big(a_{\sigma^{-1}(1)} \otimes \ldots \otimes a_{\sigma^{-1}(k)}\big) \otimes \big(a_{\sigma^{-1}(k+1)} \otimes \ldots \otimes a_{\sigma^{-1}(n)}\big).$$

Since this bialgebra is connected and cocommutative, it follows by the structure theorem of Milnor and Moore [35] that it is isomorphic to the universal enveloping algebra $U(L)$ of a connected graded Lie algebra $L$. Ree shows that this bialgebra is isomorphic to the universal enveloping algebra $U(\mathsf{T}_\mathcal{L}(V))$ of the free Lie algebra $\mathsf{T}_\mathcal{L}(V)$ of $V$, and under this isomorphism, the ascending filtration on $U$ is given by powers of the free Lie algebra $\mathsf{T}_\mathcal{L}(V)$, thought of as a subspace of $\mathsf{T}_1^+(V)$:

$$F_p U = \sum_{q \leq p-1} \mathsf{T}_\mathcal{L}(V)^q.$$

This filtration induces an ascending filtration $F_p \mathbf{e}_1$ of the $\mathbb{S}$-module underlying the operad $\mathbf{e}_1$, in which $F_p \mathbf{e}_1(k)$ consists of elements of the subspace $F_p$ of the free associative algebra generated by $\{a_1, \ldots, a_n\}$ in which each letter occurs just once. This may be identified with words which may be written as a product of at most $p-1$ Lie words.

The analogous filtrations of the $\mathbb{S}$-modules underlying $\mathbf{e}_n$, $n > 1$, may be defined by

$$F_p \mathbf{e}_n(k) = \bigoplus_{i \geq (n-1)(k-p+1)} H_i(\mathbb{F}_n(k)).$$

Note that $F_p \mathbf{e}_n = 0$ for $p < 2$. The reindexed filtration

$$G_p \mathbf{e}_n = F_{p+2} \mathbf{e}_n$$

gives $\mathbf{e}_n$ the structure of a filtered operad: the structure map $\mu$ preserves the filtration, in the sense that it sends $G_p(\mathbf{e}_n \circ \mathbf{e}_n)$ to $G_p \mathbf{e}_n$. The associated graded operad $\mathrm{gr}_G \mathbf{e}_n$ is naturally isomorphic to $\mathbf{e}_n$ for $n > 1$, while for $n = 1$ it is isomorphic to the Poisson operad $\mathbf{p}$.

Dualizing the results of Ree, we see that

$$\Sigma Q \mathsf{B} V \cong \mathsf{C}_\infty(V),$$

and that there is a bijection between $\infty$-codifferentials on $\mathsf{C}_\infty(V)$ and differentials of the bialgebra $\mathsf{B}V$ with shuffle product. This proves the following result.

**Proposition 5.6.** *A $C_\infty$-algebra is equivalent to a graded vector space $A$ together with a differential on the bialgebra $\mathsf{B}A$, or equivalently, an $A_\infty$-algebra whose multiplication cochain is a Harrison cochain.*

An antipode on a bialgebra is a map $S : A \to A$ of degree 0 such that the composition

$$A \xrightarrow{\Delta} A \otimes A \xrightarrow{A \otimes S} A \otimes A \xrightarrow{\mu} A$$

is the identity. The following lemma is easily proved by induction with respect to the filtration degree.

**Lemma 5.7.** *If the ascending filtration $F_i A$ of a bialgebra $A$ is complete, there is a unique antipode $S$ on $A$, and all derivations $\delta$ of $A$ commute with $S$.*

*Proof.* The proof is by induction on the ascending filtration $F_i A$ of $A$; $S$ is trivial to define on $F_0 A$, since $F_0 A = 0$. Assume that $S$ has been constructed on $F_{i-1} A$, and that $x \in F_i A$. Then

$$\Delta x = x \otimes 1 + \sum_i y_i \otimes z_i + 1 \otimes x,$$

where $y_i$ and $z_i$ lie in $F_{i-1}A$, and

(∗) $$S(x) = -x - \sum_i y_i S(z_i).$$

Since the ascending filtration $F_i A$ is complete, the antipode $S$ is completely determined.

If $\delta$ is a derivation on $A$, we will show that $\delta S(x) = S(\delta x)$ by induction on $x \in F_i A$. For $F_0 A$, this is clear. Applying the derivation $\delta$ to Eq. (∗) with $x \in F_i A$, and using the induction hypothesis that $S(\delta y) = \delta S(y)$ for $y \in F_{i-1}A$, we see that

$$\delta S(x) = -\delta x - \sum_i (\delta y_i) S(z_i) - \sum_i (-1)^{|y_i|} y_i \delta S(z_i).$$

On the other hand, since $\delta$ is compatible with $\Delta$,

$$\Delta(\delta x) = (\delta x) \otimes 1 + \sum_i (\delta y_i) \otimes z_i + \sum_i (-1)^{|y_i|} y_i \otimes (\delta z_i) + 1 \otimes (\delta x),$$

from which we see that

$$S(\delta x) = -\delta x - \sum_i (\delta y_i) S(z_i) - \sum_i (-1)^{|y_i|} y_i S(\delta z_i).$$

But by induction, $\delta S(z_i) = S(\delta z_i)$, and we see that $\delta S(x) = S(\delta x)$. □

The antipode of the Hopf algebra $\mathsf{B}V$ with shuffle product determined by Lemma 5.7 is given by the formula

$$S[a_1|\ldots|a_n] = \pm [a_n|\ldots|a_1];$$

the sign is given by the sign rule for the permutation $\begin{pmatrix} 1 & \ldots & n \\ n & \ldots & 1 \end{pmatrix}$, and equals $(-1)^{n(n-1)/2}$ if all of the elements $a_i$ have even degree. The commutativity of a homotopy commutative algebra is reflected by the fact that the differential $\delta$ commutes with the antipode.

We may now prove a theorem of Barr [5] and Quillen [41], which identifies the Harrison homology of a commutative algebra in characteristic zero with its André-Quillen homology.

**Theorem 5.8.** *Let $A$ be a commutative algebra over a field of characteristic zero. There is a natural homotopy equivalence of complexes*

$$\mathsf{L}\varepsilon_* A \cong (\mathsf{C}_\infty(A), \delta),$$

*where $\delta$ is the $\infty$-codifferential on $\mathsf{C}_\infty(A) \cong Q\mathsf{B}A$ induced by the Hochschild codifferential on $\mathsf{B}A$.*

*Proof.* This follows immediately from the identification of $\mathsf{B}_\infty(A)$ with $\Sigma Q\mathsf{B}A$, with differential induced by $\delta$. □

Barr's proof of this theorem is an application of what is now known as the Hodge filtration of the shuffle bialgebra, whose existence is a reflection of Ree's theorem. The descending filtration $F^p \mathsf{B}V$ induces a descending filtration of the $\mathbb{S}$-module $\mathbf{z}_1$, and Ree's theorem identifies $\mathbf{z}_\infty$ with $F^1 \mathbf{z}_1 / F^2 \mathbf{z}_1$. Over a field of characteristic zero, the Poincaré-Birkhoff-Witt theorem shows the existence of a family of commuting idempotents on $\mathbf{z}_1$ splitting this filtration (Solomon [48]; see also Hanlon [24] and Reutenauer [44]). The induced idempotents on the bar coalgebra $\mathsf{B}A$ commute with the Hochschild differential if $A$ is a $\mathsf{C}_\infty$-algebra, and we obtain the Hodge decomposition of $\mathsf{B}A$ (Gerstenhaber-Schack [17] and Loday [31]). The simplest of these idempotents, which Barr constructed by hand, projects from the Hochschild complex to the Harrison complex, permitting us to show that the Harrison complex of a free commutative algebra is a resolution of its space of indecomposables.

**5.4. The lexicographical decomposition of the configuration space of $\mathbb{R}^n$.** In this section, we show how the Fulton-MacPherson operad $\mathsf{F}_n$ of Chapter 3 may be given a regular cell decomposition. This permits the definition of dg-operads $\mathcal{E}_n$ which are almost free and resolve the Koszul operads $\mathbf{e}_n$ studied in Chapter 3. The operad $\mathcal{E}_\infty(k) = \operatorname*{colim}_{n \to \infty} \mathcal{E}_n(k)$ acts on the singular cochains $S^\bullet(M)$ of a topological space $M$, and each cell in $\mathsf{F}_\infty(k)$ may be thought of as a $k$-linear cochain operation: for example, $\mathsf{F}_\infty(2) \cong \mathring{\mathsf{F}}_\infty(2) \cong S^\infty$ has its standard $\mathbb{S}_2$-equivariant cell decomposition, and the associated operations are Steenrod's $\cup_i$-products.

**Definition 5.9.** *A preorder $\pi$ on a set $S$ is a reflexive transitive relation $\leq$ on $S$ such that if $a, b \in S$, either $a \leq b$ or $b \leq a$.*

A preorder determines an equivalence relation on $S$ ($a \sim b$ iff $a \leq b$ and $b \leq a$) and induces a total order on the quotient $S/\sim$. Denote by $|\pi|$ the number of equivalence classes. Preorders on $S$ form a poset: $\pi_1 \prec \pi_2$ if $a \leq_2 b$ implies that $a \leq_1 b$ for all $a, b \in S$. The maximal elements of this poset are the total orders of $S$, for which the equivalence relation is discrete. If $f : S \to T$ is a surjective map and $\pi$ is a preorder on $T$, there is an induced preorder $f^*\pi$ on $S$; this map preserves the partial order on the set of preorders.

A map $f : S \to \mathbb{R}^n$ determines a flag of preorders $\pi_1 \prec \cdots \prec \pi_n$ of $S$ in the following manner. Let $f_k$, $1 \leq k \leq n$, be the composition of $f$ with projection $(x_1, \ldots, x_n) \mapsto (x_1, \ldots, x_k)$ from $\mathbb{R}^n$ to $\mathbb{R}^k$. The image of $f_k$ is a finite set, totally ordered by the lexicographical ordering of $\mathbb{R}^k$: this is the total order such that $(x_1, \ldots, x_k) < (y_1, \ldots, y_k)$ if there exists $j$ such that $x_i = y_i$ for $i < j$ and $x_j < y_j$. The preorder $\pi_k$ is the pull-back by $f_k$ of this total order. Note that if $f$ is an embedding, $\pi_n$ is a total order on $S$.

We may now decompose the configuration space $\mathbb{F}_n(S)$ into convex cells. (In the special case $n = 2$, the decomposition which we describe may be found in Fox-Neuwirth [14], though the description of the boundary given there is incorrect.) Denote by $(\pi_1 \prec \cdots \prec \pi_n) \subset \operatorname{Map}(S, \mathbb{R}^n)$ the convex set consisting of those maps $f : S \to \mathbb{R}^n$ such that the induced flag of preorders on $S$ is $\pi_1 \prec \cdots \prec \pi_n$. This cell is contained in $\mathbb{F}_n(S)$ if and only if $\pi_n$ is a total order.

**Lemma 5.10.** *The set $(\pi_1 \prec \cdots \prec \pi_n) \subset \operatorname{Map}(\mathbb{R}^n, S)$ has dimension $\sum_{i=1}^n |\pi_i|$. The group of bijections $\mathbb{S}(S)$ acts transitively on the cells, through its action on the poset of preorders of $S$.*

The translation group preserves the lexicographical order of $\mathbb{R}^n$, and thus preserves the cells $(\pi_1 \prec \cdots \prec \pi_n)$. We denote the quotient of the cell $[\pi_1 \prec \cdots \prec \pi_n]$ by the translation group by $\mu(\pi_1 \prec \cdots \prec \pi_n)$: it is a cell of dimension $\sum_{i=1}^n |\pi_i| - n$. Denote by $\mathcal{Z}_n(S)$ the free $\mathbb{S}(S)$-module spanned by this collection of cells. As $S$ varies, we obtain a free $\mathbb{S}$-module $\mathcal{Z}_n$; note that $\mathcal{Z}_n(1)$ is one-dimensional, spanned by a unique 0-dimensional cell, while $\mathcal{Z}_n(0) = 0$.

A flag of preorders $\pi_1 \prec \cdots \prec \pi_n$ such that $\pi_n$ is a total order may be denoted in the following way. First, we write the elements of $S$ in the order determined by $\pi_n$, and surround them by brackets. Between each neighbouring pair of elements $a$ and $b$, we place $n - i + 1$ bars, where $i$ is the number of preorders in the flag $\pi_1 \prec \cdots \prec \pi_n$ such that $a \sim_i b$. The total number of bars is then the dimension of the associated cell in $\mathcal{Z}_n$. For example, $\mathbb{F}_2(3)$ is decomposed into cells $[\sigma_1|\sigma_2|\sigma_3]$, $[\sigma_1||\sigma_2|\sigma_3]$, $[\sigma_1|\sigma_2||\sigma_3]$ and $[\sigma_1||\sigma_2||\sigma_3]$, where $\sigma \in \mathbb{S}_3$.

The generating function for the number of cells $c_n(k, d)$ in $\mathbb{F}_n(k)$ of dimension $d$ is

$$\sum_{d=k-1}^\infty x^d c_n(k, d) = k!(x + \cdots + x^n)^{k-1},$$

Summing over $k$, we see that the generating function for the number of cells of $\mathcal{Z}_n$ is

$$\sum_{k=1}^\infty \frac{t^k}{k!} \sum_{d=k-1}^\infty x^d c_n(k, d) = \frac{t}{1 - t(x + \cdots + x^n)}.$$

In the limit $n \to \infty$, this formula converges to

$$\sum_{k=1}^{\infty} \frac{t^k}{k!} \sum_{d=k-1}^{\infty} x^d c_{\infty}(k,d) = \frac{t - tx}{1 - (1+t)x}.$$

The cases $n = 1$ and $n = 2$ are of special interest. If $n = 1$, a flag $\pi_1 \prec \cdots \prec \pi_n$ reduces to a single datum, a total order on $S$, and $\mathcal{Z}_1(k)$ has $k!$ cells, each of dimension $k - 1$. If $n = 2$, a flag $\pi_1 \prec \cdots \prec \pi_n$ reduces to a pair of data, a total order on $S$ and a preorder $\pi$ refining this total order. For each $1 \leq j \leq k$, there are $k!\binom{k-1}{j-1}$ cells in $\mathcal{Z}_2(k)$ of dimension $k + j - 2$, labelled by decompositions $k_1 + \cdots + k_j = k$ with $k_i \geq 1$.

Let $\mathring{\mathsf{F}}_n(S)$ be the quotient of the configuration space $\mathbb{F}_n(S)$ by the group of translations and dilatations $G(n)$. This space has a decomposition into cells $\nu(\pi_1 \prec \cdots \prec \pi_n)$, the quotients of the cells $\mu(\pi_1 \prec \cdots \prec \pi_n)$ by the dilatation group $\mathbb{R}_+^\times$. As $S$ varies, the spaces $\mathring{\mathsf{F}}_n(S)$ form an $\mathbb{S}$-module: we define $\mathring{\mathsf{F}}_n(0)$ and $\mathring{\mathsf{F}}_n(1)$ to be empty.

As explained in Chapter 3, Fulton and MacPherson construct a compactification $\mathsf{F}_n$ of the $\mathbb{S}$-space $\mathring{\mathsf{F}}_n$, such that $\mathsf{F}_n(S)$ is a manifold with corners containing $\mathring{\mathsf{F}}_n(S)$ as its unique open stratum. The $\mathbb{S}$-space $\mathsf{F}_n$ is obtained by gluing the free operad $\mathbb{T}\mathring{\mathsf{F}}_n$. In particular, the cell decomposition of $\mathring{\mathsf{F}}_n$ induces a cell decomposition of $\mathsf{F}_n$. The following lemma is clear from Fulton and MacPherson's description of the gluing maps of their compactification.

**Lemma 5.11.** *This cell decomposition of $\mathsf{F}_n$ is a regular cell complex, on which $\mathbb{S}$ acts freely. The structure maps of the operad $\mathsf{F}_n$ are cellular.*

The cells of $\mathsf{F}_n(S)$ are labelled by the following data:

(1) a tree $\mathcal{S} \in \mathcal{T}(S)$;
(2) a total order on $\mathrm{in}(v)$ for each vertex (so that $\mathcal{S}$ is a *planar* tree);
(3) a partition of the vertices of $\mathcal{S}$ into two sets, called the microscopic and macroscopic vertices.

In addition, it is required that the root is a macroscopic vertex, and each ascending chain of microscopic vertices is of length less than $n$.

Let us first consider the case in which there is a unique macroscopic vertex, the root of $\mathcal{S}$. The vertices connected to the root by an ascending chain of length $i + 1$ determine a preorder $\pi_i$ on $S$, since they are totally ordered and determine an evident partition of $S$ labelled by the set of vertices themselves. In this way, we obtain a flag of preorders $\pi_1 \prec \cdots \prec \pi_n$, such that $\pi_n$ is a total order. It is clear that we obtain in this way a bijection between planar trees on the set $S$ such that an ascending chain of vertices (possibly including the root) has length at most $n$ and cells of $\mathring{\mathsf{F}}_n(S)$.

This representation may be extended to cells of $\mathsf{F}_n(S)$, since such a cell is labelled by a tree on $S$, together with a choice of cell in $\mathring{\mathsf{F}}_n(\in(v))$ for each vertex $v$ of this tree. By inserting the corresponding planar tree in place of the vertex in the original tree, in such a way that the root is macroscopic and the remaining vertices are microscopic, we obtain the representation whose existence was asserted.

**Definition 5.12.** *Let $\mathcal{E}_n$ be the almost free dg-operad obtained by taking the cellular chain complex of the cellular operad $\mathsf{F}_n$. We call an $\mathcal{E}_n$-algebra a homotopy $n$-algebra.*

The $\mathbb{S}$-module $\mathcal{Z}_n$ spanned by the cells $\mu(\pi_1 \prec \cdots \prec \pi_n)$ is augmented by the map $\varepsilon : \mathcal{Z}_n \to \mathbb{1}$ which projects onto the unique cell in $\mathcal{Z}_n(1)$. We see that the free operad underlying $\mathcal{E}_n$ is $\mathbb{T}\Sigma^{-1}\bar{\mathcal{Z}}_n$, where

$$\bar{\mathcal{Z}}_n(k) = \begin{cases} \mathcal{Z}_n(k), & k > 1, \\ 0, & k \leq 1, \end{cases}$$

is the kernel of the augmentation, and that the graded space of generators $\Sigma^{-1}\mathcal{Z}_n$ corresponds to the cells of $\mathring{\mathsf{F}}_n \subset \mathsf{F}_n$. It would be very satisfactory if $\mathcal{Z}_n$ was a cooperad and $\mathcal{E}_n$ was its cobar operad $\mathcal{B}^*\mathcal{Z}_n$. Unfortunately, except for $n = 1$, where $\mathcal{Z}_1 = \mathbf{z}_1$, this is not the case. However, in the same way that a homotopy coalgebra $C$ is a chain complex $C$ with coaugmentation $\eta : \mathbb{1} \to C$ together with a differential $\delta$ on the free algebra $\mathsf{T}\Sigma^{-1}\bar{C}$, where $\bar{C} = \mathrm{coker}(\eta)$, we may define a homotopy cooperad $\mathbf{z}$ to be an $\mathbb{S}$-module with coaugmentation $\eta : \mathbb{1} \to \mathbf{z}$ together with a differential on the free operad $\mathbb{T}\Sigma^{-1}\bar{\mathbf{z}}$, where $\bar{\mathbf{z}} = \ker(\varepsilon)$.

It is then clear that the $\mathbb{S}$-module $\mathcal{Z}_n$ is a homotopy cooperad, by the existence of the almost free operad $\mathcal{E}_n$. We may think of $\mathcal{E}_n$ as the cobar operad of the homotopy cooperad $\mathcal{Z}_n$: when $n = 1$, $\mathcal{Z}_1 = \mathbf{z}_1$ is a true cooperad, and $\mathcal{E}_1$ is isomorphic to the operad $\mathrm{A}_\infty$.

Since the free homotopy 2-algebra $\mathsf{T}(\mathcal{E}_2, V)$ is given by the formula

$$\mathsf{T}(\mathcal{E}_2, V) = \bigoplus_{k=1}^{\infty} H_\bullet(\mathbb{B}_k, V^{(k)}),$$

we call a homotopy 2-algebra a braid algebra. A braid algebra has products $m_{k_1,\ldots,k_q}$ of degree $k_1 + \cdots + k_q + q - 3$ for each sequence $k_i \geq 1$, where the only non-vanishing product with $k_1 + \cdots + k_q = 1$ is $m_1$, the differential of $A$. As we saw in Section 5.2, a $\mathrm{B}_\infty$-algebra has products $m_k$ and $m_{k,\ell}$. In fact, the $\mathrm{B}_\infty$-operad is the quotient of $\mathcal{E}_2$, in which the products $m_{k_1,\ldots,k_q}$, $q > 2$, and their differentials, are set to zero.

Observe that there is a natural embedding of dg-operads $\mathcal{E}_n \hookrightarrow \mathcal{E}_{n+1}$ induced by the inclusion $\mathsf{F}_n \hookrightarrow \mathsf{F}_{n+1}$, and we denote by $\mathcal{E}_\infty$ the colimit $\varinjlim_{n \to \infty} \mathcal{E}_n$. The cellular basis of $\mathcal{E}_\infty$ is labelled by planar trees with a choice of macroscopic vertices as above (in particular, the root is macroscopic), but without any condition on the length of ascending chains of microscopic vertices.

There is a map from the operad $\mathcal{E}_n$ to $\mathbf{e}_n$, induced by sending the cell $[a|b]$ to the operation $x_a x_b$ in $\mathbf{e}_n(2)$, the cell $[a \overset{n}{|} b]$ with $n$ bars to the operation $\{x_a, x_b\} = (-1)^{|x_a|}[x_a, x_b]$ in $\mathbf{e}_n(2)$, and all other generators of $\mathcal{E}_n$ to zero. This map is a weak equivalence, since $H_\bullet(\mathcal{E}_n(k)) \cong \mathbf{e}_n(k)$ for all $k \geq 1$, and $\mathbf{e}_n(2)$ generates $\mathbf{e}_n$ as an operad. Thus, $\mathcal{E}_n$ is an almost free resolution of the operad $\mathbf{e}_n$.

Using the results of Smirnov [46] and Hinich and Schechtman [25], it now follows that the singular cochain functor $S^\bullet(X)$ lifts to a functor from the category of topological spaces to the category of homotopy $\infty$-algebras. Let $\mathcal{E}$ be the Eilenberg-Zilber operad, such that $\mathcal{E}(k)$ is the complex of natural transformations from the cosimplicial chain complex $\mathbb{Z} \otimes \Delta^\bullet$ of chains on the cosimplicial space $\Delta^\bullet$ to its $k$-th tensor power $(\mathbb{Z} \otimes \Delta^\bullet)^{(k)}$. The cup-product induces a surjective map $\mathcal{E} \to \mathbf{e}_\infty$, which is a weak equivalence. It is easily seen that the singular cochain functor $S^\bullet(-)$ lifts to a functor from the category of topological spaces to the category of $\mathcal{E}$-algebras. (Note that the operad $\mathcal{E}$ is not bounded below, but this does not cause any difficulty with the argument.)

Since the operad $\mathcal{E}_\infty$ is almost free, and the map $\mathcal{E} \to \mathbf{e}_\infty$ is an acyclic fibration, there is a lift $\mathcal{E}_\infty \to \mathcal{E}$ in the diagram

$$\begin{array}{ccc} & & \mathcal{E} \\ & \nearrow & \downarrow \\ \mathcal{E}_\infty & \longrightarrow & \mathbf{e}_\infty \end{array}$$

and this map is again a weak equivalence. This shows that there is a natural transformation from the category of $\mathcal{E}$-algebras to the category of homotopy $\infty$-algebras.

The space $\mathsf{F}_n(2)$ is diffeomorphic to $S^{n-1}$, and its cell decomposition is the usual $\mathbb{S}_2$-invariant one, into cells $[1 \overset{i}{|} 2]$ and $[2 \overset{i}{|} 1]$ of dimension $i - 1$, for $1 \leq i \leq n$. The above argument may be modified to ensure that the map of operads $\mathcal{E}_\infty \to \mathcal{E}$ is chosen in such a way that the element $[1 \overset{i}{|} 2]$ with $i$ bars of $\mathcal{E}_\infty(2)$ maps to Steenrod's operaton $\cup_i$, and that the operations $m_{k_1,\ldots,k_q}$ generating $\mathcal{E}_2$ maps to the operations constructed by Baues (and in particular vanish except for $m_{1,k}$). It might be interesting to construct a weak equivalence of operads $\mathcal{E}_\infty \to \mathcal{E}$ explicitly, extending Baues's argument to higher levels of commutativity.

The above argument may be applied with the Eilenberg-Zilber operad $\mathcal{E}$ replaced by the operad $C_\infty$. The resulting weak equivalence $\mathcal{E}_\infty \to C_\infty$ is given by an explicit formula, in which all of the generators of $\mathcal{E}_\infty$ other than $m_k$, $k \geq 1$, are sent to zero.

Recall from Section 3.3 that there is a map of operads $\Lambda \mathbf{e}_{n+1} \to \mathbf{e}_n$ induced by the Gysin map of the embedding $\mathsf{F}_n \hookrightarrow \mathsf{F}_{n+1}$. This map lifts to a map of operads $\Lambda \mathcal{E}_{n+1} \to \mathcal{E}_n$, defined in the following way. It suffices to define the map on the $\mathbb{S}$-module of generators $\Lambda \Sigma^{-1} \bar{\mathcal{Z}}_{n+1}$ of $\mathcal{E}_{n+1}$. Given a planar tree representing a basis element of $\mathcal{Z}_{n+1}$, its image is non-zero only if the valence of the root equals 1, in which case its image in $\Sigma^{-1} \bar{\mathcal{Z}}_n$ is represented by the tree obtained by removing the root: the root of the new planar tree is the parent of the old one. For example, the cell $[1 \overset{i}{\mid} 2]$ of $\Lambda\Sigma^{-1}\mathcal{Z}_{n+1}(2)$ with $i$ bars maps to the cell $[1 \overset{(i-1)}{\mid} 2]$ of $\mathcal{E}_{n+1}(2)$ with $i-1$ bars.

Geometrically, the map $\Lambda \mathcal{E}_{n+1} \to \mathcal{E}_n$ corresponds to the operation of projecting cells of $\mathsf{F}_{n+1}(k)$ onto $\{(x_1, \ldots, x_{n+1}) \in \mathbb{R}^n \mid x_1 = 0\}$; if the resulting projection does not correspond to a cell in $\mathsf{F}_n(k)$, because two points have the same projection, the element of $\Lambda \mathcal{E}_{n+1}(k)$ spanned by the cell of $\mathsf{F}_{n+1}(k)$ is mapped to zero in $\mathcal{E}_n(k)$. This representation makes it clear that the map thus defined intertwines the differentials of the operads $\Lambda \mathcal{E}_{n+1}$ and $\mathcal{E}_n$.

Taking $n \to \infty$, we obtain a map of operads $\omega : \mathcal{E}_\infty \to \Lambda^{-1}\mathcal{E}_\infty$. In another paper, we will prove the following version of a theorem of Smirnov [46]. This theorem strengthens the theorem of Adams [1] identifying the cohomology of $\Omega X$, for $X$ a simply connected topological space, with the Hochschild homology of $S^\bullet(X)$.

**Theorem 5.13.** *The functors $\mathsf{L}\omega_* S^\bullet(X)$ and $\Sigma^{-1} S^\bullet(\Omega X)$ from the homotopy category of simply connected topological spaces to the homotopy category of $\mathcal{E}_\infty$-algebras are equivalent.*

We close this section with a discussion of the differential of $\mathcal{E}_n$. This differential is of course determined by its restriction to $\Sigma^{-1} \mathcal{Z}_n \subset \mathcal{E}_n$. The differential induced on $\Sigma^{-1}\mathcal{Z}_n$ by projection from $\mathcal{E}_n$ to $\Sigma^{-1}\mathcal{Z}_n$ may be identified with the boundary of the relative cell complex $(\mathsf{F}_n(k), \partial \mathsf{F}_n(k))$. To determine the signs, we orient the cells of $\mathring{\mathsf{F}}_n(k)$, ordering their coordinates as follows: first the coordinates $x_1$ of the equivalence classes of $\pi_1$ in increasing order, next the coordinates $x_2$ of the equivalence classes of $\pi_2$, also in increasing order, and so on. For example, this rule leads to the following formula for the boundary of the cell $[1||2||3]$ in $\mathcal{Z}_2(3)$:

$$\partial[1||2||3] = [1|2||3] - [2|1||3] - [1||2|3] + [1||3|2].$$

Similarly, we have

$$\partial[1|2||3] = [1|2|3] - [1|3|2] + [3|1|2],$$
$$\partial[1||2|3] = [1|2|3] - [2|1|3] + [2|3|1].$$

Using these formulas, it is easily seen that $\partial^2[1||2||3] = 0$.

There is an elegant way to describe the differential on $\mathcal{Z}_n$ in terms of the functor $\mathsf{B}_1 \cong \Sigma^{-1}\bar{\mathsf{B}}$ on the category of commutative dg-algebras. The functor $\mathsf{B}_1^n \cong \Sigma^{-n}\bar{\mathsf{B}}^n$ from commutative differential graded algebras to chain complexes obtained by iterating this functor $n$ times is central to Eilenberg and MacLane's approach to the homology of $K(\pi, n)$ [13]. The following result follows straightforwardly from our description of the boundary in $\mathcal{Z}_n$.

**Proposition 5.14.** *There is a natural equivalence of endofunctors on the category of chain complexes*

$$\mathsf{T}(\mathcal{Z}_n, V) \cong \mathsf{B}_1^n V \cong \Sigma^{-n}\bar{\mathsf{B}}^n V.$$

This relationship between Eilenberg and MacLane's functor $\Sigma^{-n}\bar{\mathsf{B}}^n$ and the theory of configuration spaces does not appear to have been observed before.

The formula for the full differential $\partial : \mathcal{E}_n \to \mathcal{E}_n$ may also be described in terms of planar trees, but we leave its explicit description to another paper.

## 6. $n$-ALGEBRAS AND ITERATED INTEGRALS OF DOUBLE LOOP SPACES

In this chapter, we show how almost free resolutions may be applied to the calculation of the total left derived functor $\mathsf{L}\varepsilon_*$ of the indecomposables of an $n$-algebra. Recall from Section 4.3 that this functor may be represented by the almost free $n$-coalgebra $\mathsf{B}_n(A) = \mathsf{B}(\mathbf{e}_n, A)$. We call the homology of $\mathsf{B}_n(A)$ the $n$-homology of $A$, and denote it by $\mathsf{H}(n, A)$.

In Section 6.1, we recall the theory of Hopf operads, due to Ginzburg and Kapranov (unpublished). Just as a Hopf algebra is an algebra in the monoidal category of coalgebras, so a Hopf operad is an operad $\mathbf{a}$ in the monoidal category of coalgebras: we show that if $\mathbf{a}$ is a Hopf operad, there is a natural $\mathbf{a}$-algebra structure on the tensor product of two $\mathbf{a}$-algebras. In the same way that the homology of a topological monoid is a Hopf algebra, the homology of a topological operad is a Hopf operad: this shows, for example, that the operads $\mathbf{e}_n$ are Hopf operads.

In Section 6.2, we prove a Künneth theorem: if $A_1$ and $A_2$ are $n$-algebras, the $n$-homology of $A_1 \boxtimes A_2$ is given by the formula

$$\Sigma^n \mathsf{H}(n, A_1 \boxtimes A_2) \simeq \Sigma^n \mathsf{H}(n, A_1) \boxtimes \Sigma^n \mathsf{H}(n, A_2);$$

here $V \boxtimes W = V \oplus W \oplus (V \otimes W)$. This is proved by explicitly constructing an almost free resolution of $A_1 \boxtimes A_2$ from almost free resolutions of $A_1$ and $A_2$.

In Section 6.3, we calculate the $n$-homology of a commutative algebra $A$ considered as an $n$-algebra: we prove that

$$\Sigma^n \mathsf{B}_n(A) \cong \mathsf{T}_\infty(\Sigma^n \mathsf{B}_\infty(A)).$$

We also show how, given a minimal model for $A$ as a commutative algebra, we may construct a minimal model for $A$ as an $n$-algebra.

The results of this chapter may be applied to the de Rham functor $\mathcal{A}^\bullet(M)$ of a pointed manifold $M$, where $\mathcal{A}^\bullet(M)$ is algebra of differential forms $\mathcal{A}^\bullet(M)$ which vanish at the base-point. If $M$ is a compact $n$-connected manifold, the cochain algebra $\mathcal{A}^\bullet(M)$ is weakly equivalent to an $n$-connected cochain algebra $\mathcal{A}^\bullet_{[n]}(M)$: if $\mathcal{V}^{n+1}$ is a complement to the subspace $d\mathcal{A}^n(M) \subset \mathcal{A}^{n+1}(M)$, then

$$\mathcal{A}^i_{[n]}(M) = \begin{cases} 0, & i \leq n, \\ \mathcal{V}^{n+1}, & i = n+1, \\ \mathcal{A}^i(M), & i > n+1. \end{cases}$$

This allows us to replace the unbounded complex $\mathsf{B}_n(\mathcal{A}^\bullet(M))$ by the weakly equivalent cochain complex $\mathsf{B}_n(\mathcal{A}^\bullet_{[n]}(M))$. However, we will speak of the $n$-homology $\mathsf{H}(n, \mathcal{A}^\bullet(M))$ rather than $\mathsf{H}(n, \mathcal{A}^\bullet_{[n]}(M))$.

Let $\tilde{\pi}_\bullet(M)$ be the cokernel of the map $\mathbb{Z} \to \pi_\bullet(M)$ induced by inclusion of a base-point in $M$, and let $\tilde{\pi}^\bullet(M, \mathbb{C}) = \operatorname{Hom}(\tilde{\pi}_\bullet(M), \mathbb{C})$ be the reduced cohomotopy of $M$ with coefficients in $\mathbb{C}$. Quillen's and Sullivan's work in rational homotopy theory shows that, for compact simply connected manifolds, the André-Quillen homology of $\mathcal{A}^\bullet(M)$ is naturally equivalent to $\tilde{\pi}^\bullet(M, \mathbb{C})$. This is an $\infty$-coalgebra, since $\pi_\bullet(M)$ carries a Lie bracket of degree 1, the Whitehead product, and the isomorphism between the André-Quillen homology of $\mathcal{A}^\bullet(M)$ and $\tilde{\pi}^\bullet(M, \mathbb{C})$ is an isomorphism of $\infty$-coalgebras.

If $M$ is a pointed compact manifold, the $n$-fold loop space $\Omega^n M$ is the space of differentiable maps from $\mathbb{R}^n$ to $M$ equal to the basepoint of $M$ outside a bounded set in $\mathbb{R}^n$. The following theorem generalizes results of Adams [1] and Chen [10] in the case $n = 1$.

**Theorem 6.1.** *The n-fold suspension of the n-homology of $\mathcal{A}^\bullet(M)$ is isomorphic to the reduced cohomology of the n-fold loop space of $M$,*

$$\Sigma^n \mathsf{H}(n, \mathcal{A}^\bullet(M)) \cong \tilde{H}^\bullet(\Omega^n M, \mathbb{C}).$$

*Proof.* By the results of Section 6.3,

$$\Sigma^n \mathsf{H}(n, \mathcal{A}^\bullet_{[n]}(M)) \cong \mathsf{T}_\infty(\Sigma^n \mathsf{H}(\infty, \mathcal{A}^\bullet_{[n]}(M)))$$
$$\cong \mathsf{T}_\infty(\Sigma^n \tilde{\pi}^\bullet(M, \mathbb{C})).$$

This is isomorphic to $\tilde{H}^\bullet(\Omega^n M, \mathbb{C})$, since for any connected $H$-space $X$,

$$\tilde{H}^\bullet(X, \mathbb{C}) \cong \mathsf{T}_\infty(\tilde{\pi}^\bullet(X, \mathbb{C})),$$

while $\tilde{\pi}_\bullet(\Omega^n M) \cong \Sigma^n \tilde{\pi}_\bullet(M)$, since $M$ is $n$-connected. □

This isomorphism preserves the natural co-Hopf $\mathsf{e}_n^*$-coalgebra structures on both sides: the $\mathsf{e}_n^*$-coalgebra structure of $\tilde{H}^\bullet(\Omega^n M, \mathbb{C})$ corresponds to the $n$-algebra structure of $\tilde{H}_\bullet(\Omega^n M, \mathbb{C})$ induced by the action of the little $n$-cubes operad $\mathcal{F}_n$ on $\Omega^n M$, while the $\mathsf{e}_n^*$-coalgebra structure of $\Sigma^n \mathsf{H}(n, \mathcal{A}^\bullet(M))$ is induced by the $n$-coalgebra structure of $\mathsf{B}_n(\mathcal{A}^\bullet(M))$. In fact, these are commutative co-Hopf $\mathsf{e}_n^*$-coalgebras: it is shown in Section 6.3 that $\Sigma^n \mathsf{B}_n(A)$ is a commutative co-Hopf $\mathsf{e}_n^*$-coalgebra for any commutative algebra $A$, while $\tilde{H}^\bullet(\Omega^n M, \mathbb{C})$ is a commutative co-Hopf $\mathsf{e}_n^*$-coalgebra with the cup product as its product.

Chen has realized the isomorphism $\Sigma^n \mathsf{H}(n, \mathcal{A}^\bullet(M)) \cong \tilde{H}^\bullet(\Omega^n M)$, for $n = 1$, by defining an explicit map, the iterated integral

$$\Sigma \mathsf{B}_1(\mathcal{A}^\bullet(M)) \to \mathcal{A}^\bullet(\Omega M),$$

which is a homomorphism of commutative algebras (and indeed of commutative Hopf algebras), and a weak equivalence if $M$ is simply connected. (As we saw in Section 5.1, $\Sigma \mathsf{B}_1(\mathcal{A}^\bullet(M))$ is isomorphic to the bar complex $\mathsf{B}\mathcal{A}^\bullet(M)$ in the usual sense.) In Section 6.4, we define an analogous iterated integral map

$$\Sigma^2 \mathsf{B}_2(\mathcal{A}^\bullet(M)) \to \mathcal{A}^\bullet(\Omega^2 M),$$

which is a homomorphism of commutative co-Hopf $\mathsf{e}_2^*$-coalgebras, and a weak equivalence if $M$ is 2-connected. An iterated integral map for $n > 2$ would be substantially more complicated to define, since the configuration spaces $\mathbb{F}_n(k)$ are formal only for $n = 1, 2, \infty$ (Kontsevich, unpublished).

If $M$ and $N$ are two $n$-connected manifolds, there is a weak equivalence of commutative algebras $\mathcal{A}^\bullet(M) \boxtimes \mathcal{A}^\bullet(N) \simeq \mathcal{A}^\bullet(M \times N)$, and the Künneth theorem of Section 6.2 amounts to the Künneth theorem for the cohomology of $\Omega^n(M \times N) \cong \Omega^n M \times \Omega^n N$:

$$\Sigma^n \mathsf{B}_n(\mathcal{A}^\bullet(M \times N)) \simeq \Sigma^n \mathsf{B}_n(\mathcal{A}^\bullet(M) \boxtimes \mathcal{A}^\bullet(N))$$
$$\simeq \Sigma^n \mathsf{B}_n(\mathcal{A}^\bullet(M)) \boxtimes \Sigma^n \mathsf{B}_n(\mathcal{A}^\bullet(N))$$
$$\simeq \mathcal{A}^\bullet(\Omega^n M) \boxtimes \mathcal{A}^\bullet(\Omega^n N)$$
$$\simeq \mathcal{A}^\bullet(\Omega^n M \times \Omega^n N) = \mathcal{A}^\bullet(\Omega^n(M \times N)).$$

Sullivan has given a prescription for constructing a minimal model for $\mathrm{Map}_f(N, M)$, where $M$ and $N$ are differentiable manifolds and $\mathrm{Map}_f(N, M)$ is the space of continuous maps in the component of a map $f : N \to M$ (Section 11 of [50]). Applied with $N = S^2$, we obtain a model for $\Omega^2 M$ quite different from our model $\Sigma^2 \mathsf{B}_2(\mathcal{A}^\bullet(M))$. For example, our model carries a co-Hopf $\mathsf{e}_2^*$-coalgebra structure reflecting the fact that $\Omega^2 M$ is a double loop space, while it is not clear how this information might be gleaned from Sullivan's model.

**6.1. Hopf operads and co-Hopf cooperads.** In any symmetric monoidal category, define the augmented tensor product:
$$V \boxtimes W = V \oplus W \oplus (V \otimes W).$$
This is again a symmetric monoidal structure. In the category of commutative algebras, the coproduct $A \coprod B$ has as its underlying chain complex the augmented tensor product $A \boxtimes B$.

If $A$ and $B$ are $n$-algebras, we may use the explicit description of $n$-algebras given in Theorem 1.6 to give their augmented tensor product $A \boxtimes B$ an $n$-algebra structure: indeed, there is a unique $n$-algebra structure such that the product of $a \in A$ with $b \in B$ is the tensor product $a \otimes b$, and the bracket of $a$ and $b$ equals zero. There is a more elementary construction of this monoidal structure on the category of $n$-algebras, not requiring the explicit structure of $n$-algebras as contained in Theorem 1.6: this makes use of Ginzburg and Kapranov's notion of a Hopf operad. In particular, this theory will apply to the operad $\mathbf{e}_n$, which is the homology of the topological operad $\mathcal{F}_n$.

Denote by $\mathrm{Alg}(\mathcal{C})$ the category of algebras in a symmetric monoidal category; this is itself a symmetric monoidal category, with respect to the tensor product $-\otimes-$ of $\mathcal{C}$. An algebra is commutative if the product commutes with the symmetry $V \otimes V \to V \otimes V$ of $\mathcal{C}$. In what follows, algebras do not have units, unless they are specifically referred to as unital.

Dually, denote by $\mathrm{Coalg}(\mathcal{C})$ the category of coalgebras over a symmetric monoidal category; this is again a symmetric monoidal category, with respect to the augmented tensor product of $\mathcal{C}$. A coalgebra is cocommutative if the coproduct commutes with the symmetry $V \otimes V \to V \otimes V$ of $\mathcal{C}$. Coalgebras will not have counits, unless they are specifically referred to as counital.

**Definition 6.2.** A Hopf operad in a symmetric monoidal category $\mathcal{C}$ is an operad $\mathbf{a}$ in the category of coalgebras $\mathrm{Coalg}(\mathcal{C})$ over $\mathcal{C}$. A Hopf algebra over a Hopf operad $\mathbf{a}$ is an $\mathbf{a}$-algebra $A$ in $\mathrm{Coalg}(\mathcal{C})$. The operad $\mathbf{a}$ and the algebra $A$ are counital (respectively cocommutative) if their underlying coalgebras are.

A co-Hopf cooperad in a symmetric monoidal category $\mathcal{C}$ is a cooperad $\mathbf{z}$ in the category of algebras $\mathrm{Alg}(\mathcal{C})$ over $\mathcal{C}$. A co-Hopf coalgebra over a co-Hopf cooperad $\mathbf{z}$ is a $\mathbf{z}$-coalgebra $C$ in $\mathrm{Alg}(\mathcal{C})$. The cooperad $\mathbf{z}$ and the coalgebra $C$ are unital (respectively commutative) if their underlying algebras are.

For example, a Hopf operad such that $\mathbf{a}(S) = 0$ for $|S| \neq 1$ is a Hopf algebra in the usual sense. The following result generalizes the fundamental property of the category of modules over a Hopf algebra.

**Proposition 6.3.** *If $\mathbf{a}$ is a Hopf operad, the category of $\mathbf{a}$-algebras is a monoidal category with respect to the augmented tensor product $A \boxtimes B$, and is symmetric if $\mathbf{a}$ is cocommutative.*

*Proof.* If $A$ and $B$ are $\mathbf{a}$-algebras, the structure map of $A \boxtimes B$ is defined in the same way as for modules over Hopf algebras, by the composition
$$\mathsf{T}(\mathbf{a}, A \boxtimes B) \to \mathsf{T}(\mathbf{a}, A) \oplus \mathsf{T}(\mathbf{a}, B) \oplus \mathsf{T}(\mathbf{a} \otimes \mathbf{a}, A \otimes B) \hookrightarrow \mathsf{T}(\mathbf{a}, A) \boxtimes \mathsf{T}(\mathbf{a}, B) \to A \boxtimes B. \quad \square$$

We also have the dual result.

**Proposition 6.4.** *If $\mathbf{z}$ is a co-Hopf cooperad, the category of $\mathbf{z}$-coalgebras is a monoidal category with respect to the augmented tensor product $A \boxtimes B$, and is symmetric if $\mathbf{z}$ is commutative.*

If the symmetric monoidal structure of $\mathcal{C}$ is cartesian (the tensor product of $\mathcal{C}$ is the product), operads in $\mathcal{C}$ have a canonical structure of a cocommutative Hopf operad, with coproduct the diagonal map. Thus, any topological operad is a cocommutative Hopf operad, and if $\mathcal{F}$ is a topological operad, the dg-operad $\mathbf{a}(S) = H_\bullet(\mathcal{F}(S), K)$ is a cocommutative Hopf operad. (This generalizes the fact that the homology of a topological monoid is a cocommutative Hopf algebra.) In particular, the operads $\mathbf{e}_n$ of Section 1.3 are cocommutative Hopf operads. This shows that the tensor product of two $n$-algebras is in a natural way an $n$-algebra. On the other hand, the Lie operad $\mathcal{L}$ is not a Hopf operad: the tensor product of two Lie algebras is not in general a Lie algebra.

**6.2. The Künneth theorem.** The augmented tensor product $A \boxtimes B$ of two commutative algebras is their coproduct in the category of commutative algebras. This allows us to calculate the André-Quillen homology of $A \boxtimes B$ using the following result.

**Proposition 6.5.** *If $\mathbf{a}$ is an exact dg-operad and $A_1$ and $A_2$ are $\mathbf{a}$-algebras, there is a natural weak equivalence from $\mathbb{B}(\mathbf{a}, A_1 \coprod A_2)$ to $\mathbb{B}(\mathbf{a}, A_1) \oplus \mathbb{B}(\mathbf{a}, A_2)$.*

*Proof.* The result is true for almost free algebras, so it suffices to resolve $A_1$ and $A_2$ by almost free algebras and observe that $\mathbb{B}(\mathbf{a})$ is a homotopy functor. $\square$

For $n < \infty$, the calculation of the $n$-homology of an augmented tensor product is more complicated. We now show how, given two almost free $n$-algebras $\mathsf{T}_n(V_1, d_1)$ and $\mathsf{T}_n(V_2, d_2)$, their augmented tensor product $\mathsf{T}_n(V_1, d_1) \boxtimes \mathsf{T}_n(V_2, d_2)$ has a natural resolution by an almost free $n$-algebra $\mathsf{T}_n(V_1 \oplus V_2 \oplus \Sigma^n(V_1 \otimes V_2), d)$.

Denote the element $\Sigma^n(v_1 \otimes v_2)$ of $\Sigma^n(V_1 \otimes V_2)$ by $(v_1, v_2)$. There is a unique extension of $(-,-)$ to a bilinear operation on $\mathsf{T}_n(V_1 \oplus V_2 \oplus \Sigma^n(V_1 \otimes V_2))$ satisfying the following conditions:

(1) $(a, b) = 0$ if both $a$ and $b$ lie in $V_1$ or $V_2$;
(2) $(a, b) = 0$ if $a$ or $b$ lie in $\Sigma^n(V_1 \otimes V_2)$;
(3) $(a, b) = (-1)^{|a||b|+n}(b, a)$;
(4) for each $a \in \mathsf{T}_n(V_1 \oplus V_2 \oplus \Sigma^n(V_1 \otimes V_2))$, the map $b \mapsto (a, b)$ is an $n$-derivation of $\mathsf{T}_n(V_1 \oplus V_2 \oplus \Sigma^n(V_1 \otimes V_2))$: that is,

$$(a, bc) = (a, b)c + (-1)^{(|a|+n)|b|}b(a, c),$$
$$(a, \{b, c\}) = \{(a, b), c\} + (-1)^{(|a|+n)|b|}\{b, (a, c)\}.$$

Let $d$ be the $n$-derivation on $\mathsf{T}_n(V_1 \oplus V_2 \oplus \Sigma^n(V_1 \otimes V_2))$ such that $dv_i = d_i v_i$ for $v_i \in V_i$, and

$$d(v_1, v_2) = \{v_1, v_2\} + (-1)^n(d_1 v_1, v_2) + (-1)^{|v_1|+n}(v_1, d_2 v_2).$$

**Lemma 6.6.** *$d$ is an $n$-differential: $d^2 = 0$.*

*Proof.* By induction on the length of $a$ and $b$, one may verify that the above formula for $d(a, b)$ continues to hold for all $a, b \in \mathsf{T}_n(V_1 \oplus V_2 \oplus \Sigma^n(V_1 \otimes V_2))$. It is then straightforward to show that $d^2(a, b) = 0$: it is true on words of length one, since $d_1^2 = d_2^2 = 0$, and the induction step follows from the calculation

$$\begin{aligned} d^2(a, b) &= d\big(\{a, b\} + (-1)^n(da, b) + (-1)^{|a|+n}(a, db)\big) \\ &= (-1)^{n-1}\big(\{da, b\} + (-1)^{|a|}\{a, db\}\big) \\ &\quad + (-1)^n\big(\{da, b\} + (-1)^{|a|+n+1}(da, db)\big) \\ &\quad + (-1)^{|a|+n}\big(\{a, db\} + (-1)^n(da, db)\big) = 0. \quad \square \end{aligned}$$

**Theorem 6.7.** *The map $\mathsf{T}_n(V_1 \oplus V_2 \oplus \Sigma^n(V_1 \otimes V_2)) \to \mathsf{T}_n(V_1) \boxtimes \mathsf{T}_n(V_2)$ defined by sending the elements $(v_1, v_2)$ and $\{v_1, v_2\}$, $v_i \in V_i$, to zero, is a weak equivalence.*

*Proof.* There is an isomorphism of chain complexes $\mathsf{T}_n(V) \cong \mathsf{T}_\infty(\Sigma^{1-n}\mathsf{T}_\mathcal{L}(\Sigma^{n-1}V))$ by Corollary 1.7. Denoting $\Sigma^{n-1}V_i$ by $W_i$, we see that it suffices to prove that the map

$$\mathsf{T}_\mathcal{L}(W_1 \oplus W_2 \oplus \Sigma(W_1 \otimes W_2), d) \to \mathsf{T}_\mathcal{L}(W_1, d_1) \oplus \mathsf{T}_\mathcal{L}(W_2, d_2)$$

is a weak equivalence, where the $\mathcal{L}$-differentials $d_1$, $d_2$ and $d$ are induced by the corresponding $n$-differentials.

The chain complex $\mathsf{T}_\mathcal{L}(W_1 \oplus W_2 \oplus \Sigma(W_1 \otimes W_2), d)$ is a direct sum of three subcomplexes

$$\mathsf{T}_\mathcal{L}(W_1 \oplus W_2 \oplus \Sigma(W_1 \otimes W_2), d) = \mathsf{T}_\mathcal{L}(W_1, d_1) \oplus \mathsf{T}_\mathcal{L}(W_2, d_2) \oplus Z,$$

where $Z$ is spanned by words $\mathrm{ad}(a_k)\ldots\mathrm{ad}(a_2)a_1$ in which either some $a_i \in \Sigma(W_1 \otimes W_2)$ or there exist $1 \leq i, j \leq k$ such that $a_i \in V_1$ and $a_j \in V_2$. We will show that the subcomplex $Z$ is contractible.

We define a descending filtration on $Z$ as follows. A word $[a_k, \ldots, [a_2, a_1] \ldots]$ lies in $F^i$ if one of the following conditions hold:

(1) $a_1, \ldots, a_i \in W_1$;
(2) $a_1, \ldots, a_i \in W_2$;
(3) $a_1, \ldots, a_{i-1} \in W_1$ and $a_i \in \Sigma^{-1}(W_1 \otimes W_2)$;
(4) $a_1, \ldots, a_{i-1} \in W_2$ and $a_i \in \Sigma^{-1}(W_1 \otimes W_2)$.

It is clear that $d$ preserves this filtration. Define an operator $h$ of degree 1 on $E^0 Z$ by

$$h[a_k, \ldots, [a_{i+1}, [a_i, \ldots [a_2, a_1] \ldots] = (-1)^{|a_k|+\cdots+|a_{i+2}|}[a_k, \ldots, [(a_{i+1}, a_i), \ldots [a_2, a_1] \ldots]$$

if $a_1, \ldots, a_i \in W_2$ and $a_{i+1} \in W_1$ or $a_1, \ldots, a_i \in W_1$ and $a_{i+1} \in W_2$, while $h$ vanishes on other words in $Z$. It is easily checked that $d^0 h + hd^0$ equals the identity, where $d^0$ is the differential on $E^0$: thus the complex $E^0 Z$ is contractible. Since the spectral sequence associated to the filtration $F^i$ is convergent, the lemma follows $\square$

**Corollary 6.8.** *If $A_1$ and $A_2$ are $n$-algebras, there is a homotopy equivalence*

$$\mathsf{B}_n(A_1 \boxtimes A_2) \simeq \mathsf{B}_n(A_1) \oplus \mathsf{B}_n(A_2) \oplus \Sigma^n(\mathsf{B}_n(A_1) \otimes \mathsf{B}_n(A_2)).$$

*Proof.* The homotopy type of each side is not changed if $A_i$ is replaced by its almost free resolution $\Omega_n(\mathsf{B}_n(A_i))$: the result is then clear, since our results give an almost free resolution of $\Omega_n(\mathsf{B}_n(A_1)) \otimes \Omega_n(\mathsf{B}_n(A_2))$ by an almost free algebra of the form $\mathsf{T}_n(\mathsf{B}_n(A_1) \oplus \mathsf{B}_n(A_2) \oplus \Sigma(\mathsf{B}_n(A_1) \otimes \mathsf{B}_n(A_2)), d)$. $\square$

This formula for $\mathsf{B}_n(A_1 \boxtimes A_2)$ may be rewritten

$$\Sigma^n \mathsf{B}_n(A_1 \boxtimes A_2) \simeq \Sigma^n \mathsf{B}_n(A_1) \boxtimes \Sigma^n \mathsf{B}_n(A_2).$$

The Künneth theorem for Hochschild homology

$$H_\bullet(A_1 \boxtimes A_2) \cong H_\bullet(A_1) \boxtimes H_\bullet(A_2),$$

may be viewed as the limiting case $n = 1$, since $\Sigma \mathsf{B}_1(A)$ is the bar complex of the associative algebra $A$.

**6.3. The $n$-homology of commutative algebras.** In this section, we calculate the $n$-homology of a commutative algebra, thought of as an $n$-algebra.

**Theorem 6.9.** *If $A$ is a commutative algebra, the $n$-fold suspension $\Sigma^n \mathsf{B}_n(A)$ of its $n$-bar complex is a commutative co-Hopf $\mathrm{e}_n^*$-coalgebra, and there is an isomorphism of commutative dg-algebras $\Sigma^n \mathsf{B}_n(A) \simeq \mathsf{T}_\infty(\Sigma^n \mathsf{B}_\infty(A))$.*

*Proof.* For $n = 1$, this theorem is a restatement of the results recalled in Section 5.3: the shuffle product gives the Hochschild bar coalgebra of a commutative algebra the structure of a differential graded commutative Hopf algebra.

We now turn to the case $n > 1$. We start by defining a commutative product on $\Sigma^n \mathsf{C}_n(V) \cong \mathsf{C}(\mathbf{e}_n^*, \Sigma^n V)$. By Corollary 1.7 and the isomorphism of functors $\mathsf{C}_\infty(V) \cong \Sigma^{-1}\mathsf{S}(\mathcal{L}^*, \Sigma V)$, there is a natural equivalence of functors
$$\Sigma^n \mathsf{C}_n(V) \cong \mathsf{T}_\infty(\Sigma^n \mathsf{C}_\infty(V)).$$
It is easily checked that the product defined on $\Sigma^n \mathsf{C}_n(V)$ by identifying it with the free commutative algebra $\mathsf{T}_\infty(\Sigma^n \mathsf{C}_\infty(V))$ makes it into a commutative co-Hopf $\mathbf{e}_n^*$-coalgebra.

It remains to prove that if $A$ is a commutative algebra, the differential $d$ of the bar complex $\mathsf{B}_n(A) = \mathsf{C}_n(A, d)$ induces a derivation of the commutative product on $\Sigma^n \mathsf{B}_n(A)$. In fact, if $A$ is a commutative algebra, there is actually an isomorphism of complexes
$$\Sigma^n \mathsf{B}_n(A) \cong \mathsf{T}_\infty(\Sigma^n \mathsf{B}_\infty(A)),$$
since the terms of the differential of $\Sigma^n \mathsf{B}_n(A)$ which do not occur in the differential of $\mathsf{T}_\infty(\Sigma^n \mathsf{B}_\infty(A))$ all involve the Lie bracket of the $n$-algebra $A$, which vanishes if $A$ is a commutative algebra. This shows that the differential of $\Sigma^n \mathsf{B}_n(A)$ is the coderivation associated to a linear map $d : \Sigma^n \mathsf{B}_\infty(A) \to A$, and is thus a derivation of the commutative product in $\Sigma^n \mathsf{B}_n(A)$. □

In the remainder of this section, we present another proof of this theorem, by constructing a natural almost free resolution in the category of $n$-algebras of an almost free commutative algebra. Given an almost free commutative algebra $\mathsf{T}_\infty(V, d)$ such that $V$ is finite dimensional, let $W$ be the graded vector space $\Sigma^{-n}\mathsf{T}_\infty(\Sigma^n V)$. We will show that there is an $n$-differential $\mathcal{D}$ on the free $n$-algebra $\mathsf{T}_n(W)$ such that the almost free $n$-algebra $\mathsf{T}_n(W, \mathcal{D})$ is a resolution of the $n$-algebra $\mathsf{T}_\infty(V, d)$.

Let $\{v_i \mid 1 \leq i \leq k\}$ be a homogeneous basis of $V$, such that $|v_i| = k_i$. If $\alpha$ is a multi-index $(\alpha_1, \ldots, \alpha_k)$, denote by $v_\alpha$ the element $\Sigma^{-n}\big((\Sigma^n v_1)^{\alpha_1} \ldots (\Sigma^n v_k)^{\alpha_k}\big)$ of $W$, of degree
$$|v_\alpha| = \sum_{i=1}^k \alpha_i(k_i + n) - n.$$
(Here, $\alpha_i$ is an arbitrary natural number if $k_i + n$ is even, and is 0 or 1 if $k_i + n$ is odd.) Denote by $\{x^i\}$ the basis of $\Sigma^n V^*$ dual to $\{v_i\}$, with $|x^i| = -k_i - n$. We introduce a generating function $v(x^1, \ldots, x^k)$ for the basis $\{v_\alpha \mid |\alpha| > 0\}$ of $W$ by the formula
$$v(x^1, \ldots, x^k) = \sum_{|\alpha| > 0} x^\alpha v_\alpha;$$
here, we use the canonical pairing between the vector spaces $V$ and $\Sigma^n V^*$. Thus, $v$ is an element of $W \otimes \mathsf{T}_\infty(\Sigma^n V^*)$ of degree $-n$. The cocommutative coproduct of $W$, of degree $-n$, is given by the formula
$$\Delta v = v \otimes v.$$

Let $f_i(v_1, \ldots, v_k)$ be the polynomial, with $|f_i| = k_i - 1$, in the basis $\{v_i\}$ such that
$$dv_i = f_i(v_1, \ldots, v_k).$$
With the notation
$$f_{i,j}(v_1, \ldots, v_k) = \frac{\partial f_i(v_1, \ldots, v_k)}{\partial v_j},$$
the formula $d^2 v_i = 0$ is equivalent to
$$\sum_{j=1}^k f_j f_{i,j} = 0.$$

Observe that $|f_{i,j}| = k_i - k_j - 1$.

Let $v_i$ be the partial derivative of the generating function $v$ with respect to $x^i$

$$v_i(x^1, \ldots, x^k) = \frac{\partial v}{\partial x^i}.$$

We see that $v_i(0)$ may be identified with the basis vector $v_i$ of $V$: thus, the partial derivatives $v_i(x^1, \ldots, x^k)$ of $v$ are a kind of prolongation of $v_i(0)$.

We now define an $n$-derivation $\mathcal{D}$ of the free $n$-algebra $\mathsf{T}_n(W)$, by the formula

$$(*) \qquad \mathcal{D}v + \tfrac{1}{2}\{v,v\} = \sum_{i=1}^{k} (-1)^{k_i + n} x^i f_i(v_1, \ldots, v_k).$$

Since $v(0) = 0$, this formula allows us to define $\mathcal{D}v_\alpha$ inductively, as a function of the generators $v_\beta$ with $|\beta| < |\alpha|$.

**Theorem 6.10.** *The $n$-derivation $\mathcal{D}$ is a differential: $\mathcal{D}^2 v = 0$. The map from $\mathsf{T}_n(W, \mathcal{D})$ to $\mathsf{T}_\infty(V, d)$ defined by sending $v_i$ to $v_i$, and $v_\alpha$ to $0$ for $|\alpha| > 1$, intertwines the differentials $\mathcal{D}$ and $d$. The almost free $n$-algebra $\mathsf{T}_n(W, \mathcal{D})$ is a resolution of the $n$-algebra $\mathsf{T}_\infty(V, d)$.*

*Proof.* We start by proving that the $n$-derivation $\mathcal{D}$ is a differential. The following formula is a simple exercise in the chain rule:

$$\begin{aligned}\mathcal{D}v_j &= -\{v, v_j\} + f_j(v_1, \ldots, v_k) \\ &\quad + \sum_{p=1}^{k}\sum_{q=1}^{k}(-1)^{(n-k_j)(n-k_p)+(n-k_j)+(n-k_p)} x^p \partial_j \partial_q v f_{p,q}.\end{aligned}$$

Evaluating at $0$, we see that $\mathcal{D}v_j(0) = f_j(v_1(0), \ldots, v_k(0))$. This shows that the map from $\mathsf{T}_n(W, \mathcal{D})$ to $\mathsf{T}_\infty(V, d)$ defined by sending $v_i(0)$ to $v_i$ and $v_\alpha$ to $0$ for $|\alpha| > 1$ intertwines $\mathcal{D}$ and $d$.

Applying $\mathcal{D}$ to formula $(*)$ defining $\mathcal{D}$, we see that

$$\mathcal{D}^2 v = \{v, \mathcal{D}v\} + \sum_{i,j=1}^{k} x^i \mathcal{D}v_j f_{i,j}(v_1, \ldots, v_k).$$

Inserting the formulas for $\mathcal{D}v$ and $\mathcal{D}v_j$ into $[v, \mathcal{D}v]$, we see that

$$\begin{aligned}\mathcal{D}^2 v &= -\tfrac{1}{2}\{v, \{v,v\}\} + \sum_{i=1}^{k} x^i \{v, f_i(v_1, \ldots, v_k)\} \\ &\quad - \sum_{i,j=1}^{k} x^i \{v, v_j\} f_{i,j}(v_1, \ldots, v_k) \\ &\quad + \sum_{i,j=1}^{k} x^i f_j(v_1, \ldots, v_k) f_{i,j}(v_1, \ldots, v_k) \\ &\quad + \sum_{i,j,p,q=1}^{k} (-1)^{(k_j+n)(k_p+n)+(k_j+n)+(k_p+n)} x^i x^p \partial_j \partial_q v f_{p,q}(v_1, \ldots, v_k) f_{i,j}(v_1, \ldots, v_k)\end{aligned}$$

The term proportional to $\{v, \{v,v\}\}$ vanishes by the Jacobi identity. The next two terms cancel each other by the Poisson relation. The penultimate term vanishes, by the identity $\sum_{j=1}^{k} f_j f_{i,j} = 0$. The last term vanishes because, on exchanging the indices $(i,j)$ with $(p,q)$, it changes sign. This completes the proof that $\mathcal{D}^2 v = 0$.

It remains to show that the map $\mathsf{T}_n(W, \mathcal{D}) \to \mathsf{T}_\infty(V, d)$ is a weak equivalence. Filter the $n$-algebra $\mathsf{T}_n(W, \mathcal{D})$ by assigning to the generator $v_\alpha$ filtration degree $|\alpha|$. The $E^0$-term of the associated spectral sequence may be identified with the $n$-algebra $\mathsf{T}_n(W, \mathcal{D}_0)$ associated to the free commutative algebra $\mathsf{T}_\infty(V)$, where $\mathcal{D}_0$ is given explicitly by the formula

$$\mathcal{D}_0 v + \tfrac{1}{2}\{v,v\} = \sum_{i=1}^{k}\sum_{j=1}^{k}(-1)^{k_i+n} v_j f_{i,j}(0).$$

Thus, we see that it suffices to show that the map $\mathsf{T}_n(W, \mathcal{D}_0) \to \mathsf{T}_\infty(V)$ is a weak equivalence in the case of a free commutative algebra $\mathsf{T}_\infty(V)$.

Recall that the cooperad $\mathcal{L}^\perp$ dual to the Lie operad $\mathcal{L}$ is isomorphic to $\Lambda^{-1}\mathbf{e}_\infty^*$; in other words, $\mathcal{L}^\perp$-coalgebras are the suspensions of cocommutative coalgebras. By Corollary 1.7, the graded vector space $\mathsf{T}_n(W)$ may be rewritten as

$$\begin{aligned}\mathsf{T}_n(W) &\cong \mathsf{T}_\infty(\Sigma^{1-n}\mathsf{T}_\mathcal{L}(\Sigma^{n-1}W))\\ &\cong \mathsf{T}_\infty(\Sigma^{-n}\mathsf{T}_\mathcal{L}(\mathsf{C}(\mathbf{e}_\infty^*,\Sigma^n V)))\\ &\cong \mathsf{T}_\infty(\Sigma^{1-n}\mathsf{T}_\mathcal{L}(\mathsf{C}(\mathcal{L}^\perp,\Sigma^{n-1}V))),\end{aligned}$$

and the map from $\mathsf{T}_n(W)$ to $\mathsf{T}_n(V)$ is the result of applying the free commutative algebra functor $\mathsf{T}_\infty(-)$ to the map

$$\Sigma^{1-n}\mathsf{T}_\mathcal{L}(\mathsf{C}(\mathcal{L}^\perp,\Sigma^{n-1}V)) \to V.$$

The $n$-differential $\mathcal{D}_0$ on $\mathsf{T}_n(W)$ is induced by a differential on $\mathsf{T}_\mathcal{L}(\mathsf{C}(\mathcal{L}^\perp,\Sigma^{n-1}V))$, which we may also denote by $\mathcal{D}_0$, and it suffices to show that the map

$$\mathsf{T}_\mathcal{L}(\mathsf{C}(\mathcal{L}^\perp,\Sigma^{n-1}V),\mathcal{D}_0) \to \Sigma^{n-1}V$$

is a weak equivalence. The differential $\mathcal{D}_0$ may be identifed with the negative of the differential in the cobar complex $\Omega(\mathcal{L}, \mathsf{C}(\mathcal{L}^\perp, \Sigma^{n-1}V))$. The weak equivalence $\mathsf{T}_\mathcal{L}(\mathsf{C}(\mathcal{L}^\perp,\Sigma^{n-1}V)) \simeq \Sigma^{n-1}V$ follows from the fact that the Lie operad $\mathcal{L}$ is Koszul. $\square$

Let us write out the almost free $n$-algebra $\mathsf{T}_n(W, \mathcal{D})$ more explicitly in the simplest case, where there is only one generator $u$, of degree $k$. The vector space $\Sigma^n V^*$ is spanned by a single element $x$ of degree $-k-n$. Let us consider the two cases $-k-n$ even and $-k-n$ odd separately.

If $-k-n$ is even, we may write

$$v(x) = \sum_{i=0}^{\infty} x^i v_{(i)},$$

where $|v_{(i)}| = i(k+n) - n$. Taking the coefficient of $x^i$ in $(*)$, we see that

$$\mathcal{D}v_{(i)} + \tfrac{1}{2}\sum_{k=1}^{i-1}\{v_{(k)}, v_{(i-k)}\} = 0.$$

If $-k-n$ is odd, the situation is much simpler, since then $W$ is isomorphic to $V$, the differential $\mathcal{D}$ vanishes, and the map $\mathsf{T}_n(W) \to \mathsf{T}_\infty(V)$ is an isomorphism.

**6.4. The iterated integral map for $\Omega^2 M$.** In this section, we will realize the homotopy equivalence

$$\Sigma^2 \mathsf{B}_2(\mathcal{A}^\bullet(M)) \to \mathcal{A}^\bullet(\Omega^2 M)$$

by a map of chain complexes. As explained in the introduction, the search for such an iterated integral map motivated the construction of the bar complex $\mathsf{B}_n(A)$ of $n$-algebras, and the proof that it realizes the André-Quillen homology of $A$ in the category of $n$-algebras.

Let us write the differential of $\mathsf{B}_n(A)$ more explicitly. If $A$ is an $n$-algebra, the complex underlying $\Sigma^n \mathsf{C}_n(A)$ may be identified with the free $\mathbf{e}_n^*$-coalgebra generated by $\Sigma^n A$:

$$\Sigma^n \mathsf{C}_n(A) \cong H^\bullet(\mathbb{F}_n(k)) \otimes_{\mathbb{S}_k} (\Sigma^n A)^{(k)}.$$

Given $\omega \in H^\bullet(\mathbb{F}_n(k))$ and $a_i \in \Sigma^n A$, denote the element

$$\omega \otimes_{\mathbb{S}_k} (\Sigma^n a_1 \otimes \ldots \Sigma^n a_k)$$

by $(\omega | a_1, \ldots, a_k)$.

Given a map $\pi : \{1, \ldots, k\} \to T$, where $T$ is a finite set, let $\rho_\pi : \mathbf{e}_n^*(k) \to \mathbf{e}_n^*(k)$ be the homomorphism of algebras such that $\rho_\pi(\omega_{ij}) = \omega_{ij}$ if $\pi(i) \neq \pi(j)$ and zero otherwise. Let $\delta_\pi : \mathbf{e}_n^*(k) \to \mathbf{e}_n^*(k)$ be the derivation of algebras over $\rho_\pi$ such that $\delta_\pi \omega_{ij} = 1$ if $\pi(i) = \pi(j)$ and zero otherwise. These definitions are consistent with the relations $\omega_{ij}^2 = 0$ (if $n$ is odd) and

$$\omega_{ij} \wedge \omega_{jk} + \omega_{jk} \wedge \omega_{ki} + \omega_{ki} \wedge \omega_{ij} = 0;$$

this may be seen by examining individually the three cases in which the cardinality of the set $\{\pi(i), \pi(j), \pi(k)\}$ is 1, 2 or 3.

Denote by $\rho_{ij}$ and $\delta_{ij}$ the special case of these maps where $\pi$ is the quotient map which identifies $\{i, j\} \subset \{1, \ldots, k\}$. If $A$ is an $n$-algebra, the differential of $\mathsf{B}_n(A)$ may be identified with the following differential on $\Sigma^n \mathsf{C}_n(A)$:

$$\begin{aligned}
d(\omega | a_1, \ldots, a_k) &= \sum_{i=1}^k (-1)^{|\omega| + \varepsilon_{i-1}} (\omega | a_1, \ldots, da_i, \ldots, a_k) \\
&+ \sum_{1 \leq i < j \leq k} (-1)^{|\omega| + \varepsilon_{i-1} + (\varepsilon_{j-1} - \varepsilon_i)(|a_j| + n)} (\delta_{ij} \omega | a_1, \ldots, a_i a_j, \ldots, \widehat{a}_j, \ldots, a_k) \\
&+ \sum_{1 \leq i < j \leq k} (-1)^{|\omega| + \varepsilon_{i-1} + (\varepsilon_{j-1} - \varepsilon_i)(|a_j| + n)} (\rho_{ij} \omega | a_1, \ldots, \{a_i, a_j\}, \ldots, \widehat{a}_j, \ldots, a_k),
\end{aligned}$$

where $\varepsilon_i = |a_1| + \cdots + |a_i| + in$.

Following Arnold, we may identify $\mathbf{e}_2^*(k)$ with the subalgebra of the differential forms on the configuration space $\mathbb{F}_2(k)$ generated by the one-forms

$$\omega_{ij} = \frac{1}{2\pi i} \frac{dz_i - dz_j}{z_i - z_j}, \quad 1 \leq i, j \leq k.$$

Denote by $\mathcal{A}_\bullet(\mathbb{C}^k)$ the complex of currents on the manifold $\mathbb{C}^k$: $\mathcal{A}_i(\mathbb{C}^k)$ is the dual of the space $\mathcal{A}_c^i(\mathbb{C}^k)$ of compactly supported $i$-forms while the boundary $d^*$ is the adjoint of the differential on $\mathcal{A}_c^\bullet(\mathbb{C}^k)$:

$$\langle d^* \mu, \alpha \rangle = (-1)^{|\mu|} \langle \mu, d\alpha \rangle.$$

As a first step in defining iterated integrals for $\Omega^2 M$, we construct a sequence of $\mathbb{S}_k$-equivariant maps

$$\mathrm{pv} : \mathbf{e}_2^*(k) \cong H^\bullet(\mathbb{F}_2(k), \mathbb{C}) \to \Sigma^{-2k} \mathcal{A}_\bullet(\mathbb{C}^k);$$

these may be interpreted as the principal values of the Arnold forms.

If $\varepsilon > 0$, let $\mathbb{C}_\varepsilon^k$ be the closed subset of $\mathbb{C}^k$

$$\mathbb{C}_\varepsilon^k = \{(z_1, \ldots, z_k) \mid |z_i - z_j| \geq \varepsilon \text{ for all } i, j\}.$$

Given $\omega \in \mathbf{e}_2^*(k)$, we now define the current $\mathrm{pv}\,\omega \in \mathcal{A}_\bullet(\mathbb{C}^k)$ as the limit

$$\langle \mathrm{pv}\,\omega, \alpha \rangle = \lim_{\varepsilon \to 0} \int_{\mathbb{C}_\varepsilon^k} \omega \wedge \alpha, \quad \text{for all } \alpha \in \mathcal{A}^\bullet(\mathbb{C}^k).$$

It is clear that this map is equivariant for the action of the symmetric group $\mathbb{S}_k$.

**Lemma 6.11.** $d^*(\mathrm{pv}\,\omega) = \sum_{1 \leq i < j \leq k} \mathrm{pv}(\delta_{ij} \omega)$

*Proof.* By Stokes's theorem,

$$\langle d^*(\operatorname{pv}\omega), \alpha\rangle = (-1)^{|\omega|}\langle \operatorname{pv}\omega, d\alpha\rangle$$
$$= \lim_{\varepsilon\to 0}\int_{\mathbb{C}_\varepsilon^k}(-1)^{|\omega|}\omega\wedge d\alpha$$
$$= \lim_{\varepsilon\to 0}\int_{\partial\mathbb{C}_\varepsilon^k}\omega\wedge\alpha.$$

The boundary $\partial\mathbb{C}_\varepsilon^k$ is a manifold with corners, whose interior is a disjoint union

$$\bigcup_{1\leq p\leq q\leq k}\{(z_1,\ldots,z_k)\mid |z_p - z_q| = \varepsilon,\text{ and }|z_i - z_j| > \varepsilon\text{ for }i,j\text{ such that }\{i,j\}\neq\{p,q\}\}.$$

Let $\partial_{pq}\mathbb{C}_\varepsilon^k$ be the set $\{(z_1,\ldots,z_k)\in\mathbb{F}_2(k)\mid |z_p - z_q|=\varepsilon\}$. Then

$$\int_{\partial\mathbb{C}_\varepsilon^k}\omega\wedge\alpha = \sum_{1\leq p,q\leq k}\int_{\partial_{pq}\mathbb{C}_\varepsilon^k}\omega\wedge\alpha + O(\varepsilon),$$

and we see that

$$\langle d^*(\operatorname{pv}\omega),\alpha\rangle = \sum_{1\leq p,q\leq k}\lim_{\varepsilon\to 0}\int_{\partial_{pq}\mathbb{C}_\varepsilon^k}\omega\wedge\alpha.$$

But it is easily seen that

$$\lim_{\varepsilon\to 0}\int_{\partial_{pq}\mathbb{C}_\varepsilon^k}\omega\wedge\alpha = \langle \operatorname{pv}(\delta_{pq}\omega),\alpha\rangle. \quad\square$$

Consider the diagram

$$\begin{array}{ccc}\mathbb{C}^k\times\Omega^2 M & \xrightarrow{e(k)} & M^k \\ {\scriptstyle\pi}\big\downarrow & & \\ \Omega^2 M & & \end{array}$$

where $e(k)$ is the evaluation map $e(k)(z_1,\ldots,z_k,\gamma) = (\gamma(z_1),\ldots,\gamma(z_k))$, while $\pi$ is projection along the second factor. Pulling back along the proper map $e(k)$ gives a linear map

$$\mathcal{A}^\bullet(M)^{(k)} \to \mathcal{A}^\bullet(M^k) \xrightarrow{e(k)^*} \mathcal{A}^\bullet(\mathbb{C}^k\times\Omega^2 M),$$

where the first map sends a differential form $\alpha_1\otimes\ldots\otimes\alpha_k\in\mathcal{A}^\bullet(M)^{(k)}$ to the external product $\alpha_1\boxtimes\cdots\boxtimes\alpha_k\in\mathcal{A}^\bullet(M^k)$. Given $\omega\in\mathbf{e}_2^*(k)$, $\rho(\omega|\alpha_1,\ldots,\alpha_k)$ is defined to be the differential form on $\Omega^2 M$ defined by integrating $e(k)^*(\alpha_1\boxtimes\cdots\boxtimes\alpha_k)$ against the current $\operatorname{pv}(\omega)\in\mathcal{A}_\bullet(\mathbb{C}^k)$ along the fibres of $\pi$. In this way, we obtain an $\mathbb{S}_k$-equivariant map

$$\mathbf{e}_2^*(k)\otimes\mathcal{A}^\bullet(M)^{(k)} \to \Sigma^{-2k}\mathcal{A}^\bullet(\Omega^2 M).$$

Summing over $k$, we obtain a map

$$\rho : \Sigma^2\mathsf{C}_2(\mathcal{A}^\bullet(M)) = \bigoplus_{k=1}^\infty \mathbf{e}_2^*(k)\otimes_{\mathbb{S}_k}\Sigma^2\mathcal{A}^\bullet(M)^{(k)} \to \mathcal{A}^\bullet(\Omega^2 M).$$

Now, $\mathcal{A}^\bullet(M)$ is a 2-algebra with vanishing bracket. Using Lemma 6.11, we obtain the following result.

**Lemma 6.12.** *The iterated integral $\rho : \Sigma^2\mathsf{B}_2(\mathcal{A}^\bullet(M)) \to \mathcal{A}^\bullet(\Omega^2 M)$ is a map of commutative dg-algebras.*

We may now prove the following theorem.

**Theorem 6.13.** *If $M$ is a 2-connected manifold, the iterated integral map $\rho$ is a weak equivalence (of commutative co-Hopf 2-algebras).*

*Proof.* Since $M$ is a 2-connected manifold, there is a minimal model $\mathsf{T}_\infty(V,d) \hookrightarrow \mathcal{A}^\bullet(M)$ for its algebra of differential forms, that is, a resolution of $\mathcal{A}^\bullet(M)$ by an almost free commutative algebra $\mathsf{T}_\infty(V,d)$ generated by a graded vector space $V$, such that $V_i = 0$ for $i > -2$. It is easy to see that the diagram

$$\begin{array}{ccc} \Sigma^2 V & \xrightarrow{\simeq} & \mathsf{B}_\infty(\Sigma^2 \mathsf{B}_2(\mathcal{A}^\bullet(M))) \\ \Big\| & & \Big\downarrow{\mathsf{B}_\infty(\rho)} \\ \Sigma^2 V & \xrightarrow{\simeq} & \mathsf{B}_\infty(\mathcal{A}^\bullet(\Omega^2 M)) \end{array}$$

commutes, proving that $\mathsf{B}_\infty(\rho)$ is a weak equivalence. To prove that $\rho$ is a weak equivalence, we invoke the Whitehead-type theorem, that if $\mathsf{B}_\infty(\rho)$ is a weak equivalence, then so is $\rho$: this is proved by applying the functor $\Omega_\infty$, since $\Omega_\infty \mathsf{B}_\infty$ is weakly equivalent to the identity. □

DEPT. OF MATHEMATICS, MIT, CAMBRIDGE, MASS. 02139 USA

*E-mail*: getzler@math.mit.edu

MATHEMATICS INSTITUTE, UNIVERSITY OF WARWICK, COVENTRY CV4 7AL, ENGLAND

*E-mail*: jdsj@maths.warwick.ac.uk